\documentclass[preprint]{aastex6}

\usepackage{amsfonts}
\usepackage[utf8x]{inputenc}
\usepackage{graphicx}	%load this before "epstopdf"
\usepackage{epstopdf}
\usepackage{bm}
\usepackage{amsmath}
\usepackage{natbib}	% use Natbib bibliography
\usepackage{xcolor}
\usepackage{caption}
\usepackage[FIGTOPCAP]{subfigure}

\newcommand{\Rm}{\mathrm}

\begin{document}
	\title{Multi-fluid approach to high-frequency waves in plasmas. II. Small-amplitude regime in partially ionized media}
	\shorttitle{High-frequency waves in partially ionized plasmas}
	\shortauthors{Martínez-Gómez et al.}
	
	\author{David Martínez-Gómez\altaffilmark{1,2}, Roberto Soler\altaffilmark{1,2}, and Jaume Terradas\altaffilmark{1,2}}
	\altaffiltext{1}{Departament de Física, Universitat de les Illes Balears, 07122, Palma de Mallorca, Spain}
	\altaffiltext{2}{Institut d'Aplicacions Computacionals de Codi Comunitari (IAC3), Universitat de les Illes Balears, 07122, Palma de Mallorca, Spain}
	\email{david.martinez@uib.es}
	
	\begin{abstract}
		The presence of neutral species in a plasma has been shown to greatly affect the properties of magnetohydrodynamic waves. For instance, the interaction between ions and neutrals through momentum transfer collisions causes the damping of Alfvén waves and alters their oscillation frequency and phase speed. When the collision frequencies are larger than the frequency of the waves, single-fluid MHD approximations can accurately describe the effects of partial ionization, since there is a strong coupling between the various species. However, at higher frequencies the single-fluid models are not applicable and more complex approaches are required. Here, we use a five-fluid model with three ionized and two neutral components, which takes into consideration Hall's current and Ohm's diffusion in addition to the friction due to collisions between different species. We apply our model to plasmas composed of hydrogen and helium and allow the ionization degree to be arbitrary. By means of the analysis of the corresponding dispersion relation and numerical simulations, we study the properties of small-amplitude perturbations. We discuss the effect of momentum transfer collisions on the ion-cyclotron resonances and compare the importance of magnetic resistivity, ion-neutral and ion-ion collisions on the wave damping at various frequency ranges. Applications to partially ionized plasmas of the solar atmosphere are performed. 
	\end{abstract}


\section{Introduction} \label{sec:intro}
	Many astrophysical and laboratory plasmas contain neutrals that produce an important deviation from the behavior predicted by the magnetohydrodynamic (MHD) theory for a fully ionized plasma \citep[see, e.g.,][]{1961CaJPh..39.1197W,1969ApJ...156..445K,1990ApJ...350..195P,2004PhPl...11.1358W}. For instance, as deduced by, e.g., \citet{1956MNRAS.116..314P} or \citet{1961CaJPh..39.1044W}, the presence of neutral species causes the damping of Alfvén waves \citep{1942Natur.150..405A}. Such damping has its origin in the exchange of momentum between ions and neutrals by means of collisions. Moreover, when the collision frequency is large compared to the oscillation frequency, the phase speed of Alfvén waves is reduced in comparison with the fully ionized case because the inertia of the neutrals has to be taken into account in addition to that of ions, as shown by, e.g., \citet{2003SoPh..214..241K}. Other circumstances where the interaction between the ionized and the neutral species may play an important role are, e.g., experiments involving discharge tubes \citep{1962JFM....13..570W,1962JFM....13..587J,1974PlPh...16..813M}, heating and gravitational collapse of interstellar gas clouds \citep{1977ApJ...213..705S,1982ApJ...263..696B}, magnetic reconnection \citep{1989ApJ...340..550Z}, driving of chromospheric spicules \citep{1992Natur.360..241H}, star formation \citep{1956MNRAS.116..503M,1992ApJ...391..199F,1993ApJ...415..680F,2008A&A...484....1P}, acceleration of solar wind \citep{1998JGR...103.6551A}, support of solar prominences against gravity \citep{2015ApJ...802L..28T}, solar coronal rain \citep{2016ApJ...818..128O} or heating of chromospheric jets due to Kelvin-Helmholtz vortices \citep{2016ApJ...830..133K}.
	
	In the research of how partial ionization affects the properties of MHD waves, it has been a common practice to resort to single-fluid models, like those used by \citet{1965RvPP....1..205B}, \citet{2001ApJ...558..859D}, \citet{2004A&A...422.1073K}, \citet{2007A&A...461..731F} or \citet{2009ApJ...699.1553S}. Such models assume that there is a strong coupling between all the components of the plasma and, typically, include the effects of partial ionization through additional terms in a generalized Ohm's law. The single-fluid approximation is appropriate when the frequency of the waves is lower than the collision frequencies between ionized and neutral particles. However, it becomes inaccurate when waves have frequencies of the order of or larger than the collision frequencies. In that range, each species may have a different dynamics and the single-fluid approach fails to properly describe the behavior of the waves in the plasma. Hence, a more general model is required.
	
	The multi-fluid theory \citep[see, e.g.,][]{1956MNRAS.116..114C,1969ApJ...156..445K,1973MitchnerKruger,2011A&A...529A..82Z,2014PhPl...21i2901K} treats several of the components of a plasma as separate fluids, independently of their coupling degree. Thus, it has a larger range of applicability in comparison with the single-fluid approximation and it can reveal some effects that are overlooked by the simpler models. For instance, by means of a two-fluid model,  \citet{1969ApJ...156..445K} and \citet{1990ApJ...350..195P} showed the existence of cutoffs for Alfvén waves excited by an impulsive driver in weakly ionized plasmas, although in a later work \citet{2015A&A...573A..79S} found that, due to the effect of Hall's current \citep{1960RSPTA.252..397L} and electron inertia, those are not strict cutoffs but regions where Alfvén waves are overdamped.
	
	Several versions of the multi-fluid theory have been applied to many different environments. For instance, two-fluid models that separate the plasmas in an ionized component and a neutral one have been extensively used in the investigation of waves in the solar atmosphere \citep[e.g.,][]{2009ApJ...707..662S,2013ApJS..209...16S} or in interstellar molecular clouds \citep{1996ApJ...465..775B,2011MNRAS.415.1751M}. A three-fluid approximation with two oppositely charged ionized fluids was employed by \cite{2008A&A...484....1P} in their investigation of the process of star formation and evolution of protostellar jets, while two different neutral components in solar plasmas have been considered by, e.g., \citet{2011A&A...534A..93Z}.
	
	In the present paper we improve the multi-fluid model described in \citet{2016ApJ...832..101M} (from now on, Paper I) for the case of fully ionized plasmas by adding the equations corresponding to two neutral components. Hence, we have here a five-fluid approximation that may consider the presence of up to three different ionized species plus two neutral fluids, which is particularly appropriate for the study of plasmas composed of hydrogen and helium. We assume that all the species interact with each other by means of momentum transfer collisions, while the ionized species also interact through electromagnetic fields. In Paper I, we checked that the inclusion of Hall's current is fundamental for the correct description of the properties of high-frequency waves in fully ionized plasmas: it produces a separation in the behavior of the two types of circular polarizations. But when partial ionization is considered, Hall's current may play an important role at much lower frequencies, as it was shown by \citet{2006astro.ph..8008P,2008MNRAS.385.2269P} for the case of weakly ionized plasmas. Another improvement that we make with respect to our previous work is the addition of the terms associated with electron collisions to the equations of momentum and to Ohm's law, thus taking into account the effects of resistivity.
	
	Although the model is presented in a way that can be applied to many diverse environments, we focus on the investigation of plasmas of the solar atmosphere, which are mainly composed of protons (which will be denoted by the subscript $p$), neutral hydrogen ($\Rm{H}$), neutral helium ($\Rm{He}$), singly ionized helium ($\Rm{He} \ \textsc{ii}$) and doubly ionized helium ($\Rm{He} \ \textsc{iii}$), in addition to electrons ($e$). We choose three specific regions of the solar atmosphere with different degrees of ionization: the higher chromosphere (where ions are more abundant than neutrals), a quiescent prominence (where the abundances of ions and neutrals are similar) and the lower chromosphere (which is weakly ionized).
	
	We study small-amplitude perturbations for a wide range of frequencies and two possible methods of excitation: by means of an impulsive driver and by means of a periodic one. In the first place, we explore the properties of the normal modes through the derivation and analysis of the corresponding dispersion relation. Then, we use an updated version of the numerical code MolMHD \citep{2009JCoPh.228.2266B} to compute the full temporal evolution of the perturbations. 
 
	Although we focus our attention on the study of waves that are in the linear regime, i.e., the amplitude of the velocity perturbations are much lower than the Alfvén speed, we also briefly examine a nonlinear effect: the heating of the plasma due to the frictional dissipation of the energy of the waves. The study of the energy balance in plasmas of the solar atmosphere is of great interest \citep{2012RSPTA.370.3217P,2015ASSL..415..157G,2015ASSL..415..103H} and the dissipation of MHD waves may play a significant role in the heating of those plasmas \citep{2015RSPTA.37340261A}. \citet{2006AdSpR..37..447K} compared several MHD wave damping mechanisms and found that for Alfvén waves the energy dissipation due to ion-neutral collisions is usually more important than other mechanisms like the damping due to viscosity or thermal conduction. Likewise, the relevance of collisional friction in the heating of the chromosphere or solar prominences has been studied by, e.g., \citet{2011ApJ...735...45G,2011JGRA..116.9104S,2012ApJ...747...87K,2013ApJ...777...53T,2015ApJ...810..146S,2016ApJ...817...94A} or \citet{2016A&A...592A..28S}. Since in the present paper we restrict ourselves to the analysis of small-amplitude perturbations, it is expected that the heating which may appear would not be of a large magnitude. In a future work we will apply our five-fluid model to the exploration of the role of nonlinear waves in the heating of partially ionized plasmas.
	
	The present paper is organized in a similar way as Paper I. To start with, we describe th model equations and highlight the main differences with those used in the fully ionized case of Paper I. Then, two different but complementary approaches are followed. First, the dispersion relation of linear waves is derived and its solutions are explored. Later, the full temporal evolution of the equations is solved numerically with the MolMHD code and compared with the predictions of the dispersion relation. 
	
\section{Description of the model} \label{sec:model}
	The 5-moment transport equations \citep{1977RvGSP..15..429S} that govern the mass, momentum and energy temporal evolution for each species of a multi-component plasma can be found in Section 2 of Paper I. Due to the inclusion of neutral species in the present investigation, the heat transfer term in the equation of energy is now expressed as
	\begin{equation} \label{eq:qterm}
		Q_{s}^{st} \equiv \frac{2 \alpha_{st}}{m_{s}+m_{t}}\left[\frac{A_{st}}{2}k_{\Rm{B}}\left(T_{t}-T_{s}\right)\Psi_{st}+\frac1{2}m_{t}\left(\bm{V_{t}}-\bm{V_{s}}\right)^2\Phi_{st}\right],
	\end{equation}
	where $A_{st}=4$ for collisions between electrons and neutral species and $A_{st}=3$ for the remaining types of collisions \citep{1986MNRAS.220..133D}, $\alpha_{st}=\alpha_{ts}$ is the friction coefficient for collisions between species $s$ and $t$; $m_{s}$, $T_{s}$ and $\bm{V_{s}}$ are the mass, temperature and velocity of species $s$, respectively, and $k_{\Rm{B}}$ is the Boltzmann constant. The functions $\Phi_{st}$ and $\Psi_{st}$ depend on the drift speed, $|\bm{V_{s}}-\bm{V_{t}}|$, and on the reduced thermal speed, $V_{therm} \equiv \sqrt{2k_{\Rm{B}}\left(m_{t}T_{s}+m_{s}T_{t}\right)/\left(m_{s}m_{t}\right)}$, and can be taken as equal to unity when the drift speed is much smaller than the thermal speed \citep{1977RvGSP..15..429S}. The friction coefficients have different expressions depending on whether the collisions involve only ionized species or also neutral species \citep{1965RvPP....1..205B,Callen2006a}. For the case of collisions between two ionized species $s$ and $t$, the friction coefficient is given by
	\begin{equation} \label{eq:friction_ions}
		\alpha_{st}=\frac{n_{s}n_{t}Z_{s}^2Z_{t}^2e^4\ln{\Lambda_{st}}}{6\pi\sqrt{2\pi}\epsilon_{0}^2m_{st}\left(k_{\Rm{B}}T_{s}/m_{s}+k_{\Rm{B}}T_{t}/m_{t}\right)^{3/2}},
	\end{equation} 
	where $n_{s}$ is the number density of species $s$, $e$ is the elementary electric charge, $Z_{s}$ is the signed charge number, $\epsilon_{0}$ is the vacuum electrical permittivity, and $m_{\Rm{st}} = m_{\Rm{s}}m_{t}/ \left(m_{s}+m_{t}\right)$ is the reduced mass. The function $\ln \Lambda_{st}$ is known as Coulomb's logarithm \citep[see, e.g.,][]{1962pfig.book.....S,2013AA...554A..22V} and is given by 
	\begin{equation}\label{eq:coulomb}
		\ln \Lambda_{st}=\ln\left[\frac{12\pi\epsilon_{0}^{3/2}k_{\Rm{B}}^{3/2}\left(T_{s}+T_{t}\right)}{|Z_{s}Z_{t}|e^3}\left(\frac{T_{s}T_{t}}{Z_{s}^2n_{s}T_{t}+Z_{t}^2n_{t}T_{s}}\right)^{1/2}\right].
	\end{equation} 
	For collisions between neutral species $n$ and another species $s$, that can be neutral or ionized, the friction coefficient is
	\begin{equation} \label{eq:friction_neutrals}
		\alpha_{sn}=n_{s}n_{n}m_{sn}\frac{4}{3}\left[\frac{8}{\pi}\left(\frac{k_{\Rm{B}}T_{s}}{m_{s}}+\frac{k_{\Rm{B}}T_{n}}{m_{n}}\right)\right]^{1/2}\sigma_{sn},
	\end{equation}
	where $\sigma_{sn}$ is the collisional cross-section. Values of this parameter for the collisions studied in this work can be found in Table \ref{tab:cross-sections}. We note that some of the cross-sections have not been calculated using quantum mechanical computations but by means of the classical hard-sphere model, which may underestimate the correct values by one or two orders of magnitude, as shown by \citet{2013AA...554A..22V}. However, we do not expect the use of larger cross-sections for $\sigma_{\Rm{HHe} \ \textsc{ii}}$, $\sigma_{\Rm{HHe} \ \textsc{iii}}$, and $\sigma_{\Rm{HeHe} \ \textsc{iii}}$ would significantly modify our results, because the dominant ion in the plasmas we are going to study is proton.
	
	\begin{table}[h]
		\captionof{table}{Cross-sections of collisions with neutral species}
		\centering
		\begin{tabular}{c c c}
			\tableline \tableline
			& Value ($\Rm{m^{-2}}$) & Model \\ \tableline
			$\sigma_{p\Rm{H}}$ & $10^{-18}$ & \citet{2013AA...554A..22V} \\
			$\sigma_{p\Rm{He}}$ & $10^{-19}$ & \citet{2013AA...554A..22V} \\
			$\sigma_{e\Rm{H}}$ & $1.5\times 10^{-19}$ & \citet{2013AA...554A..22V} \\
			$\sigma_{e\Rm{He}}$ & $5\times 10^{-20}$ & \citet{2013AA...554A..22V} \\	
			$\sigma_{\Rm{HHe}}$ & $1.5\times 10^{-19}$ & \citet{2012ApJ...756...57L} \\
			$\sigma_{\Rm{HHe \ \textsc{ii}}}$ & $2 \times 10^{-20}$ & Hard sphere \\
			$\sigma_{\Rm{HHe \ \textsc{iii}}}$ & $1 \times 10^{-20}$ & Hard sphere \\
			$\sigma_{\Rm{HeHe \ \textsc{ii}}}$ & $5 \times 10^{-19}$ & \citet{1999JPhB...32.4919D} \\
			$\sigma_{\Rm{HeHe \ \textsc{iii}}}$ & $3 \times 10^{-21}$ & Hard sphere \\						
			\hline
		\end{tabular}
		\label{tab:cross-sections}
	\end{table}
	
	The system given by the 5-moment transport equations is not closed and must be completed with the induction equation for the magnetic field, $\bm{B}$,
	\begin{equation} \label{eq:faraday}
		\frac{\partial \bm{B}}{\partial t}=-\nabla \times \bm{E}.
	\end{equation}
	The electric field, $\bm{E}$, is given by a generalized Ohm's law that can be obtained from the momentum equation of electrons \citep[see, e.g.,][]{2011A&A...529A..82Z,2014PhPl...21i2901K} by assuming that the variations of the momentum of electrons are negligible (an assumption that is justified by the small mass of the electrons compared to the masses of the other species in the plasma). The battery term, related to the gradient of electronic pressure, has an important influence only when there are very strong gradients of pressure and density, which would not be the case of the situations we are interested in. So, we do not take into account such term. In addition, we also neglect the effect of gravity, since we are going to consider wavelengths that are shorter than the scale height. On the contrary, we retain the term associated to Hall's current and those related to resistivity, i.e., to collisions between electrons and the rest of species. Hence, Ohm's law can be written as 
	\begin{equation} \label{eq:ohm's_law}
		\bm{E}=-\bm{V}\times \bm{B}+\frac{\bm{j}\times \bm{B}}{e n_{e}}+\eta \bm{j}+\frac1{e n_{e}}\sum_{t\ne e}\alpha_{et}\left(\bm{V_{t}}-\bm{V}\right),
	\end{equation}
	where the parameter $\eta$ is known as the coefficient of resistivity or magnetic diffusivity and is given by
	\begin{equation}
		\eta=\frac{\alpha_{ep}+\alpha_{e\Rm{H}}+\alpha_{e\Rm{He}}+\alpha_{e\Rm{He}\textsc{ii}}+\alpha_{e\Rm{He}\textsc{iii}}}{\left(e n_{e}\right)^2},
	\end{equation}
	the current density can be expressed as $\bm{j}=e n_{e}\left(\bm{V}-\bm{V_{e}}\right)$, and $\bm{V}$ is a weighted mean velocity of the ions, given by $\bm{V}=\left(\sum_{i}^{M}Z_{i}n_{i}\bm{V_{i}}\right)/n_{e}$, with $M$ the number of ionized species. The condition of quasi-neutrality for a plasma states that $\sum_{s} Z_{s}n_{s}\approx 0$. Thus, the electron number density is $n_{e}\approx \sum_{i}^{M}Z_{i}n_{i}$.
	
	As already explained in Paper I, the 5-moment transport approximation does not take into account the effects of anisotropic pressures, thermal diffusion and thermal conduction. The inclusion of those effects would require the use of higher-order moment approximations, like the ones detailed in, e.g., \citet{1977RvGSP..15..429S}.
	
\section{Analysis of the dispersion relation} \label{sec:dr}
	We are interested in the study of the properties of small-amplitude incompressible perturbations in a homogeneous plasma. Thus, to obtain the corresponding dispersion relation we can ignore the various continuity and pressure equations, retaining only the equations for momentum and electromagnetic fields, and assume a uniform static background. The linearization process leads to the following equations
	\begin{equation} \label{eq:vs}
		\frac{\partial \bm{V_{s}}}{\partial t}=-\nabla P_{1,s} +\frac{Z_{s}e}{m_{s}}\left[\left(\bm{V_{s}}-\bm{V}\right)\times \bm{B_{0}}+\frac{\left(\nabla \times \bm{B_{1}}\right)\times \bm{B_{0}}}{e n_{e}\mu_{0}}+\eta \frac{\nabla \times \bm{B_{1}}}{\mu_{0}}\right]+\frac{Z_{s}}{n_{e}m_{s}}\sum_{t\ne e}\alpha_{et}\left(\bm{V_{t}}-\bm{V}\right)+\sum_{t\ne s}\nu_{st}\left(\bm{V_{t}}-\bm{V_{s}}\right),
	\end{equation}
	
	\begin{equation} \label{eq:faraday2}
		\frac{\partial \bm{B_{1}}}{\partial t}=\nabla \times \left[\bm{V}\times \bm{B_{0}}-\frac{\left(\nabla \times \bm{B_{1}}\right)\times \bm{B_{0}}}{e n_{e} \mu_{0}}-\eta \frac{\nabla \times \bm{B_{1}}}{\mu_{0}}-\frac1{e n_{e}}\sum_{t\ne e}\alpha_{et}\left(\bm{V_{t}}-\bm{V}\right)\right],
	\end{equation}
	where $\nu_{st}=\alpha_{st}/\rho_{s}$ is the collision frequency between two species $s$ and $t$, $\rho_{s}=m_{s}n_{s}$ is the mass density of species $s$, $\bm{B_{0}}$ is the background magnetic field, $\bm{B_{1}}$ is the perturbation of magnetic field, all the velocities are perturbed quantities, Ampère's law has been used to rewrite the current density as a function of the magnetic field, and $P_{1,s}$ is the pressure perturbation of species $s$. We note that the pressure term has merely been included in Equation (\ref{eq:vs}) for consistency, but it vanishes for the incompressible perturbations considered here. Hence, the pressure term is absent from the subsequent equations. In addition, the last term in Equation (\ref{eq:vs}) can be expanded as follows:
	\begin{equation} \label{eq:col_vs}
		\sum_{t\ne s}\nu_{st}\left(\bm{V_{t}}-\bm{V_{s}}\right)=\sum_{t\ne s,e}\nu_{st}\left(\bm{V_{t}}-\bm{V_{s}}\right)+\nu_{se}\left(\bm{V_{e}}-\bm{V_{s}}\right),
	\end{equation}
	which, since $\bm{V_{e}}=\bm{V}-\bm{j}/e n_{e}$ and by applying again Ampère's law, can be rewritten as
	\begin{equation} \label{eq:col_vs2}
		\sum_{t\ne s}\nu_{st}\left(\bm{V_{t}}-\bm{V_{s}}\right)=\sum_{t\ne s,e}\nu_{st}\left(\bm{V_{t}}-\bm{V_{s}}\right)+\nu_{se}\left(\bm{V}-\bm{V_{s}}\right)-\frac{\nu_{se}\nabla \times \bm{B_{1}}}{e n_{e}\mu_{0}}.
	\end{equation}
	Hence, when resistivity is taken into account, even neutral species are affected by the magnetic field due to their interaction with electrons \citep[see, e.g.,][]{2011A&A...529A..82Z,2015A&A...573A..79S}.
	
	Then, we express the perturbations as proportional to $\exp\left(-i \omega t\right)$ to perform a normal mode analysis, where $\omega$ is the frequency. We also impose that they are proportional to $\exp\left(i \bm{k}\cdot \bm{r}\right)$ to perform a Fourier analysis in space, with $\bm{k}$ being the wave vector and $\bm{r}$ the position vector. We choose a reference frame in which $\bm{B_{0}}=\left(B_{x},0,0\right)$ and $\bm{k}=\left(k_{x},0,0\right)$ and, as in Paper I, define the following circularly polarized variables \citep[see, e.g.,][]{1992wapl.book.....S,2001paw..book.....C}
	\begin{equation} \label{eq:polarized}
		V_{s,\pm}=V_{s,y}\pm iV_{s,z}, \ \ B_{1,\pm}=B_{1,y}\pm iB_{1,z}.
	\end{equation}
	The sign $+$ corresponds to the left-hand polarization ($L$-mode), while the sign $-$ corresponds to the right-hand polarization ($R$-mode). Thanks to these transformations, Equations (\ref{eq:vs}) and (\ref{eq:faraday2}) can be written as
	\begin{eqnarray} \label{eq:vs_polarized}
		\omega V_{s,\pm}&=&\Omega_{s}\left[\pm \left(V_{s,\pm}-V_{\pm}\right)-\frac{k_{x}}{e n_{e}\mu_{0}}B_{1,\pm}\right] \mp i\frac{\eta}{\mu_{0}}\frac{Z_{s}e k_{x}}{m_{s}}B_{1,\pm}+i\frac{Z_{s}}{n_{e}m_{s}}\sum_{t\ne e}\alpha_{et}\left(V_{t,\pm}-V_{\pm}\right) \nonumber \\ 
		&+&i \sum_{t\ne s,e}\nu_{st}\left(V_{t,\pm}-V_{s,\pm}\right)+i\nu_{se}\left(V_{\pm}-V_{s,\pm}\right) \pm i\frac{\nu_{se}k_{x}}{e n_{e}\mu_{0}}B_{1,\pm},
	\end{eqnarray}
	\begin{equation} \label{eq:b_polarized}
		\omega B_{1,\pm}=-k_{x}B_{x}V_{\pm} \mp \frac{k_{x}^2 B_{x}}{e n_{e} \mu_{0}}B_{1,\pm}-i \frac{\eta k_{x}^2}{\mu_{0}}B_{1,\pm} \pm i\frac{k_{x}}{e n_{e}}\sum_{t\ne e}\alpha_{et}\left(V_{t,\pm}-V_{\pm}\right),
	\end{equation}
	where $\Omega_{s}=Z_{s}eB_{x}/m_{s}$ is the cyclotron frequency of species $s$. The two different polarizations are uncoupled, so that we have two independent system of equations. To obtain the dispersion relation, we express each system in matrix form,
	\begin{equation} \label{eq:matrix}
		A_{\pm}\cdot \bm{u_{\pm}}=0,
	\end{equation}
	where $\bm{u_{\pm}}=\left(V_{p,\pm},V_{\Rm{H},\pm},V_{\Rm{He},\pm},V_{\Rm{He} \ \textsc{ii},\pm},V_{\Rm{He} \ \textsc{iii},\pm},B_{1,\pm}\right)^t$ are the vectors of unknowns and $A_{\pm}$ are the coefficient matrices, whose dimensions are $6\times 6$ (the list of coefficients can be found in Appendix \ref{app:coefficients}). The dispersion relation is the result of solving the characteristic equation of each matrix,
	\begin{equation} \label{eq:dr}
		\mathcal{D}_{\pm}\left(\omega,k_{x}\right)\equiv \det A_{\pm}=0.
	\end{equation}
	The expression resulting from the previous operation is too long to be shown here and must be solved numerically. By means of these dispersion relations we are able to study the properties of waves excited both by an impulsive driver or by a periodic driver. We note, however, that the prescribed temporal dependence $\exp(-i \omega t)$ removes the presence of any transitory effect, i.e., we are assuming the stationary state of the wave propagation. The complete temporal dependence will be solved in Section \ref{sec:sims} by means of full numerical simulations.
	
\subsection{Waves excited by an impulsive driver} \label{sec:dr_impulsive}
	To analyze the behavior of waves generated by an impulsive driver, we solve Equation (\ref{eq:dr}) as a function of a real wavenumber, $k_{x}$, and assume that the frequency may be complex, $\omega=\omega_{R}+i\omega_{I}$. If $\omega_{I}<0$, the perturbation is damped in time, while if $\omega_{I}>0$, the amplitude of the perturbation grows exponentially with time, but our model does not include any physical mechanism that may cause such growth.
	
	As subjects of our study we choose three solar plasmas with different degrees of ionization, which allows us to get a more general insight into the effects of partial ionization on the propagation of waves. In the first place, we analyze a region in the upper chromosphere, at a height of 2016 km over the photosphere, where the number density of ions exceeds by an order of magnitude the number density of neutrals. Then, we turn to the analysis of a cool prominence, where neutrals and ions have similar densities. And the third environment of our interest is a region in the lower chromosphere located 500 km above the photosphere. The plasma in this last region is very weakly ionized. The parameters corresponding to each one of those environments are shown in Tables \ref{tab:plasmas} and \ref{tab:col_freq}. The collision frequencies of Table \ref{tab:col_freq} are computed from the friction coefficients given by Equations (\ref{eq:friction_ions}) and (\ref{eq:friction_neutrals}). We note that Table \ref{tab:col_freq} only shows half of the total of collision frequencies involved in the problem; the remaining collision frequencies can be computed taking into account that $\alpha_{st}=\alpha_{ts}$, so that $\rho_{s}\nu_{st}=\rho_{t}\nu_{ts}$. The values of the magnetic field for the chromospheric regions are obtained from the semi-empirical model of \citet{2006A&A...450..805L}, which represents the magnetic field strength in a chromospheric expanding flux tube as
	\begin{equation} \label{eq:b_chrom}
		|\bm{B_{0}}|=B_{ph}\left(\frac{\rho}{\rho_{ph}}\right)^{0.3},
	\end{equation}
	where $\rho$ is the total density at the given height, and $B_{ph}$ and $\rho_{ph}$ are the magnetic field and the total density, respectively, at the photospheric level. We use the values $B_{ph}\approx 1500 \ \Rm{G}$ and $\rho_{ph} \approx 2 \times 10^{-4} \ \Rm{kg \ m^{-3}}$. The Alfvén speed, $c_{\Rm{A}}$, is given by
	\begin{equation} \label{eq:cA}
		c_{\Rm{A}}=\frac{|\bm{B_{0}}|}{\sqrt{\mu_{0} \rho_{i}}},
	\end{equation}
	where $\rho_{i}$ is the sum of the densities of the ionized species.
	
	\begin{minipage}{0.5\textwidth}
		\captionof{table}{Parameters of different solar plasmas} \label{tab:plasmas}
		\centering
		\begin{tabular}{c c c c}
			\tableline \tableline
			Region & \small{I} & \small{II} & \small{III} \\
			\tableline
			$n_{p} \ (\Rm{m^{-3}})$ & $7 \times 10^{16}$ & $1.4 \times 10^{16}$ & $1.9 \times 10^{16}$ \\
			$n_{\Rm{H}} \ (\Rm{m^{-3}})$ & $6\times 10^{15}$ & $2 \times 10^{16}$ & $2.7 \times 10^{21}$ \\
			$n_{\Rm{He}} \ (\Rm{m^{-3}})$ & $10^{15}$ & $2\times 10^{15}$ & $2.7 \times 10^{20}$ \\
			$n_{\Rm{He} \ \textsc{ii}} \ (\Rm{m^{-3}})$ & $6\times 10^{15}$ & -- & $6.5 \times 10^{11}$ \\ 
			$n_{\Rm{He} \ \textsc{iii}} \ (\Rm{m^{-3}})$ & $10^{15}$ & -- & $7.2$ \\
			$T \ (\Rm{K})$ & $20000$ & $10000$ & $4700$ \\
			$B_{0} \ (\Rm{G})$ & $22$ & $10$ & $480$  \\
			$c_{\Rm{A}} \ (\Rm{km \ s^{-1}})$ & $153$ & $184$ & $7600$ \\
			$\Omega_{p} \ (\Rm{rad \ s^{-1}})$ & $210740$ & $95800$ & $4.6 \times 10^{6}$ \\
			$\Omega_{\Rm{He} \ \textsc{ii}} \ (\Rm{rad \ s^{-1}})$ & $52685$ & $23950$ & $1.15 \times 10^{6}$ \\
			$\Omega_{\Rm{He} \ \textsc{iii}} \ (\Rm{rad \ s^{-1}})$ & $105370$ & $47900$ & $2.3 \times 10^{6}$ \\
			\tableline
		\end{tabular} \\
		\scriptsize{Regions I and III correspond to the chromosphere at heights 2016 km and 500 km above the photosphere, respectively; temperatures and number densities are taken from the Model F of \citet{1993ApJ...406..319F}. Region II represents a prominence at an altitude of 10000 km and gas pressure of $P_{g}=0.005 \ \Rm{Pa}$ according to \citet{2015A&A...579A..16H}. The values of the magnetic fields are typical values of the given plasmas, see, e.g., \citet{2006A&A...450..805L}.}
	\end{minipage}
	\begin{minipage}{0.5\textwidth}
		\centering
		\captionof{table}{Collision frequencies} \label{tab:col_freq}
		\begin{tabular}{c c c c}
			\tableline \tableline
			Region & \small{I} & \small{II} & \small{III} \\
			\tableline
			$\nu_{p\Rm{H}} {}^{*}$ & $120$ & $270$ & $2.5 \times 10^{7}$ \\
			$\nu_{p\Rm{He}}$ & $2.5$ & $3.5$ & $320000$ \\
			$\nu_{p\Rm{He} \ \textsc{ii}}$ & $2000$ & -- & $1.7$ \\
			$\nu_{p\Rm{He} \ \textsc{iii}}$ & $1260$ & -- & $7 \times 10^{-11}$ \\ 
			$\nu_{p\Rm{e}}$ & $660$ & $330$ & $1240$ \\
			$\nu_{\Rm{HHe}}$ & $3.5$ & $5.2$ & $480000$ \\
			$\nu_{\Rm{HHe} \ \textsc{ii}}$ & $3.2$ & -- & $2 \times 10^{-4}$ \\
			$\nu_{\Rm{HHe} \ \textsc{iii}}$ & $0.2$ & -- & $8\times 10^{-16}$ \\
			$\nu_{\Rm{He}}$ & $7.5$ & $1$ & $0.9$ \\
			$\nu_{\Rm{HeHe} \ \textsc{ii}}$ & $29$ & -- & $1.5 \times 10^{-3}$ \\
			$\nu_{\Rm{HeHe} \ \textsc{iii}}$ & $0.03$ & -- & $10^{-16}$ \\
			$\nu_{\Rm{Hee}}$ & $0.6$ & $0.1$ & $0.1$ \\
			$\nu_{\Rm{He} \ \textsc{ii}\Rm{He} \ \textsc{iii}}$ & $540$ & -- & $4 \times 10^{-11}$ \\
			$\nu_{\Rm{He} \ \textsc{ii}e}$ & $170$ & -- & $320$ \\
            $\nu_{\Rm{He} \ \textsc{iii}e}$ & $640$ & -- & $1200$ \\
			\tableline
		\end{tabular} \\
		\scriptsize{*: Collision frequencies are given in Hz.}
	\end{minipage}
	
\begin{figure}
		\plottwo{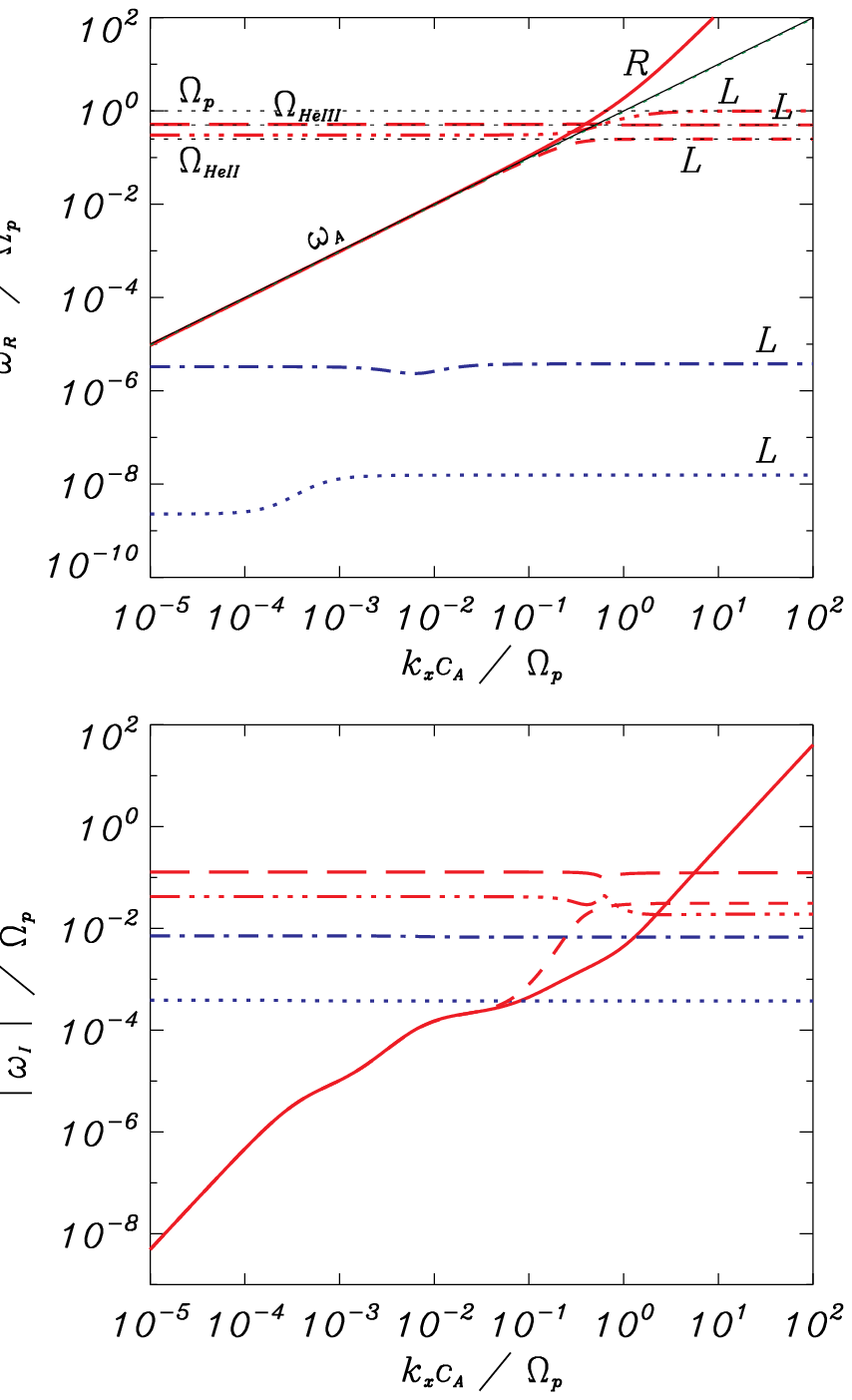}{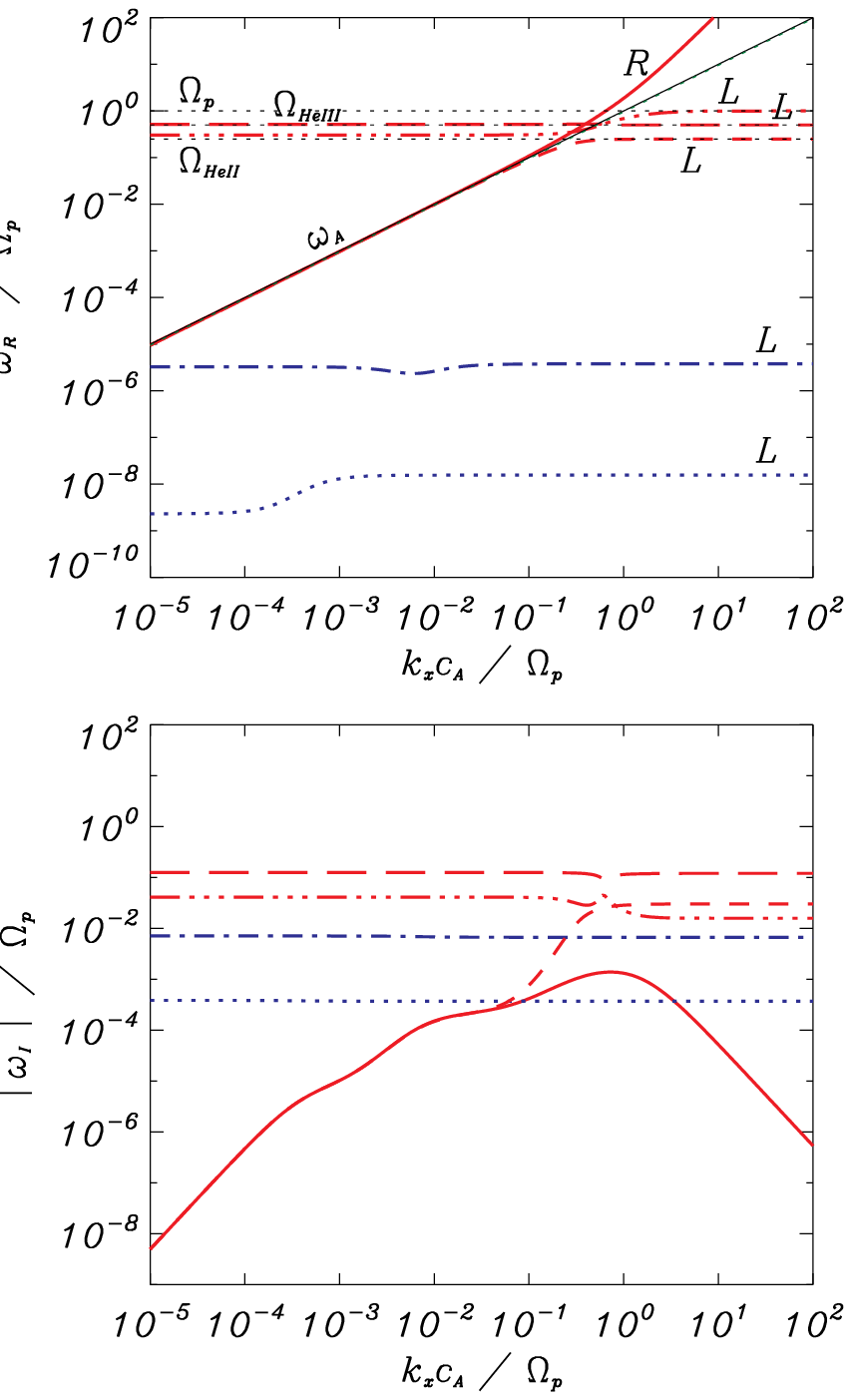}
		\caption{Solutions of Equation (\ref{eq:dr}) for the case of an impulsive driver and upper chromospheric conditions, with resistivity (left panels) and without (right). Top: normalized frequency, $\omega_{R}/\Omega_{p}$, as a function of the normalized wavenumber, $k_{x}c_{\Rm{A}}/\Omega_{p}$; bottom: absolute value of the normalized damping, $|\omega_{I}|/\Omega_{p}$, as a function of the normalized wavenumber. The blue lines are associated to the presence of neutral species. The solid black lines represent the Alfvén frequency, $\omega_{\Rm{A}}$. The horizontal dotted lines represent the cyclotron frequencies of the three ionized species, with $\Omega_{p}> \Omega_{\Rm{He}\textsc{iii}}>\Omega_{\Rm{He}\textsc{ii}}$.}
		\label{fig:chromosphere1_imp}
	\end{figure}

	The results of solving the dispersion relation with parameters corresponding to the upper chromospheric region are shown in Figure \ref{fig:chromosphere1_imp}. Only the solutions with $\omega_{R}>0$ are represented in these plots (the solutions with $\omega_{R}<0$ are symmetric with respect to the horizontal axis but with the opposite polarization). To understand the influence of resistivity, we analyze two different situations: on the left panels the effect of collisions with electrons has been included, while on the right panels it has not been taken into account. We see that the $R$ mode is greatly affected by resistivity at large wavenumbers: it causes a stronger damping of this solution. On the other hand, its effect on the $L$ modes is almost negligible compared to that of the other types of collisions. This huge contrast in the way resistivity affects each polarization can be understood as follows. Electrons are frozen to the magnetic field because we have neglected their inertia. On the contrary, ions, whose direction of gyration coincides with the rotation of the left-hand polarized waves, are not as tight to the magnetic field at high frequencies due to Hall's effect. Thus, the velocity drifts between electrons and ions, $\bm{V_{e}}-\bm{V_{i}}$, are larger when the gyration of ions is opposite to the rotation of waves, i.e., for the $R$ modes. Consequently, friction forces are larger for $R$ modes than for $L$ modes.
	
	The right panels of Figure \ref{fig:chromosphere1_imp} can be also compared with Figure 6 of Paper I, where the fully ionized case in the absence of resistivity was considered. It can be seen that in both plots the real part of the four solutions associated to the ionized species is almost identical, with the dissimilarities appearing in the imaginary part: the inclusion of collisions with neutrals produces a larger damping on the four modes. This enhanced damping is more obvious in the Alfvénic modes at low wavenumbers. Hence, we arrive to the conclusion that at low wavenumbers the damping of waves is dominated by the collisions with neutrals, while the contribution of collisions between ionized species is more important at high wavenumbers. The reason of this behavior is that waves are more efficiently damped when the collision frequency is closer to the oscillation frequency, $\nu \approx \omega$, see, e.g., \citet{2005A&A...442.1091L,2011A&A...529A..82Z,2013ApJ...767..171S}, and, as shown in Table \ref{tab:col_freq}, the interactions with neutrals have lower frequencies than the collisions between ions.  
	
	Another remarkable difference that arises in the partially ionized case with respect to the fully ionized one is the appearance of two additional modes associated to the two neutral species, represented in Figure \ref{fig:chromosphere1_imp} by blue lines. These solutions have received the names of vortex modes \citep{2011A&A...534A..93Z} or forced neutral oscillations \citep{2014PhPl...21a2110V}. When there are no collisions with neutrals, these modes do not propagate, since they have $\omega=0$. But due to the interaction between the neutrals and the ions, the frequency of these modes becomes different from zero, although with $|\omega_{R}| \ll |\omega_{I}|$; hence, such modes are heavily overdamped.

\begin{figure}
		\centering
		\includegraphics[]{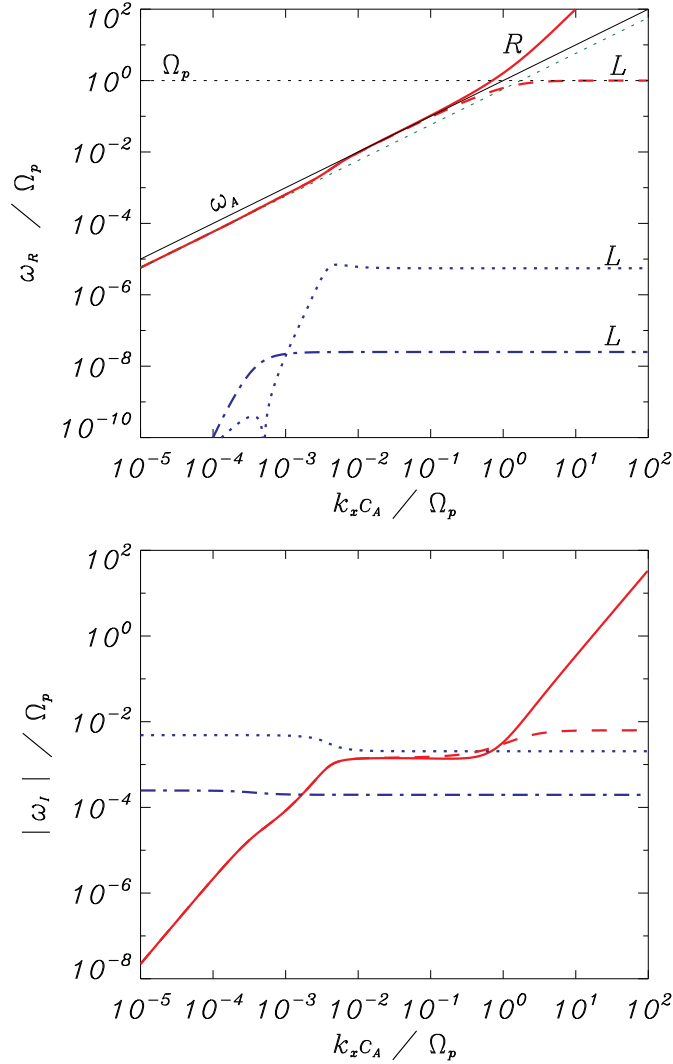}
		\caption{Solutions of Equation (\ref{eq:dr}) for the case of an impulsive driver and prominence conditions. The colors and styles are the same as in Figure \ref{fig:chromosphere1_imp}, with the addition of a green dotted line that represents the normalized modified Alfvén frequency, $\widetilde{\omega}_{\Rm{A}}/\Omega_{p}$.}
		\label{fig:prominence_imp}
	\end{figure}

	Now, we turn to the case of a plasma with properties akin to those in solar prominences. The results are shown in Figure \ref{fig:prominence_imp}. We recall that Table \ref{tab:plasmas} does not provide data about the abundances of the helium ions in this plasma. Hence, the modes associated to those species are not represented in this graphic. The effect of the large abundance of neutrals and their coupling with ions can be clearly distinguished at low wavenumbers in the top panel: the Alfvénic modes do not oscillate with the classical Alfvén frequency, $\omega_{\Rm{A}}=k_{x}c_{\Rm{A}}$, but with a smaller modified Alfvén frequency given by
	\begin{equation} \label{eq:mod_wA}
		\widetilde{\omega}_{\Rm{A}}=k_{x}\widetilde{c}_{\Rm{A}}=\frac{\omega_{\Rm{A}}}{\sqrt{1+\rho_{n}/\rho_{i}}},
	\end{equation}
	where $\rho_{n}$ is the sum of the densities of the neutral species. This equation is a generalization of Equation (27) from \citet{2013ApJ...767..171S} or Equation (19) from \citet{2013A&A...549A.113Z}, applicable when the collision frequencies between neutrals and ions are much larger than the Alfvén frequency. In that limit, the coupling between all the species is so strong that they behave as a single fluid and the inertia of neutrals must be taken into account in the description of Alfvén waves. Then, it can be seen that as the wavenumber rises and the Alfvén frequency is no longer much lower than the collision frequencies, the oscillation frequencies of these modes tend to $\omega_{\Rm{A}}$ before splitting into two different branches. The damping of these solutions is small at low wavenumbers but becomes important for larger values, particularly in the case of the $R$ mode, which is the most affected by collisions with electrons.
	
	The vortex modes shown in Figure \ref{fig:prominence_imp} have much lower oscillation frequencies than the Alfvénic modes. This is due to their excitation being indirectly caused by the magnetic field through the collisions of neutrals with the ions. On the other hand, their damping is higher at low wavenumbers. Thus, these modes are much more short-lived.
	
	Using a two-fluid model to describe Alfvén waves in a partially ionized hydrogen plasma, \citet{2013ApJ...767..171S} found approximate expressions for the damping of the evanescent or vortex mode that appears in such plasma. In the limit when the collision frequency is much smaller than the Alfvén frequency, they found that the damping is given by $\omega_{I,vort} \approx -\nu_{\Rm{H}p}$, while in the opposite limit, it is given by $\omega_{I,vort} \approx -(1+\chi)\nu_{\Rm{H}p}$, where $\chi=\rho_{\Rm{H}}/\rho_{p}$. Here, according to the bottom panel of Figure \ref{fig:prominence_imp}, we obtain a similar behavior of the two vortex modes at the mentioned limits: they tend to a constant value in each limit, with the damping being larger at low wavenumbers, where $\nu \gg k_{x}c_{\Rm{A}}$. A precise expression for those limiting values is not provided, since it involves a combination of a great number of parameters and its calculation is not as straightforward as the one obtained by \citet{2013ApJ...767..171S}.

\begin{figure}
		\centering
		\includegraphics[]{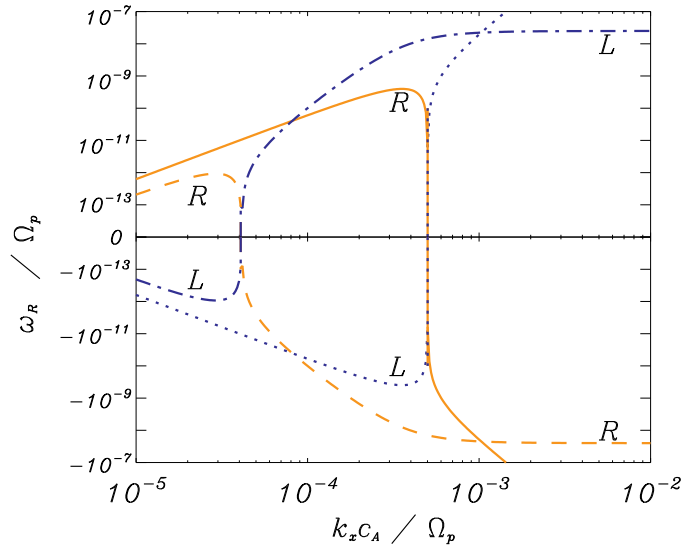}
		\caption{Magnification of the low-wavenumber and low-frequency region of the top panel of Figure \ref{fig:prominence_imp} (including also the solutions with $\omega_{R} < 0$). The blue lines represent the $L$ modes, while the orange lines represent the $R$ modes.}
		\label{fig:prominence_zoom}
	\end{figure}

	The $L$-mode represented by a dotted line in the top panel of Figure \ref{fig:prominence_imp} displays a very sharp minimum in frequency at about $k_{x}c_{\Rm{A}}/\Omega_{p} \approx 5 \times 10^{-4}$. We have checked that this behavior is not found when collisions between electrons and neutral species are not considered. Hence, it is a consequence of the electron-neutral interaction. Also, we note that this sharp minimum is just apparent and is caused by the way in which the solutions are represented in Figure \ref{fig:prominence_imp}. This is better illustrated by Figure \ref{fig:prominence_zoom}, where the solutions with $\omega_{R} < 0$ are also plotted. Due to the collisions with electrons, the $L$ vortex modes have $\omega_{R} < 0$ at low wavenumbers, and they change to $\omega_{R} > 0$ when the wavenumber increases. The contrary happens to the $R$ modes.

\begin{figure}
		\centering
		\includegraphics[]{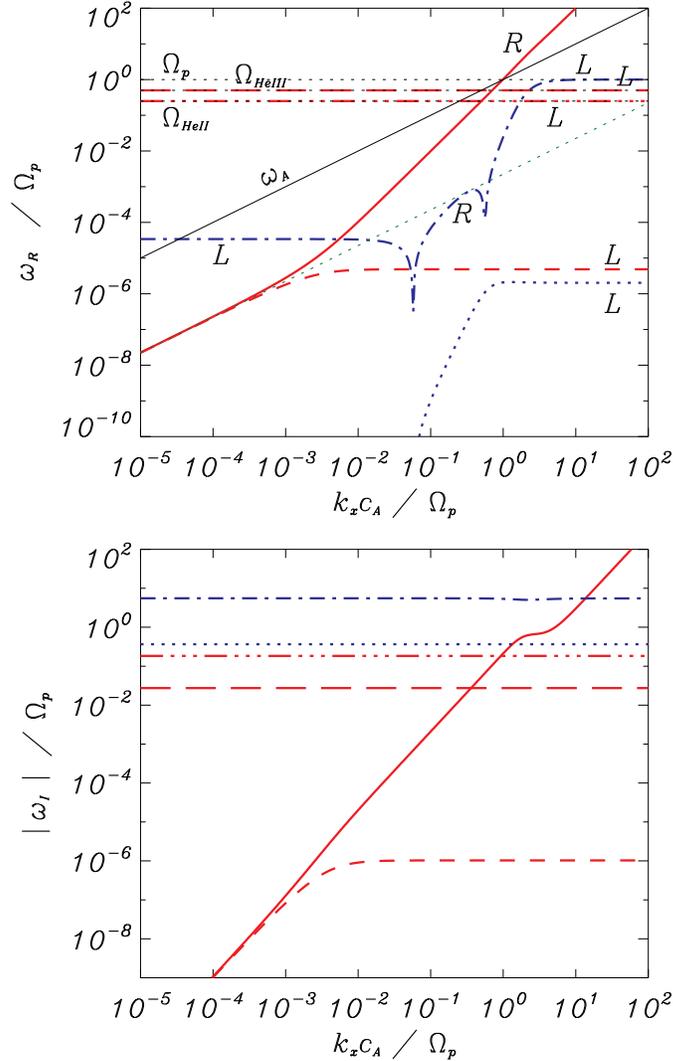}
		\caption{Solutions of Equation (\ref{eq:dr}) for the case of an impulsive driver and parameters that correspond to a region in the low chromosphere. The colors and styles are the same as in Figure \ref{fig:chromosphere1_imp}, with the addition of a green dotted line that represents the normalized modified Alfvén frequency.}
		\label{fig:lowchrom_imp}
	\end{figure}

	Next, Figure \ref{fig:lowchrom_imp} shows the results which correspond to a plasma in the low chromosphere at a height of $500$ km above the photosphere. Here, the abundance of neutral particles is several orders of magnitude larger than the abundance of ions. In addition, the large densities lead to high values of the collision frequencies. Both circumstances provoke a huge departure of the modified Alfvén frequency from $\omega_{\Rm{A}}$, as it can be checked in the top panel. Moreover, in this case, it is difficult to distinguish between the vortex, Alfvénic and ion-cyclotron modes because their properties seem to be mixed up. For instance, at low wavenumbers there is one solution that in the previous cases was referred to as a vortex mode but that here has a higher oscillation frequency than the Alfvénic modes, and that for large wavenumbers tends to the proton cyclotron frequency, as if it were an ion-cyclotron wave. Also, the $L$ mode that emerges from the Alfvénic branch does not tend to the cyclotron frequency of any ion but to the value $\Omega_{p}/(1+\chi)$, as if it were an effective cyclotron frequency \citep{1974PlPh...16..813M} for this plasma. These results point out the dramatic effect of neutrals on the behavior of the waves in conditions of very low ionization.

\begin{figure}
		(a) \hspace{5.4cm} (b) \hspace{5.4cm} (c) \\
		\includegraphics[width=0.33\hsize]{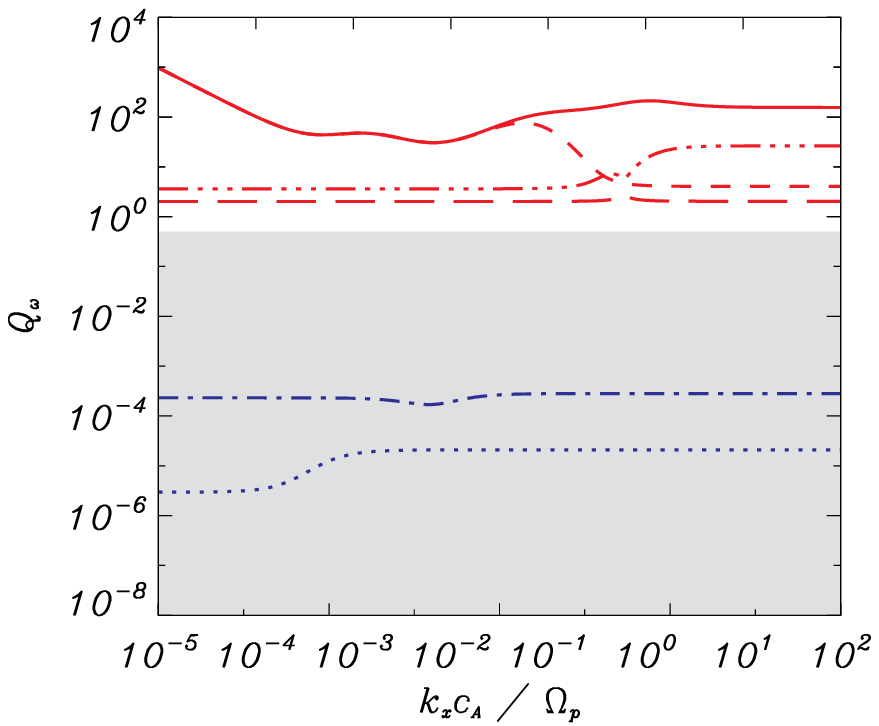}
		\includegraphics[width=0.33\hsize]{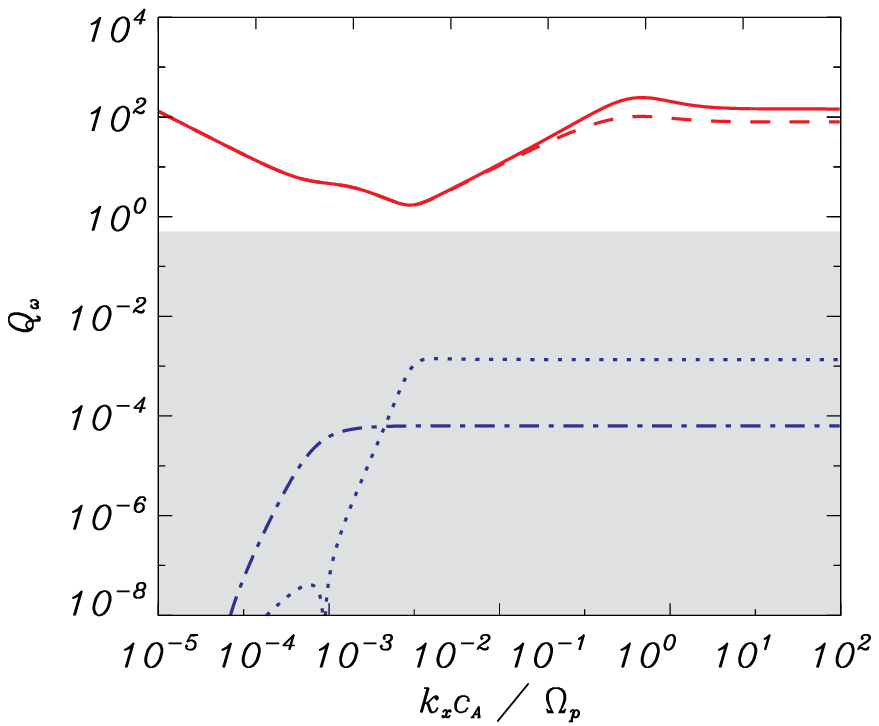} \includegraphics[width=0.33\hsize]{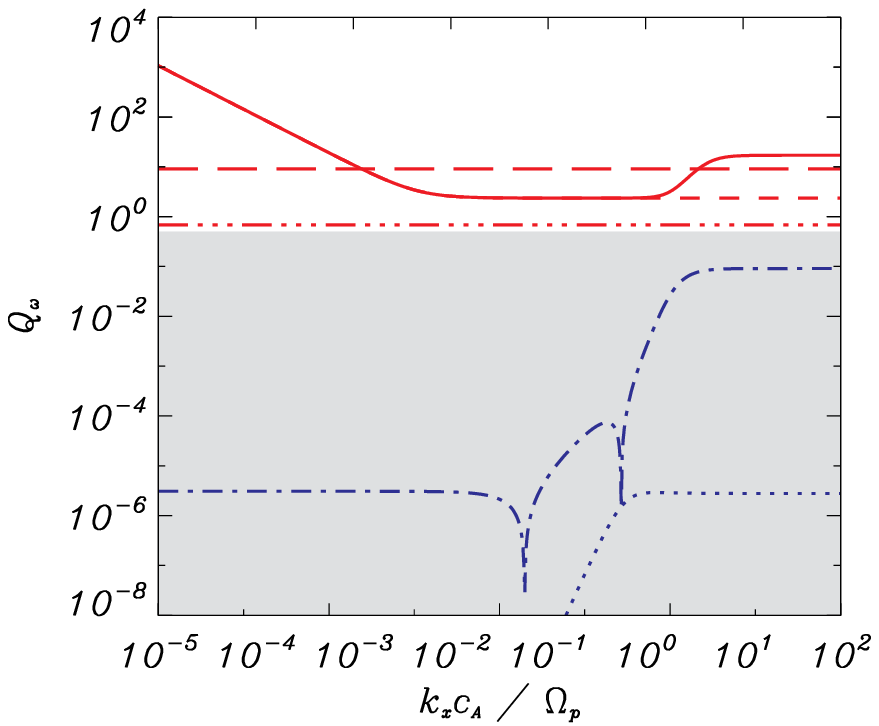}
		\caption{The panels (a), (b) and (c) represent the quality factors of the solutions shown in the left panels of Figure \ref{fig:chromosphere1_imp}, in Figure \ref{fig:prominence_imp} and in Figure \ref{fig:lowchrom_imp}, respectively. The grey areas correspond to values of $Q_{\omega} < 1/2$, where the waves are overdamped.}
		\label{fig:qfactor_imp}
	\end{figure}

	A more complete comprehension of the properties of the perturbations can be obtained through their quality factors, which can be defined as $Q_{\omega} \equiv 1/2 |\omega_{R}/\omega_{I}|$. A perturbation is underdamped, overdamped or evanescent if $Q_{\omega} > 1/2$, $Q_{\omega} < 1/2$ or $Q_{\omega} = 0$, respectively. Figure \ref{fig:qfactor_imp} shows the quality factors of (a) the solutions for the case of the higher chromosphere with resistivity, (b) the case of the prominence, and (c) the lower chromosphere. It can be seen that, in the three studied regions, the vortex modes are overdamped, specially at low wavenumbers, and, hence, the energy associated to these modes is not transported far from where the perturbation appears but it is dissipated \textit{in situ}. The remaining perturbations are underdamped. The $R$ mode (red solid line) has the larger $Q_{\omega}$ of all modes at all values of $k_{x}c_{\Rm{A}}/\Omega_{p}$, except for the case of the lower chromosphere: in panel (c) it can be seen that there is a region, between $k_{x}c_{\Rm{A}}/\Omega_{p} \approx 10^{-3}$ and $k_{x}c_{\Rm{A}}/\Omega_{p} \approx 2$, where one of the cyclotron modes has a larger value of $Q_{\omega}$. We note that none of the modes is evanescent.
	
	According to, e.g., \citet{1969ApJ...156..445K}, \citet{1990ApJ...350..195P} or \citet{2013ApJ...767..171S}, a two-fluid partially ionized plasma may present regions of wavenumbers where all modes are evanescent: oscillatory modes are suppressed in those cutoff intervals due to the strong friction caused by the collisions of ions with neutrals. Such regions appear when the condition $\chi > 8$ is fulfilled, as it is the case of the region of the lower chromosphere studied here. But later, \citet{2015A&A...573A..79S}, using a model that included the effects of Hall's term, resistivity, electron inertia and viscosity, showed that in a more realistic situation there are no strict cutoffs. They found that Alfvén waves may be underdamped or overdamped depending on the specific physical parameters of the plasma but that those modes can always be excited, due to the effect of Hall's current \citep[see, e.g.,][]{1960RSPTA.252..397L,2008MNRAS.385.2269P,2012A&A...544A.143Z} and electron inertia. The results shown in Figure \ref{fig:qfactor_imp}(c) agree with those of \citet{2015A&A...573A..79S} regarding the absence of strict cutoffs. Since here we have not considered the electron inertia, the mechanism that causes the removal of the cutoffs is Hall's current: electrons dynamics is different than that of ions and they stay more coupled to the magnetic field than ions, which are greatly affected by collisions with neutrals, allowing the propagation of Alfvén waves.
	
\begin{figure}
		\centering
		\includegraphics[]{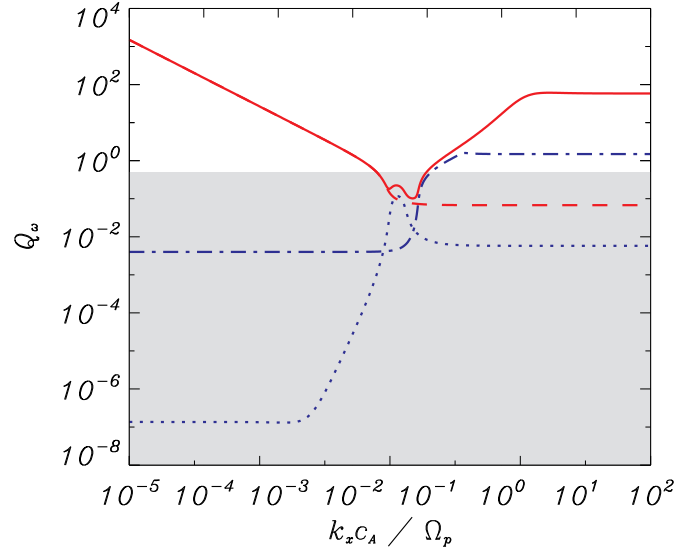}
		\caption{Quality factor of waves excited by an impulsive driver in a plasma with conditions of the chromosphere at a height of 1175 km over the photosphere (for the sake of clarity, the cyclotron modes associated to the helium ions are not plotted here).}
		\label{fig:qfactor_imp2}
	\end{figure}

	Nevertheless, the regions of the solar chromosphere analyzed by \citet{2015A&A...573A..79S} do not coincide with those studied in the present paper and, hence, a direct comparison between the results of the two investigations cannot be made. Those authors explored an altitude range from 600 km to 2000 km, while we have focused only in two heights that are outside of that range (500 km and 2016 km). Thus, to perform an appropriate comparison, we now compute the quality factor of waves at an altitude of 1175 km over the photosphere, using again the abundances given in \citet{1993ApJ...406..319F}. At such height, the magnetic field is $B_{0} \approx 110 \ \Rm{G}$ and the temperature is $T \approx 6500 \ \Rm{K}$. Figure \ref{fig:qfactor_imp2} shows that there is an interval of wavenumbers, between $k_{x}c_{\Rm{A}}/\Omega_{p} \approx 10^{-2}$ and $k_{x}c_{\Rm{A}}/\Omega{p} \approx 0.1$, where all the displayed modes are overdamped. This is in good agreement with the results represented in Figure 3(b) and (d) of \citet{2015A&A...573A..79S}. Thus, although there are no strict cut-off regions in the middle chromosphere either, an interval of wavenumbers for which the solutions are overdamped remains even when Hall's current and electron-neutral collisions are taken into account.
	
\subsection{Waves excited by a periodic driver} \label{sec:dr_periodic}
	The study of waves generated by a periodic driver is carried out by solving Equation (\ref{eq:dr}) as a function of a real frequency, $\omega$. The resulting wavenumbers may be complex, $k_{x}=k_{R}+ik_{I}$. If $\omega > 0$, the perturbations propagate along the positive direction of the $x$-axis if $k_{R} > 0$ and along the negative direction if $k_{R} < 0$, and they are spatially damped when $\Rm{sgn}(k_{I})=\Rm{sgn}(k_{R})$.
	
	We recall that when collisions between all the different species are not considered, resonances appear at the ions' cyclotron frequencies \citep{2001paw..book.....C}. At a resonance, the wavenumber tends to infinity, which leads to a null phase speed, i.e., the perturbation does not propagate. The energy of the driver is used in increasing the amplitude of the perturbation of the ion associated to the given cyclotron frequency. Moreover, after the resonances there are cutoff regions where the perturbations become evanescent. Here, we are interested in the effects that collisions with neutral species have on such resonances and cutoffs.  
	
\begin{figure}
		\plottwo{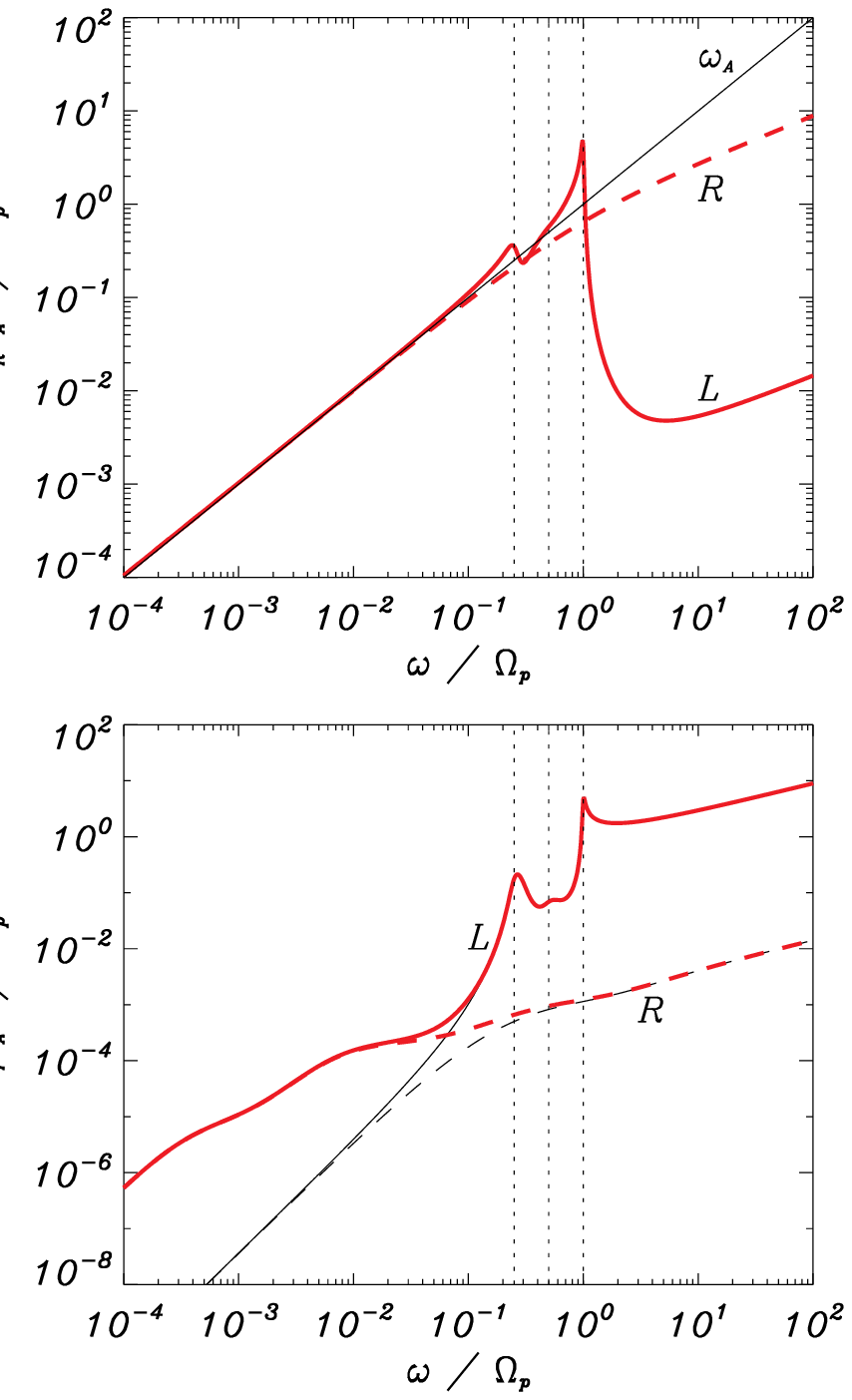}{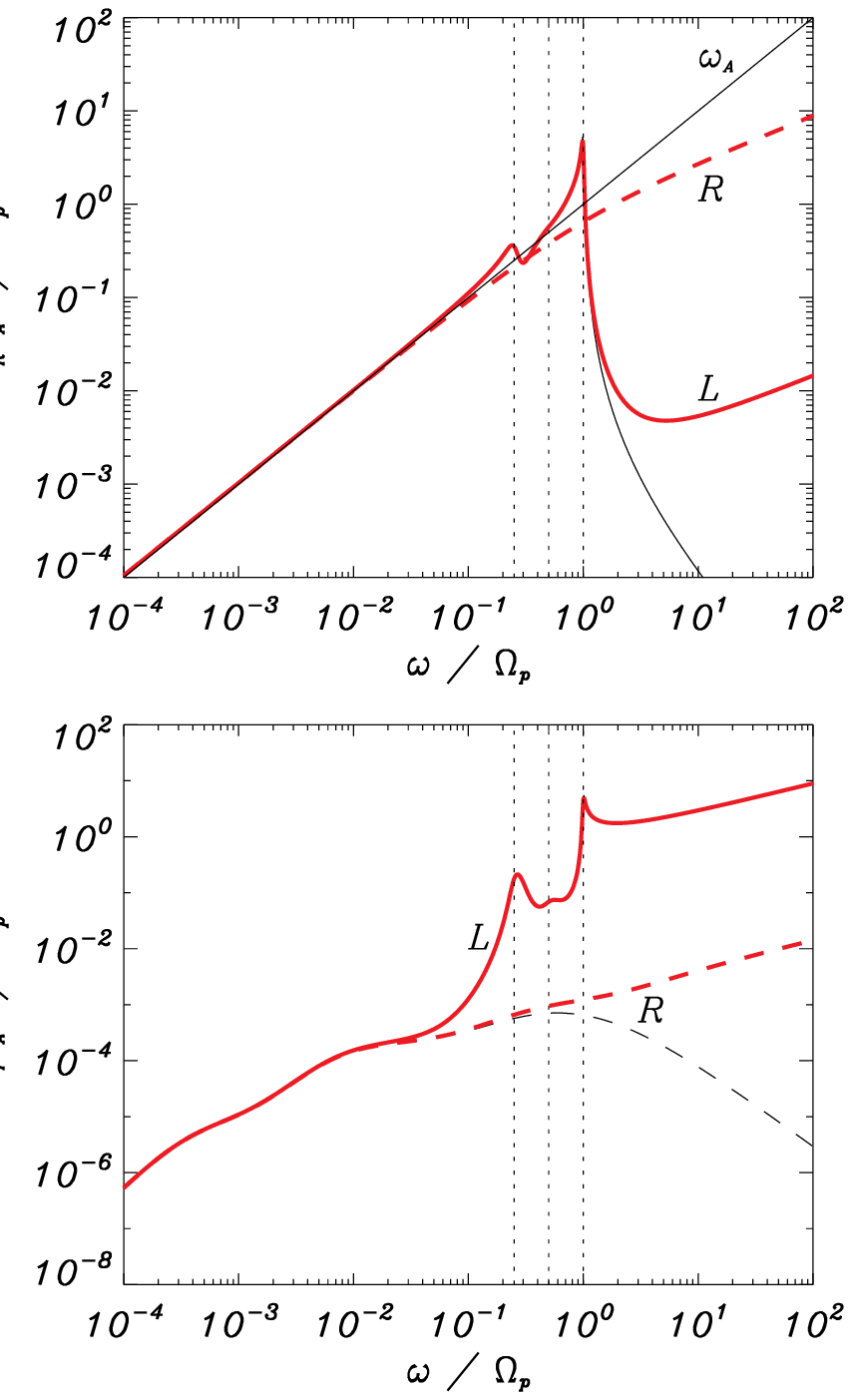}
		\caption{Solutions of Equation (\ref{eq:dr}) for the case of a periodic driver and parameters corresponding to a region in the high chromosphere (region I from Table \ref{tab:plasmas}). The wavenumber and the spatial damping are shown as functions of the frequency in the top and the bottom panels, respectively. The thin black lines on the left panels represent the solutions when the collisions with neutrals have been ignored. On the right panels, the thin black lines represent the case without resistivity. The diagonal black line corresponds to the Alfvén frequency, $\omega_{\Rm{A}}$. The red lines correspond to the case where all the effects have been taken into account. The dotted vertical lines mark the position of the frequencies of resonances, with $\Omega_{p}> \Omega_{\Rm{He} \ \textsc{iii}} > \Omega_{\Rm{He} \ \textsc{ii}}$.}
		\label{fig:highchrom_per}
	\end{figure}

	As in the previous section, we start analyzing a region in the chromosphere at an altitude of 2016 km over the photosphere. The results are shown in Figure \ref{fig:highchrom_per}, where comparisons of several situations are made. In the first place, the left panels display a comparison between the case in which all the types of collisions are considered and the one in which the interaction with neutrals is neglected. The comparison of these two scenarios informs us that the effect of neutrals is particularly important in the low-frequency range, where it produces a remarkably stronger damping than when only collisions between ionized species are taken into account. At higher frequencies, there are no appreciable differences between the two situations, which means that the collisions with neutrals are not as relevant as the collisions between ions.
	
	On the right panels of Figure \ref{fig:highchrom_per}, the case that includes all the collisions between the species is compared with the situation in which resistivity is ignored. It can be seen that the inclusion of resistivity produces a larger damping of the $R$ mode at high frequencies, while the damping of the $L$ mode is not modified. However, concerning the real part of the wavenumber, it is the ion-cyclotron mode the one who is more affected: at high frequencies, it has a larger wavenumber when collisions with electrons are involved and, hence, it propagates at a smaller speed. The real part of the wavenumber of the $R$ mode shows no variations. In addition, by inspecting any of the four panels it can be checked that all the solutions are finite at the cyclotron frequencies, which means that the resonances are removed due to the friction between species. The cutoffs regions are removed as well.
	
\begin{figure}
		\plottwo{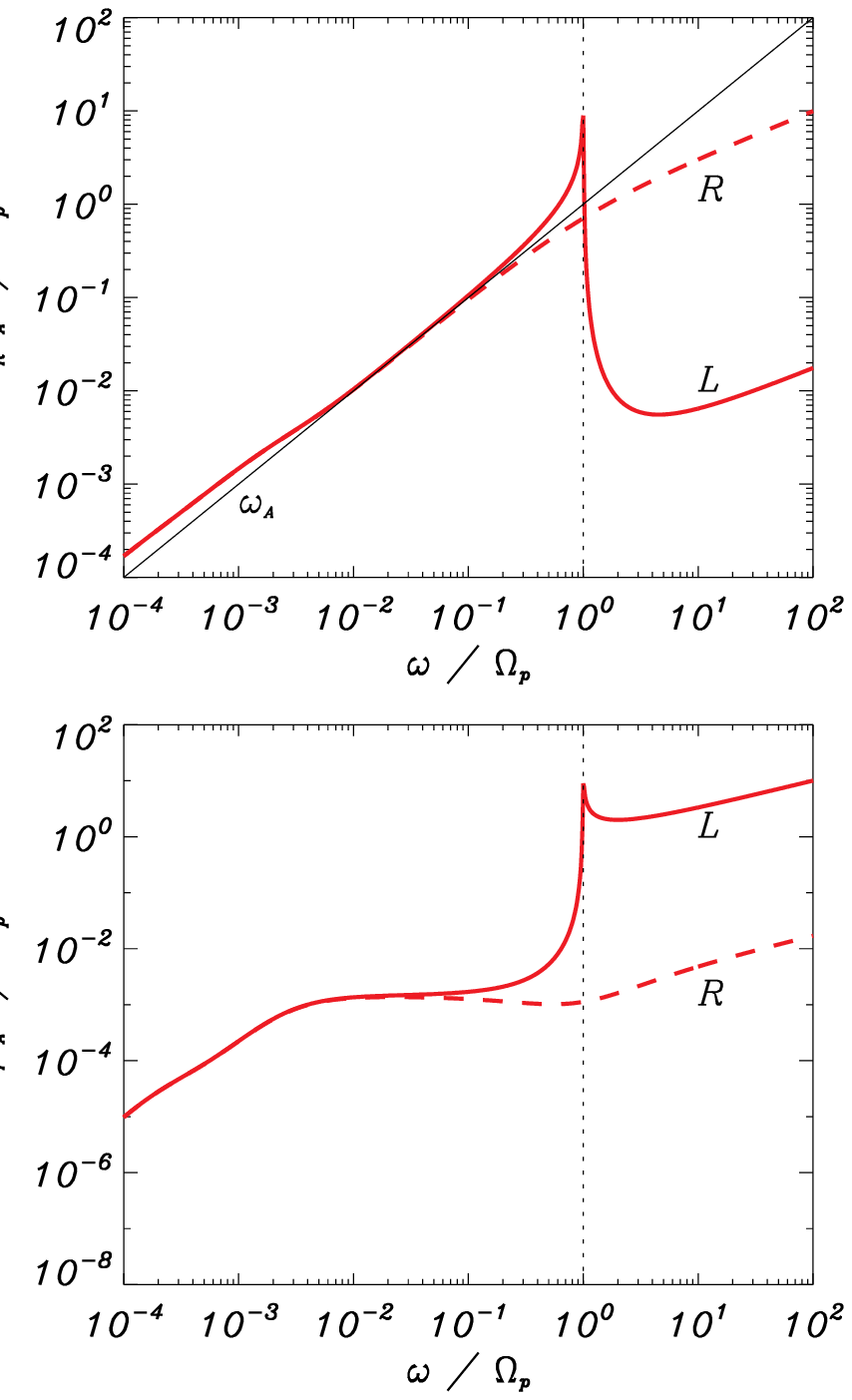}{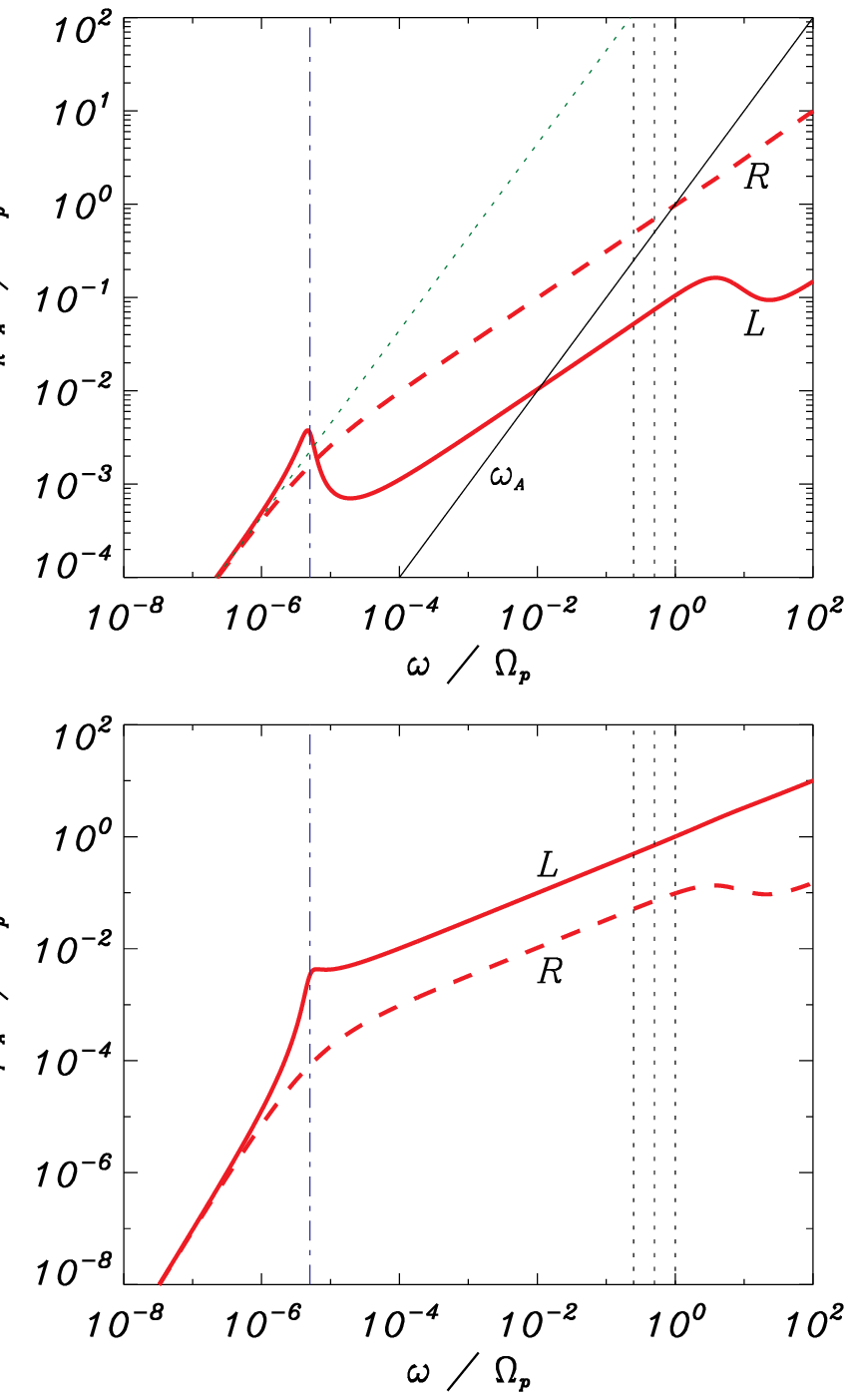}
		\caption{Normal modes of perturbations excited by a periodic driver. The left panels correspond to prominence conditions, while the right panel correspond to conditions of a region in the low chromosphere. The green dotted line on the top right panel represents the wavenumber related to the modified Alfvén speed. The vertical blue lines on the right panels represent the position of the Hall frequency, $\omega_{\Rm{H}}$.}
		\label{fig:periodic2}
	\end{figure}

	Now, we turn our attention to environments in which neutrals are the most abundant species. The results are shown in Figure \ref{fig:periodic2}, with the left and right panels corresponding to the prominence and the low chromosphere, respectively. It can be seem that, for any of the two environments, on the left area of the plots, where the frequency of the driver is smaller than the collision frequencies, all the fluids are strongly coupled and the wavenumber of the oscillation is larger than in the fully ionized case. Thus, the perturbation propagates at a slower speed, given by $\widetilde{c}_{\Rm{A}}$. Moreover, since all the species behave in that range almost as a single fluid, the spatial damping of the perturbation is smaller than at higher frequencies, where the coupling is not as strong and the frictional force increases due to the velocity drifts between the species.
	
	As the frequency of the driver is increased, the separation between the $L$ and the $R$ modes becomes evident. For the case of the prominence, the separation occurs when the frequency of the driver approaches the ion cyclotron frequency, similar behavior as those shown in Figure \ref{fig:highchrom_per} or in the fully ionized plasmas studied in Paper I. However, the right panels of Figure \ref{fig:periodic2} show that when the abundance of neutrals is much higher than the abundance of ions and there is a very strong coupling between them, the split appears at values much lower than the cyclotron frequencies. 
	
	By means of a single-fluid description of three-component plasmas (electrons, ions and neutrals), \citet{2008MNRAS.385.2269P} and \citet{2015MNRAS.447.3604P} found that the effect of Hall's current becomes important and, hence, there is a clear distinction in the properties of the $L$ and $R$ waves when the frequency $\omega$ is of the order of or larger than the so-called Hall frequency, which is defined as
	\begin{equation} \label{eq:Hall_freq}
		\omega_{\Rm{H}}=\frac{\rho_{i}}{\rho_{i}+\rho_{n}}\Omega_{i}.
	\end{equation}
	We note that this parameter coincides with the effective cyclotron frequency mentioned in Section \ref{sec:dr_impulsive}. In a fully ionized plasma, the Hall frequency matches the ion cyclotron frequency. In contrast, when a weakly ionized plasma is considered, e.g., in the lower solar chromosphere, $\omega_{H} \ll \Omega_{i}$ and Hall diffusion has a great influence on the dynamics of the plasma even at very low frequencies. The results shown in the right panels of Figure \ref{fig:periodic2} are consistent with the findings of \citet{2008MNRAS.385.2269P} and \citet{2015MNRAS.447.3604P} and it can be checked that the position of the peak that appears in both the top and the bottom panels corresponds to the Hall frequency, $\omega_{\Rm{H}} \approx 5 \times 10^{-6} \Omega_{p}$.

\begin{figure}
		(a) \hspace{5.4cm} (b) \hspace{5.4cm} (c) \\
		\includegraphics[width=0.33\hsize]{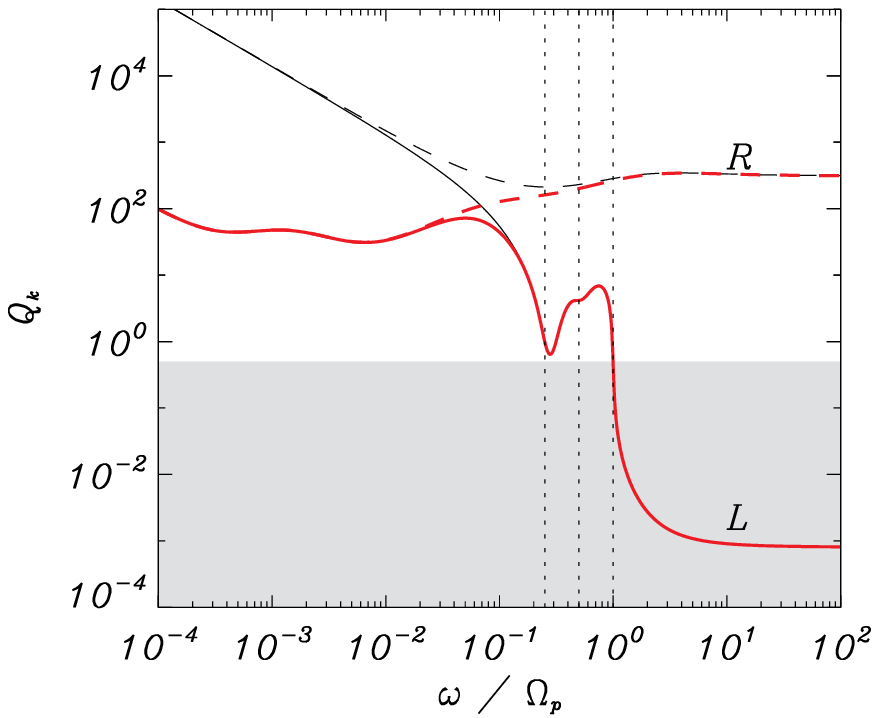}
		\includegraphics[width=0.33\hsize]{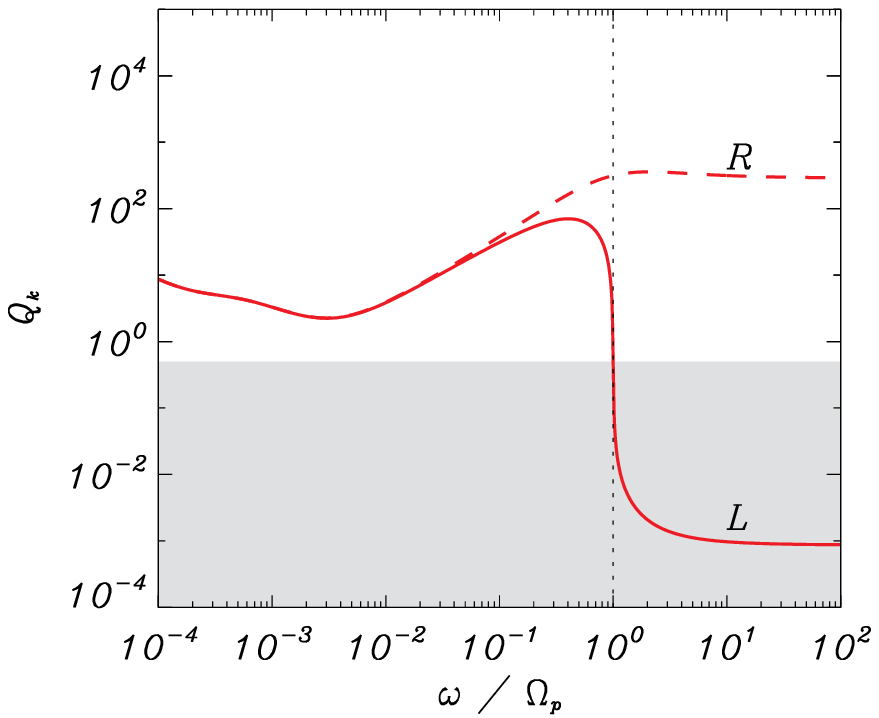}
		\includegraphics[width=0.33\hsize]{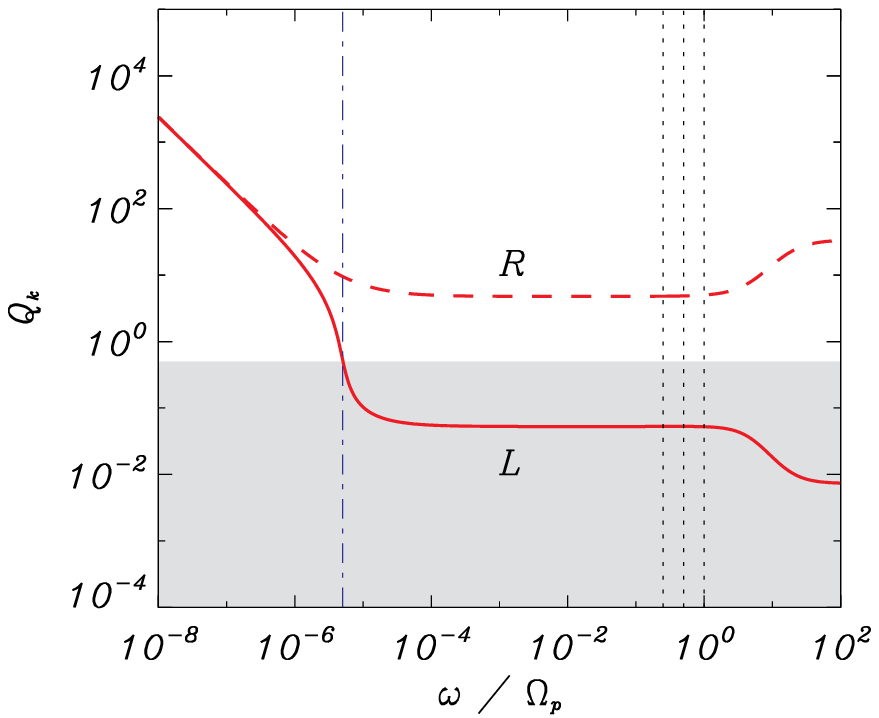}
		\caption{Quality factor, $Q_{k}$, of waves excited by a periodic driver. The left panel corresponds to the results shown on the left panels of Figure \ref{fig:highchrom_per}. The middle and the right panels correspond to the results shown on the left and right panels of Figure \ref{fig:periodic2}, respectively.}
		\label{fig:qfactor_per}
	\end{figure}

	To conclude this section, we analyze the quality factor, $Q_{k} \equiv 1/2 |k_{R}/k_{I}|$, of the solutions described in the previous paragraphs. Figure \ref{fig:qfactor_per}(a), (b) and (c) represent the quality factor of the normal modes shown in the left panels of Figure \ref{fig:highchrom_per}, and the left and right panels of Figure \ref{fig:periodic2}, respectively. Figure \ref{fig:qfactor_per}(a) indicates that the low frequency modes have a smaller $Q_{k}$ when collisions with neutrals are taken into account. In both cases, $Q_{k}$ is of the order of or larger than $\approx 100$, which means that the perturbations are underdamped. At higher frequencies, the $L$ and $R$ display very contrasting behaviors. The quality factor of the former tends to the critical value $Q = 1/2$ when the frequency of the driver approaches the lower ion cyclotron frequency and finally crosses that boundary at the higher ion cyclotron frequency, meaning that it becomes overdamped. On the other hand, the $\bm{R}$ mode stays in the underdamped area. Similar behavior is found in Figure \ref{fig:qfactor_per}(b), although the Alfvénic modes have a lower $Q_{k}$ due to a larger presence of neutrals in this plasma that produces a greater friction and dissipation of the energy carried by the perturbation. Figure \ref{fig:qfactor_per}(c), which represents the weakly ionized environment of the low chromosphere, shows that the point where the $L$ mode becomes overdamped coincides with the Hall frequency given by Equation (\ref{eq:Hall_freq}) and that there are approximately two orders of magnitude of difference between the quality factors of the two polarizations even at low frequencies. 
	
\begin{figure}
		\centering
		\includegraphics[]{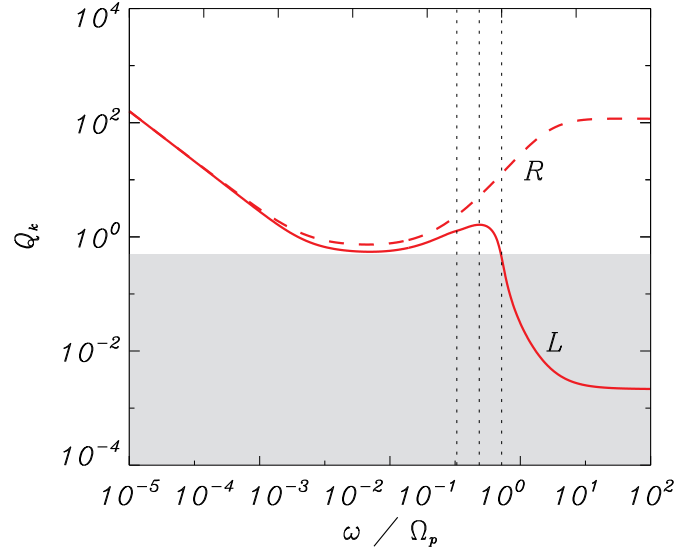}
		\caption{Quality factor of waves excited by a periodic driver in a plasma with conditions of the chromosphere at a height of 1175 km over the photosphere.}
		\label{fig:qfactor_per2}
	\end{figure}

	Again, we are interested in comparing our results with those detailed in \citet{2015A&A...573A..79S}. Thus, we choose the same region of the chromosphere as in the previous section to perform the comparison. The results are plotted in Figure \ref{fig:qfactor_per2}. In slight contrast with what is shown in Figure 4(b) and (d) of \citet{2015A&A...573A..79S}, we find small differences in the quality factor of the two polarizations for frequencies lower than $0.1 \Omega_{p}$. This slight discrepancy may be caused by some of the effects that \citet{2015A&A...573A..79S} included in their model, like the viscosity of each species or the electron inertia, which have been neglected in the present paper. Nevertheless, the two investigations agree with respect to the fact that neither mode is overdamped. At higher frequencies, we find again larger dissimilarities in the behavior of each mode but we cannot compare this frequency range with the investigation of \citet{2015A&A...573A..79S} because they did not explore it. 

\section{Numerical simulations} \label{sec:sims}
	In this section, we explain the results of using the numerical code MolMHD (see Paper I and references therein for more details) to compute the temporal evolution of small-amplitude perturbations on homogeneous partially ionized plasmas, and compare then to the analyses described in the previous sections. We recall that the code is nonlinear but that, as in Paper I, we focus mainly on the linear effects. All the simulations described in this section are one-dimensional.
	
\subsection{Impulsive driver} \label{sec:sim_impulsive}
	Here, the propagation of waves excited by an impulsive driver is studied by imposing an initial perturbation with a Gaussian profile,
	\begin{equation} \label{eq:gauss}
		f(x,t=0)=f_{0} \exp \left(-\left(\frac{x-x_{0}}{\sqrt{2}\sigma_{x}}\right)^2\right),
	\end{equation}
	where $f_{0}$ is the amplitude of the perturbation, $x_{0}$ the position of the peak of the Gaussian function, and $\sigma_{x}$ the root-mean-square width, which is related to the full width at half maximum (FWHM) by the formula $FWHM=2\sqrt{2 \ln{2}} \sigma_{x}$. The perturbation is superimposed to an static medium with a background magnetic field given by $\bm{B_{0}}(x)=(B_{0},0,0)$.
	
\begin{figure}
		\hspace{0.15cm} (a) \hspace{5.4cm} (b) \hspace{5.4cm} (c) \\
		\includegraphics[width=0.33\hsize]{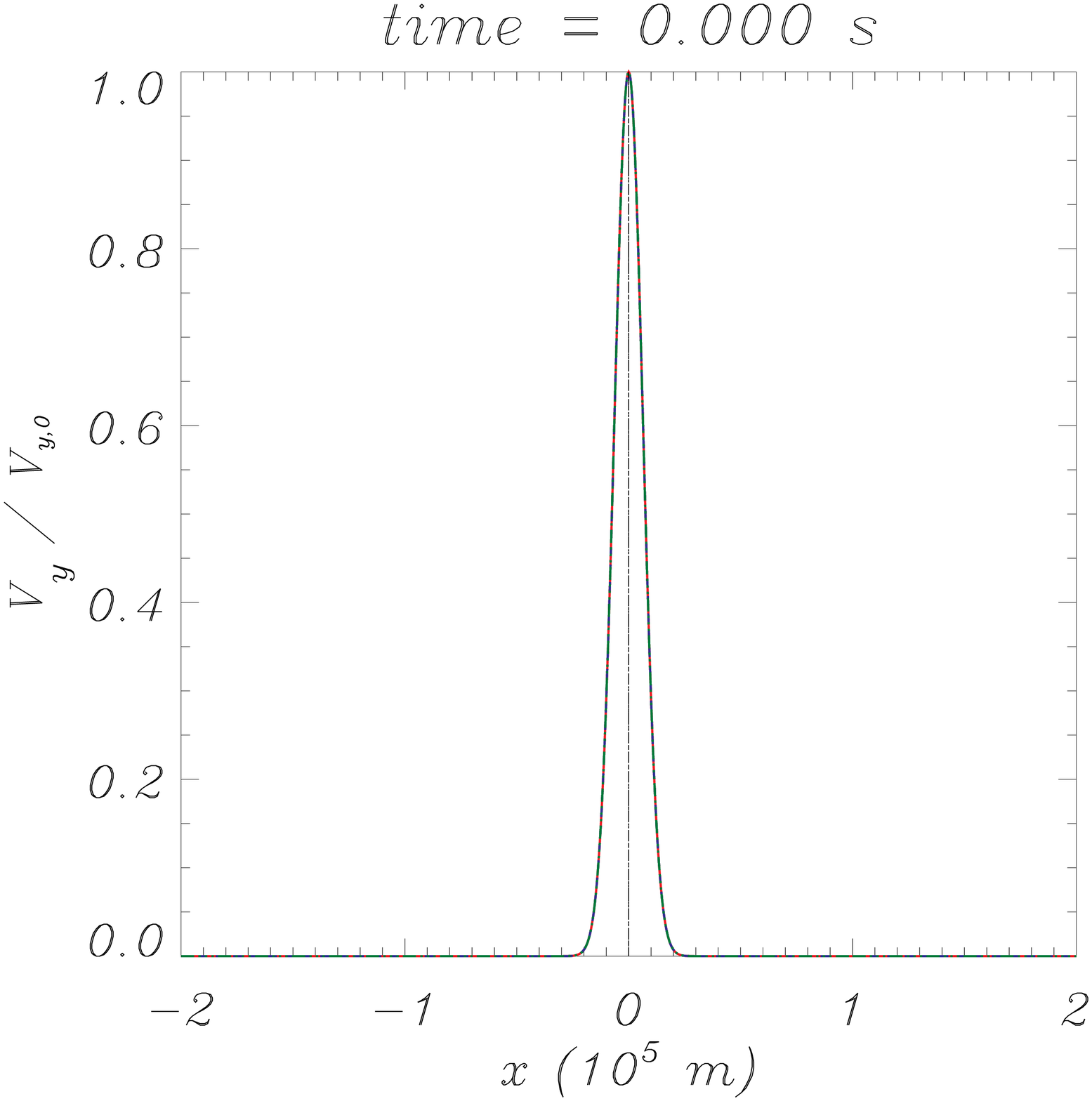}
		\includegraphics[width=0.33\hsize]{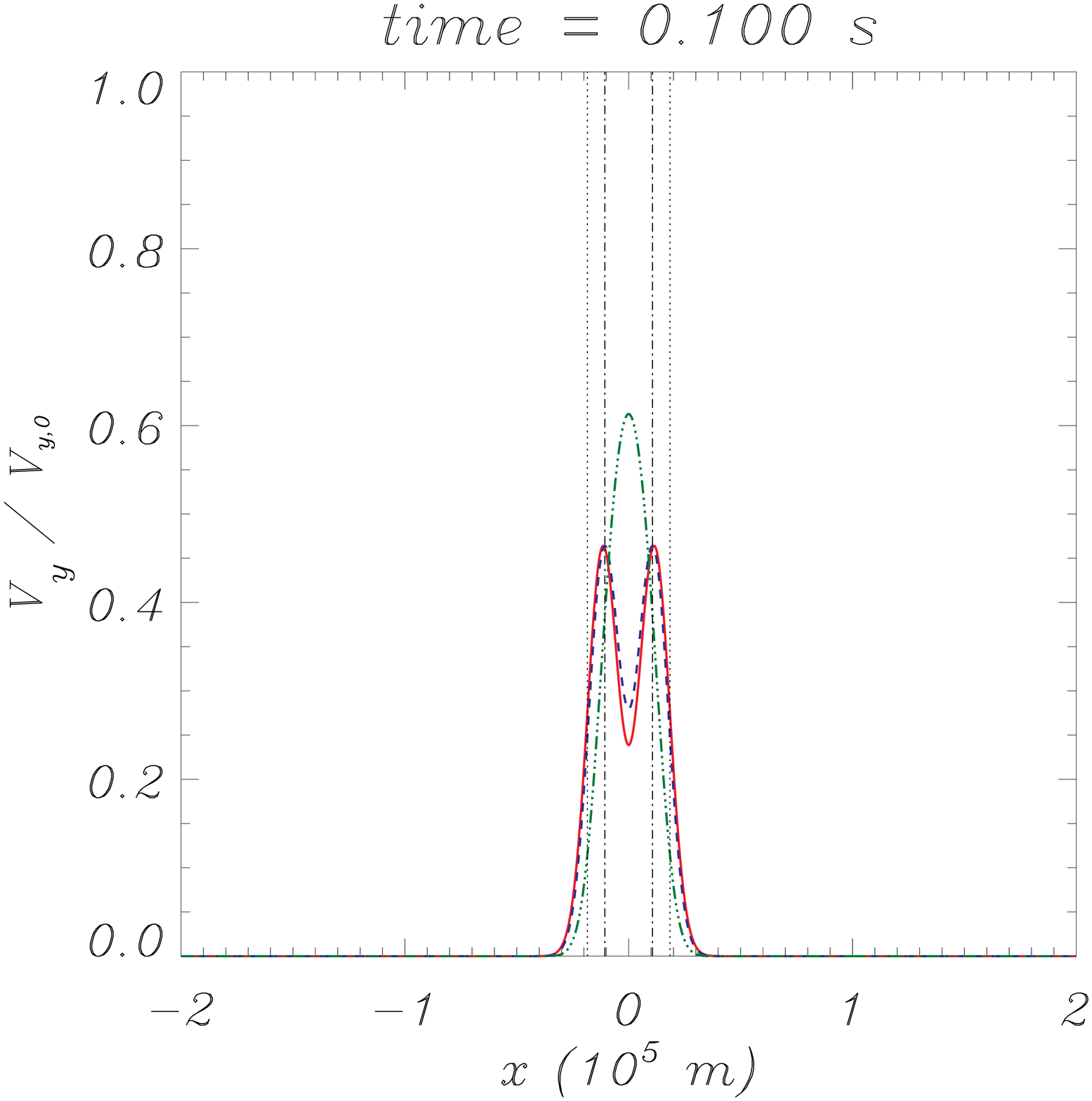}
		\includegraphics[width=0.33\hsize]{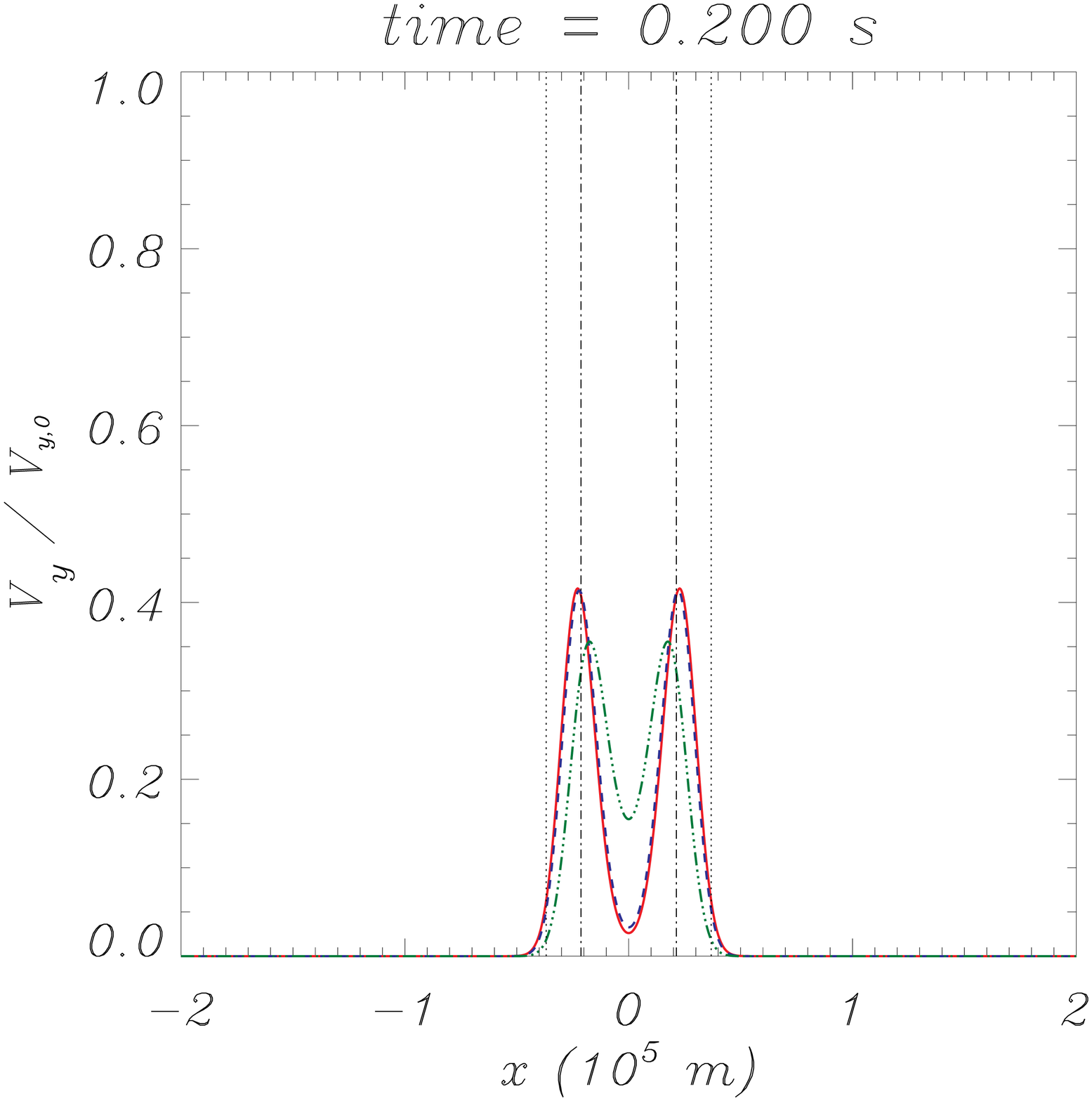} \vspace{-0.8cm} \\
		
		\hspace{0.15cm} (d) \hspace{5.4cm} (e) \hspace{5.4cm} (f) \\
	    \includegraphics[width=0.33\hsize]{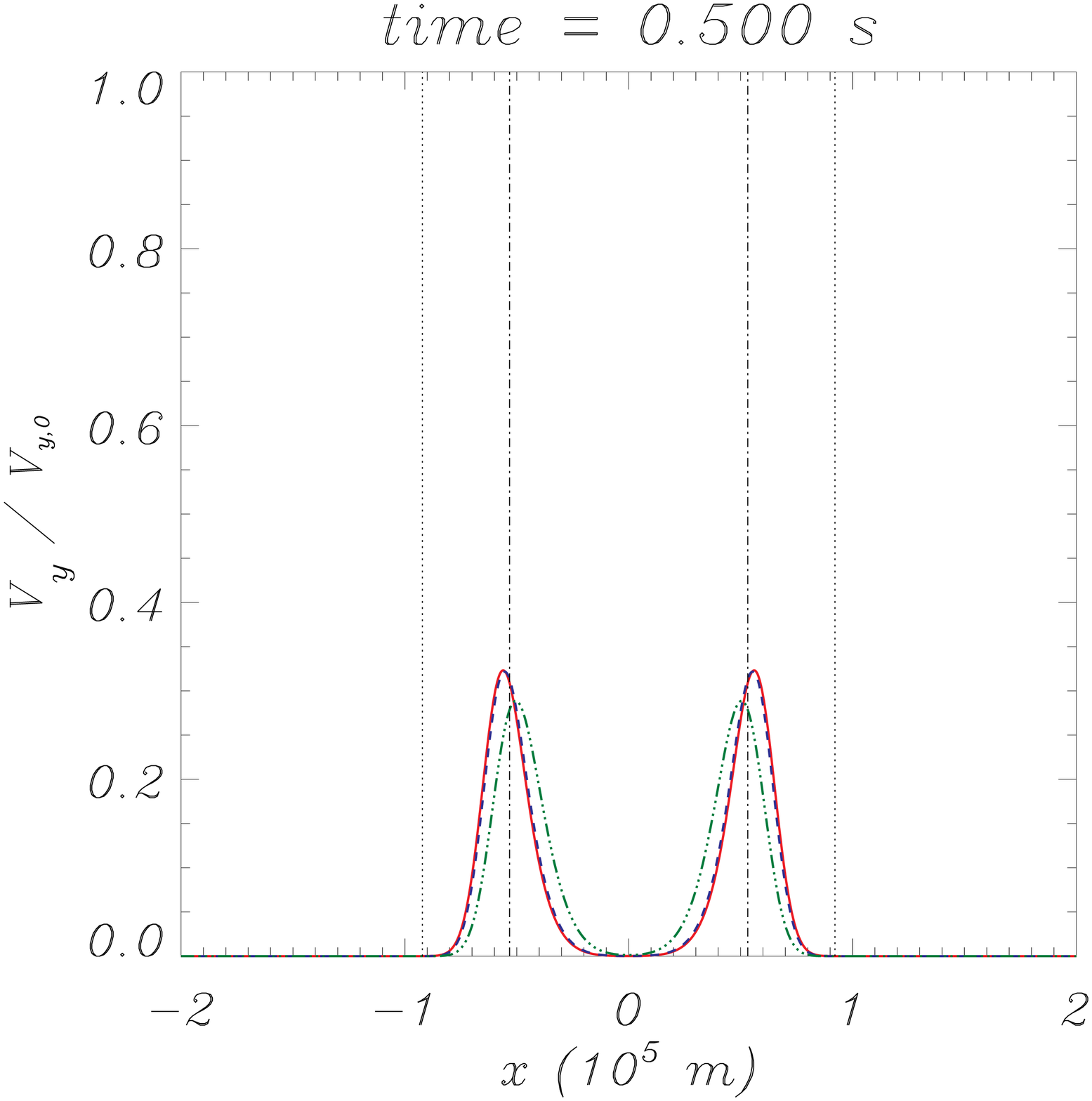}
		\includegraphics[width=0.33\hsize]{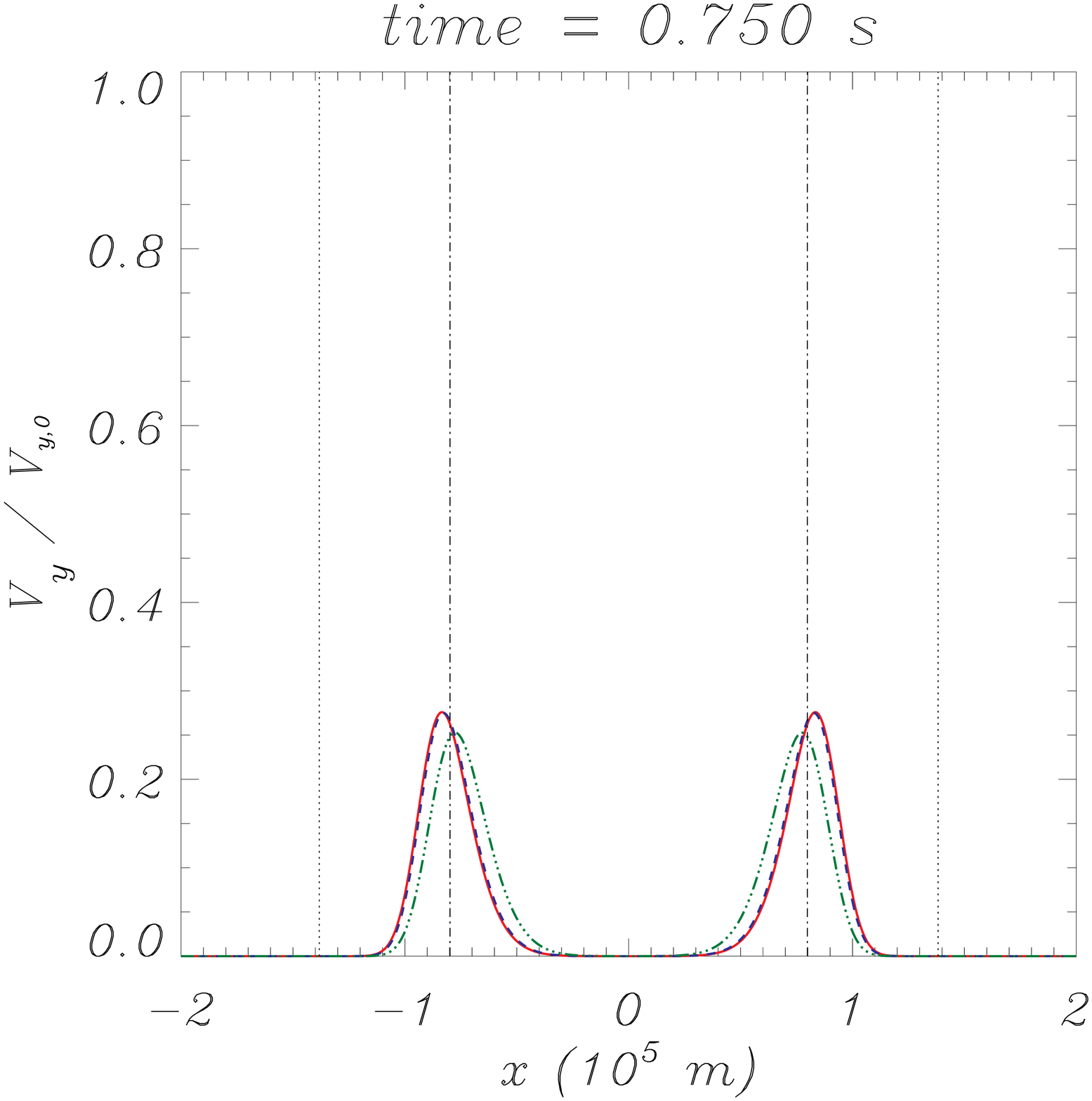}
		\includegraphics[width=0.33\hsize]{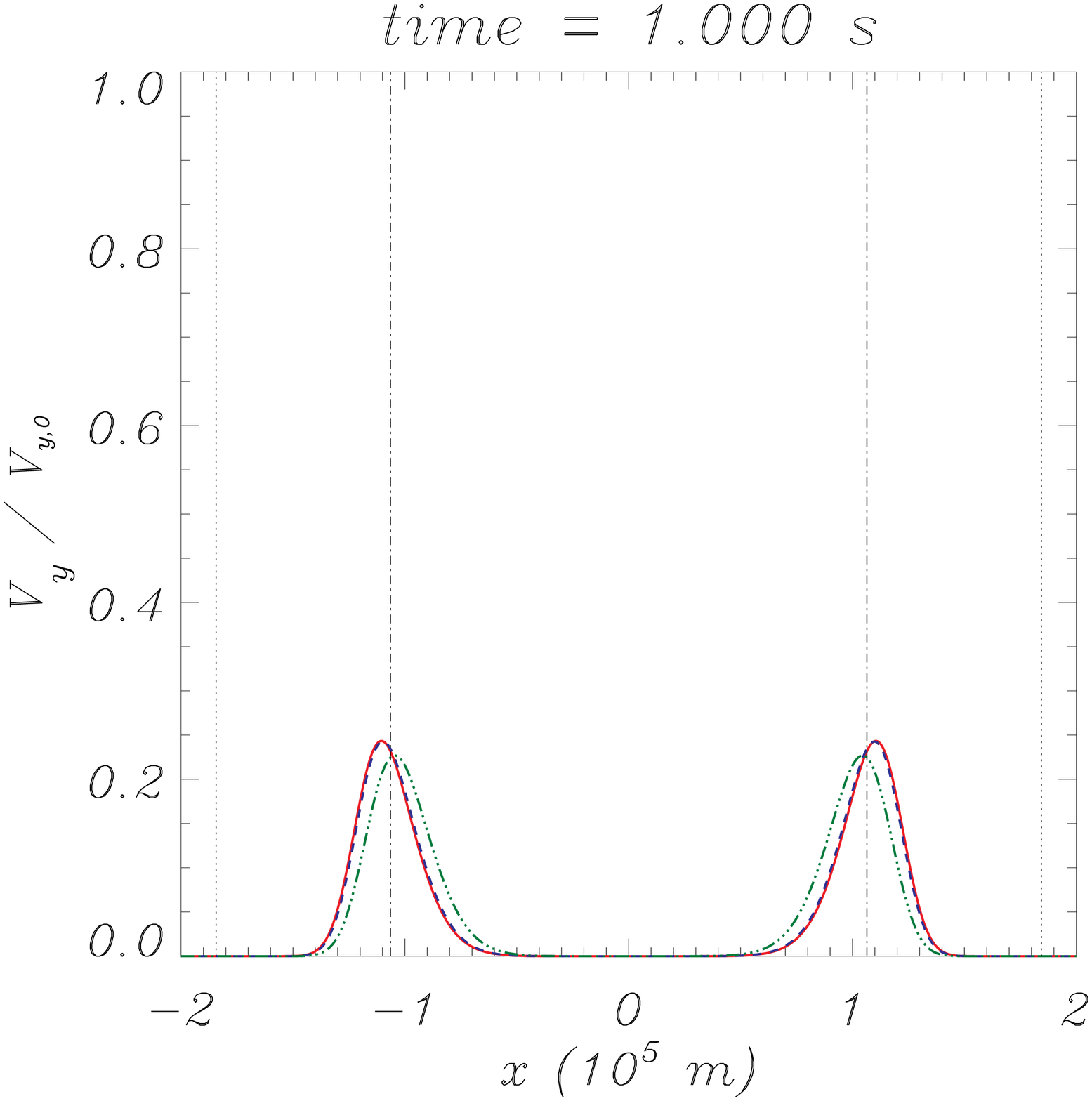}
		\caption{Simulation of waves generated by an impulsive driver with a Gaussian profile for prominence conditions. The red solid line corresponds to the protons, the blue dashed line shows the motion of neutral hydrogen and the green dotted-dashed line corresponds to the neutral helium. The dotted vertical line represents the position of points moving at Alfvén speed, while the vertical dotted-dashed line represents a motion at the modified Alfvén speed, $\widetilde{c}_{\Rm{A}}$.\\
		(An animation of this figure is available)}
		\label{fig:sim_gauss_prom}
	\end{figure}

	In the first place, we perform a simulation with the set of parameters from Table \ref{tab:plasmas} which corresponds to a prominence. The initial perturbation is applied to the $y$-component of the velocity of every species, $V_{s,y}(x,t=0)$. Since we are interested in the linear regime, the amplitude of the perturbation should be much lower than the Alfvén speed. Thus, we choose the value $V_{y,0}=10^{-3}c_{\Rm{A}}$. The initial position of the peak is $x = 0$, while the domain of the simulation is $x \in \left[-l,l\right]$, with $l = 2 \times 10^5 \ \Rm{m}$, and $FWHM=1.5 \times 10^{4} \ \Rm{m}$. The results of this simulation are shown in Figure {\ref{fig:sim_gauss_prom}}. It can be seen that protons and hydrogen are strongly coupled and that they behave almost as a single fluid. On the other hand, the coupling is weaker in the case of helium. Thus, the perturbation of the proton-hydrogen fluid is propagating approximately at the modified Alfvén speed, $\widetilde{c}_{\Rm{A}}$, while the helium fluid is trailing behind at a slightly slower speed. As time advances, the initial perturbation splits in two smaller Gaussian-like functions that propagate in opposite directions. In a case without collisions, the amplitude of each bulge would be one half of the original and it would remain constant during the whole simulation. But here, their amplitudes decrease with time; moreover, the shapes of the perturbations are not symmetric with respect to the position of their peaks and their FWHMs increase with time. Such departure from the collisionless behavior is caused by the loss of kinetic energy and the dispersion of the normal modes due to the friction.

\begin{figure}
		\hspace{0.15cm} (a) \hspace{5.4cm} (b) \hspace{5.4cm} (c) \\
		\includegraphics[width=0.33\hsize]{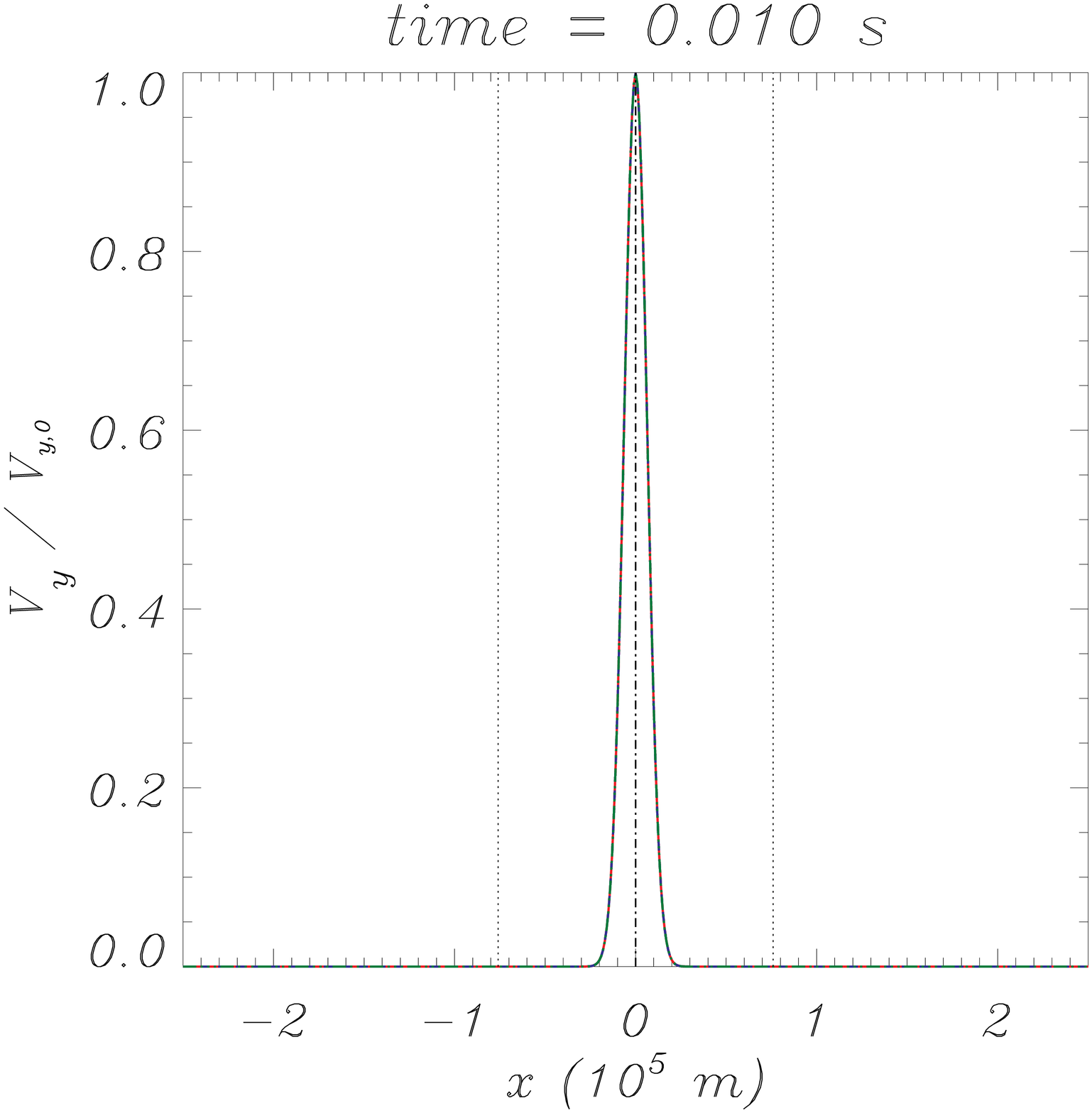}
		\includegraphics[width=0.33\hsize]{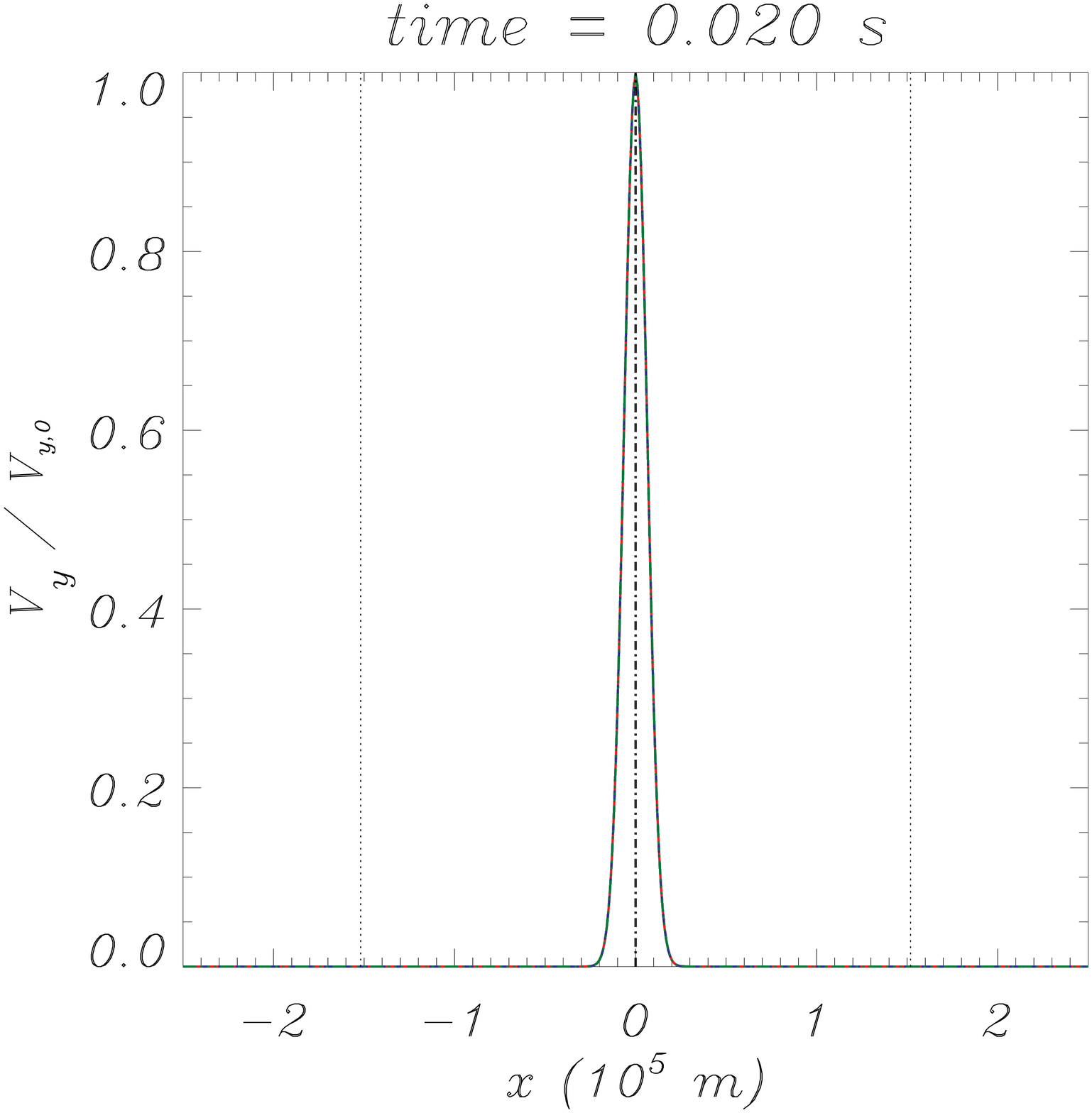}
		\includegraphics[width=0.33\hsize]{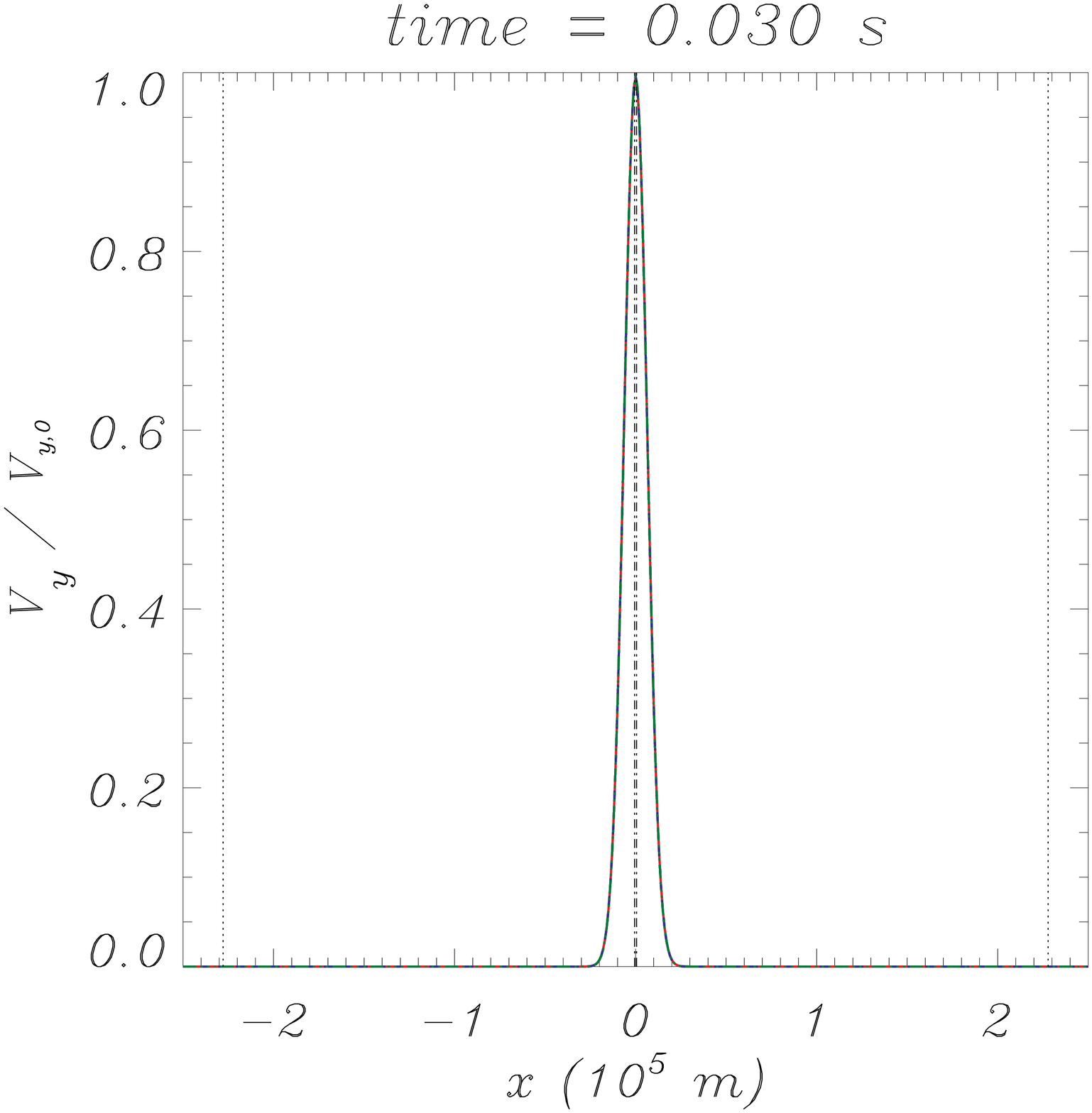} \vspace{-0.8cm} \\
		
		\hspace{0.15cm} (d) \hspace{5.4cm} (e) \hspace{5.4cm} (f) \\
		\includegraphics[width=0.33\hsize]{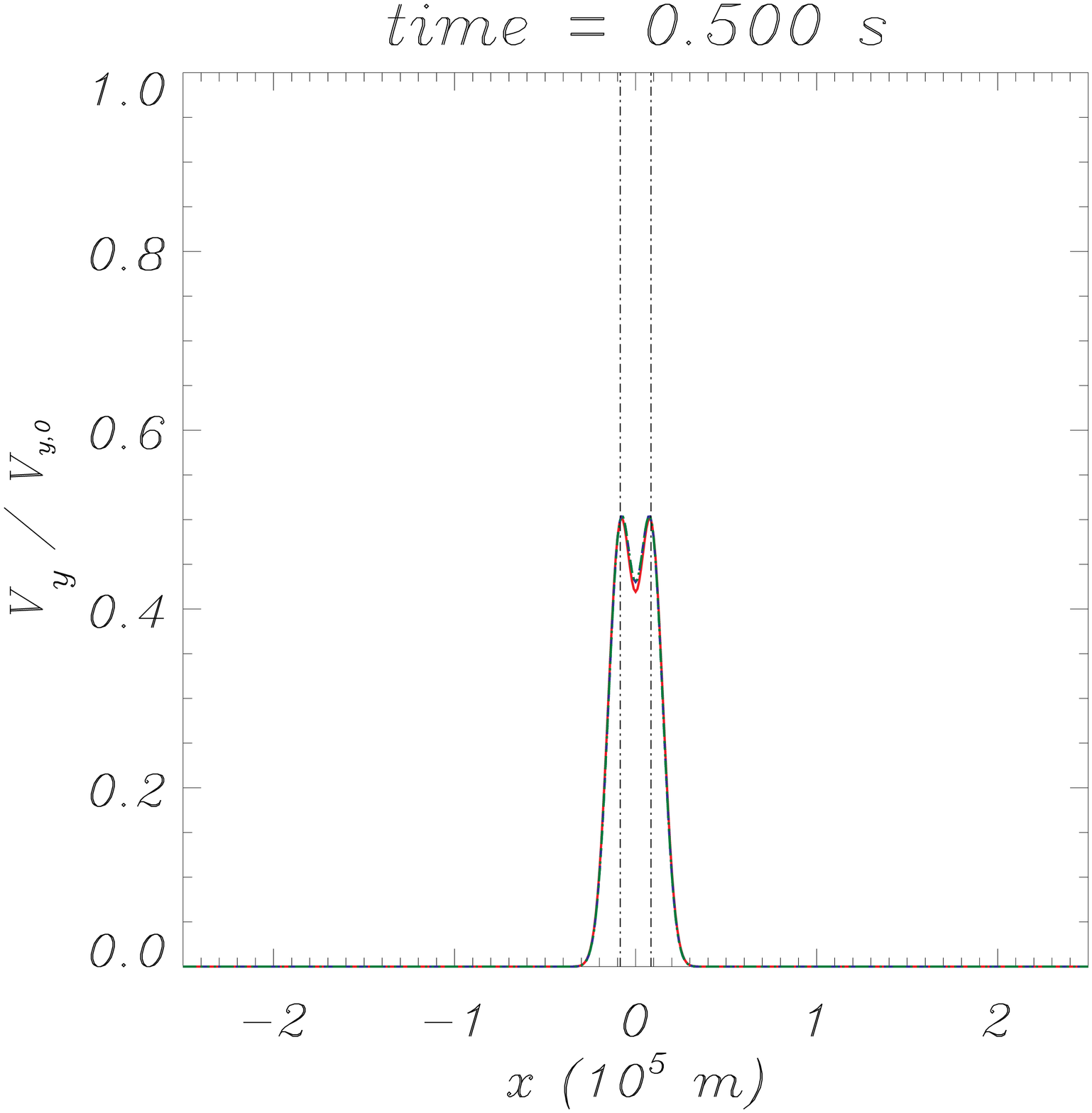}
		\includegraphics[width=0.33\hsize]{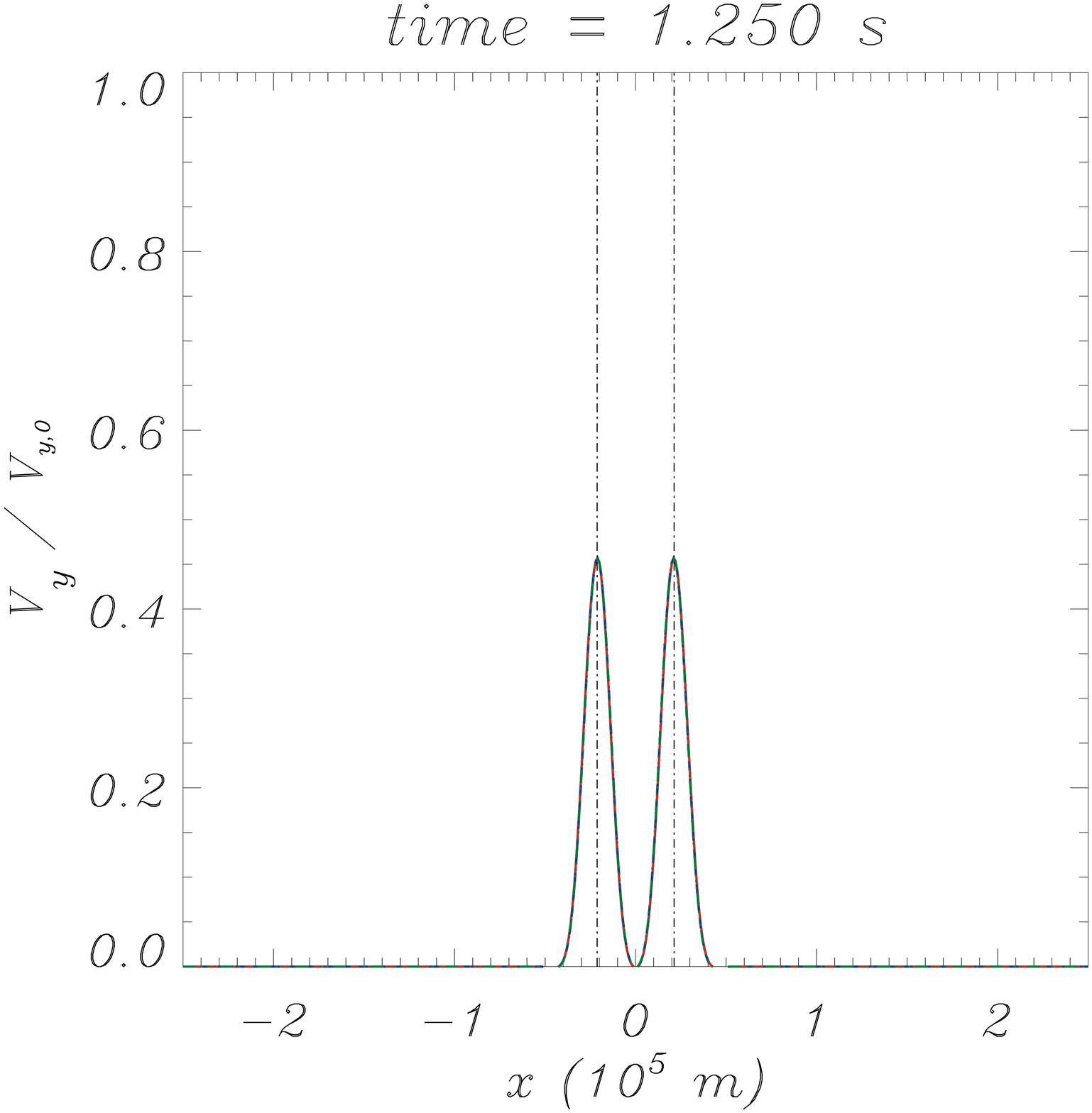}
		\includegraphics[width=0.33\hsize]{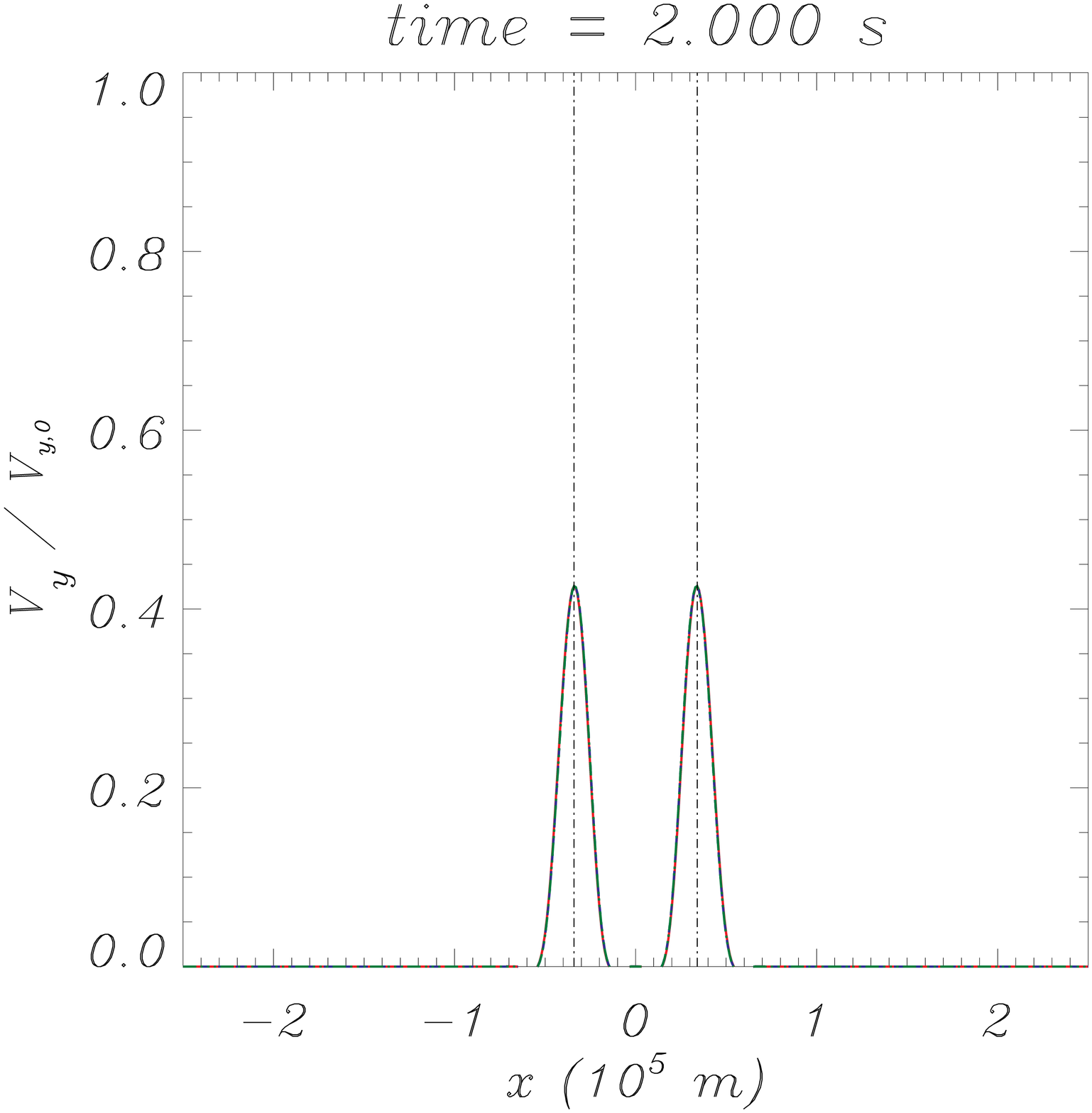}
		\caption{Simulation of waves generated by an impulsive driver with a Gaussian profile for conditions of the lower chromosphere.
		\\
		(An animation of this figure is available)}
		\label{fig:sim_gauss_lowchrom}
	\end{figure}

	Now, we simulate the temporal evolution of a perturbation in the lower chromosphere. For this problem, we use a characteristic length of $l=2.5 \times 10^{5} \ \Rm{m}$. The results are shown in Figure \ref{fig:sim_gauss_lowchrom}. We note that the motions of the singly and doubly ionized helium fluids are not represented here because a) their abundances are negligible compared to those of protons, neutral hydrogen and neutral helium, and b) they are strongly coupled to the proton fluid due to the action of the magnetic field. The simulation illustrates the high level of coupling that exists between the neutrals and ions in the chosen region: their perturbations propagate together at the modified Alfvén speed ($\widetilde{c}_{\Rm{A}} \approx 17 \ \Rm{km \ s^{-1}}$), which is much smaller than the Alfvén speed ($c_{\Rm{A}} \approx 7600 \ \Rm{km \ s^{-1}}$) due to the density of neutrals being much higher than that of ions.

\begin{figure}
		\plottwo{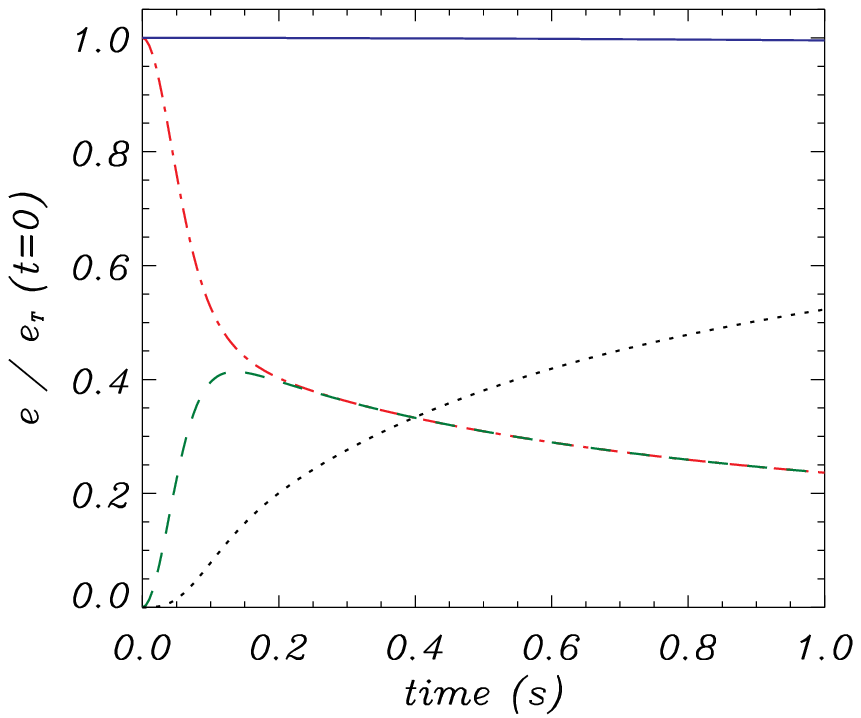}{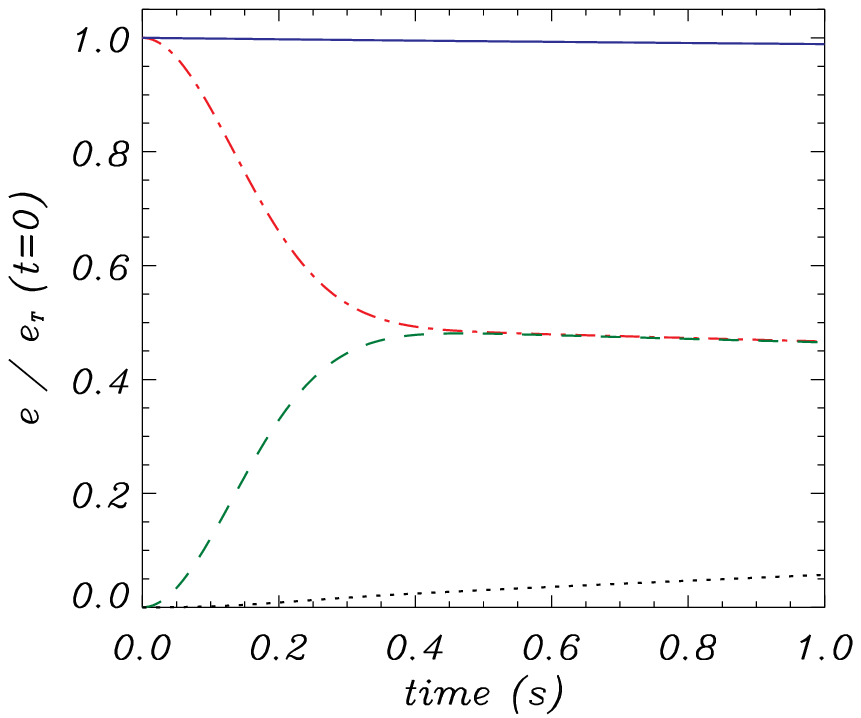}
		\caption{Temporal evolution of the total energy density, $e_{T}$ (blue solid lines), kinetic energy density (red dotted-dashed lines), magnetic energy density (green dashed lines) and internal energy density (black dotted lines). The left and right panels correspond to the simulations shown in Figures (\ref{fig:sim_gauss_prom}) and (\ref{fig:sim_gauss_lowchrom}), respectively.}
		\label{fig:sim_energy}
	\end{figure}

	It is interesting to study the evolution of the different components of the energy density during the simulations. The kinetic, magnetic and internal components of the energy density are defined as
	\begin{equation} \label{eq:EK}
		e_{K}=\frac{1}{2l}\int_{-l}^{l} \frac1{2}\sum_{s}\rho_{s}(x)V_{s}(x)^2 \, dx,
	\end{equation}
	\begin{equation} \label{eq:EB}
		e_{B}=\frac{1}{2l}\int_{-l}^{l} \frac1{2}\frac{B_{1}(x)^2}{\mu_{0}} \, dx,
	\end{equation}
	\begin{equation}
		e_{P}=\frac{1}{2l}\int_{-l}^{l} \sum_{s}\frac{P_{1,s}(x)}{\gamma -1} \, dx,
	\end{equation}
	respectively. We note that for the magnetic and the internal energy we only take into account the perturbed values and not the background ones. The behavior of these variables during the simulations shown in Figures \ref{fig:sim_gauss_prom} and \ref{fig:sim_gauss_lowchrom} is represented in the left and the right panels of Figure \ref{fig:sim_energy}, respectively. It can be seen that the kinetic energy of the initial perturbation is transformed into magnetic and internal energy and that from certain time step the kinetic and the magnetic components have the same magnitude, i.e., there is equipartition of magnetic and kinetic energy. The relative increase of internal energy is larger for the case of the prominence (left panel) than for the low chromosphere (right). This is due to a greater friction force caused by larger velocity drifts. In the prominence, the species are not as strongly coupled as in the lower chromosphere and, thus, there are higher differences in their velocities. We recall that the energy transfer due to elastic collisions, Equation (\ref{eq:qterm}), is directly proportional to the collision frequencies but it has a quadratic dependence on the velocity drifts. Nevertheless, this heat transfer is a nonlinear effect and, since the perturbations studied here are in the linear regime, the increase of internal energy due to the perturbed pressures is negligible compared with the internal energy associated to the total background pressure.

\subsection{Periodic driver} \label{sec:sim_periodic}
	The simulations of waves excited by a periodic driver are performed by imposing at a determined point of the domain a perturbation given by a periodic function of time. As shown in Paper I, depending on the chosen driver, it may possible to excite both the $L$ and $R$ modes or only one of them.
	
\begin{figure}
		\hspace{1.8cm} (a) \hspace{2.2cm} $\omega=10^{-4} \ \Omega_{p}$ \hspace{2.5cm} (b) \hspace{2.2cm} $\omega=10^{-3} \ \Omega_{p}$ \vspace{-0.8cm} \\
		\begin{center}
		\subfigure{\includegraphics[width=0.4\hsize]{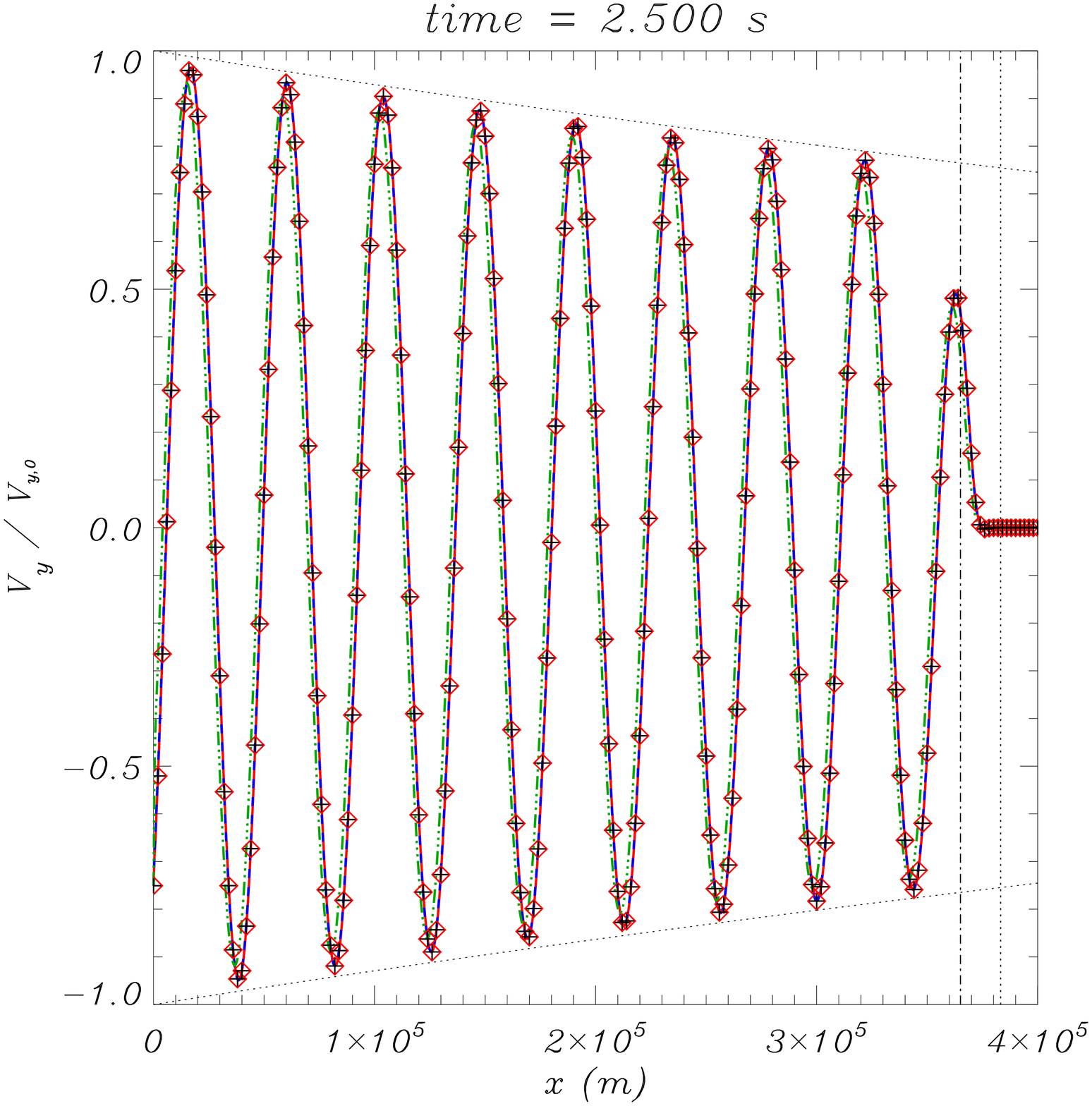} \label{subfig:shp2_a}}
		\subfigure{\includegraphics[width=0.4\hsize]{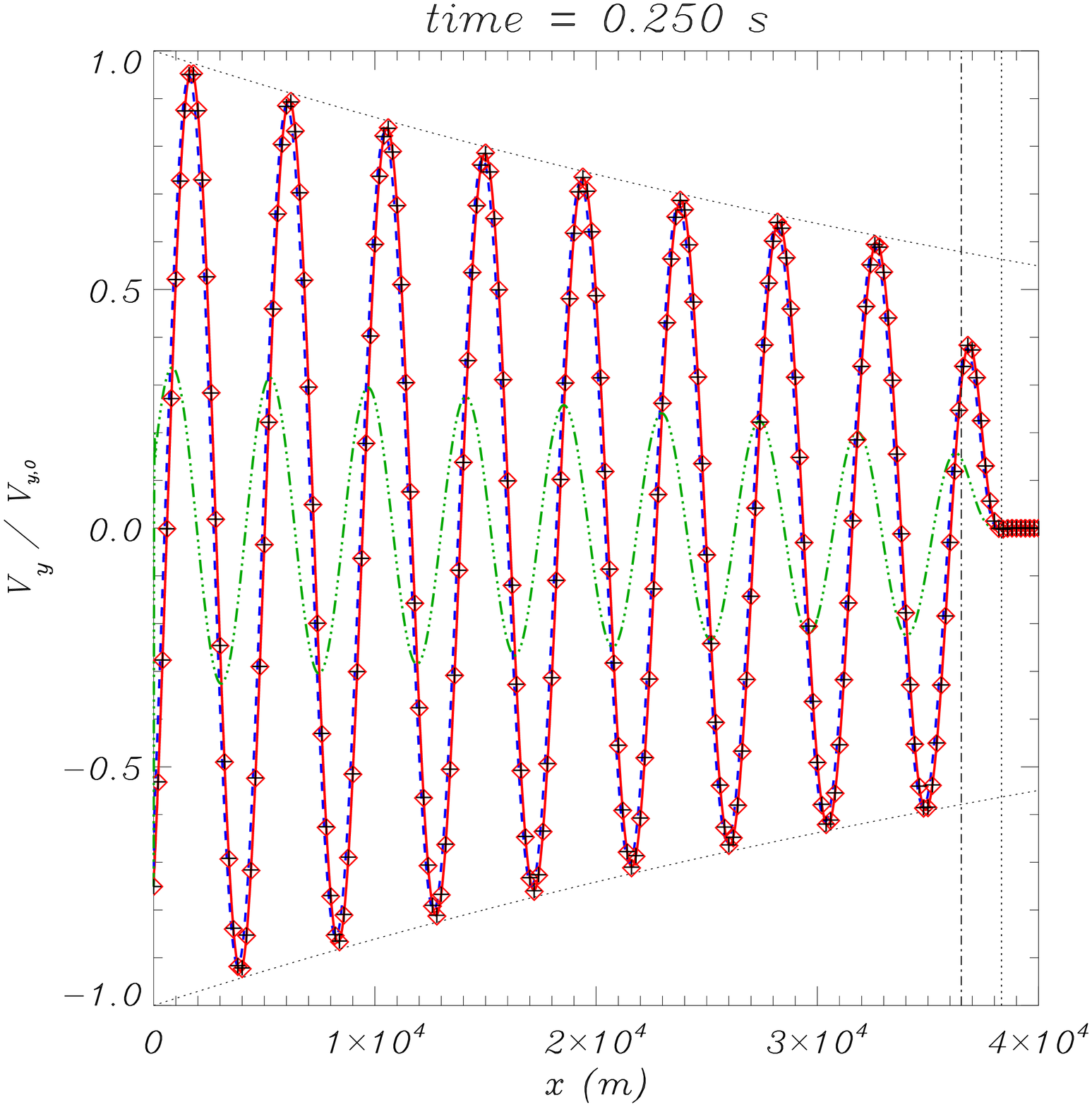} \label{subfig:shp2_b}}
		\end{center} \vspace{-0.3cm}
		\hspace{1.8cm} (c) \hspace{2.2cm} $\omega=10^{-2} \ \Omega_{p}$ \hspace{2.5cm} (d) \hspace{2.2cm} $\omega=10^{-1} \ \Omega_{p}$ \vspace{-0.8cm} \\
		\begin{center}
		\subfigure{\includegraphics[width=0.4\hsize]{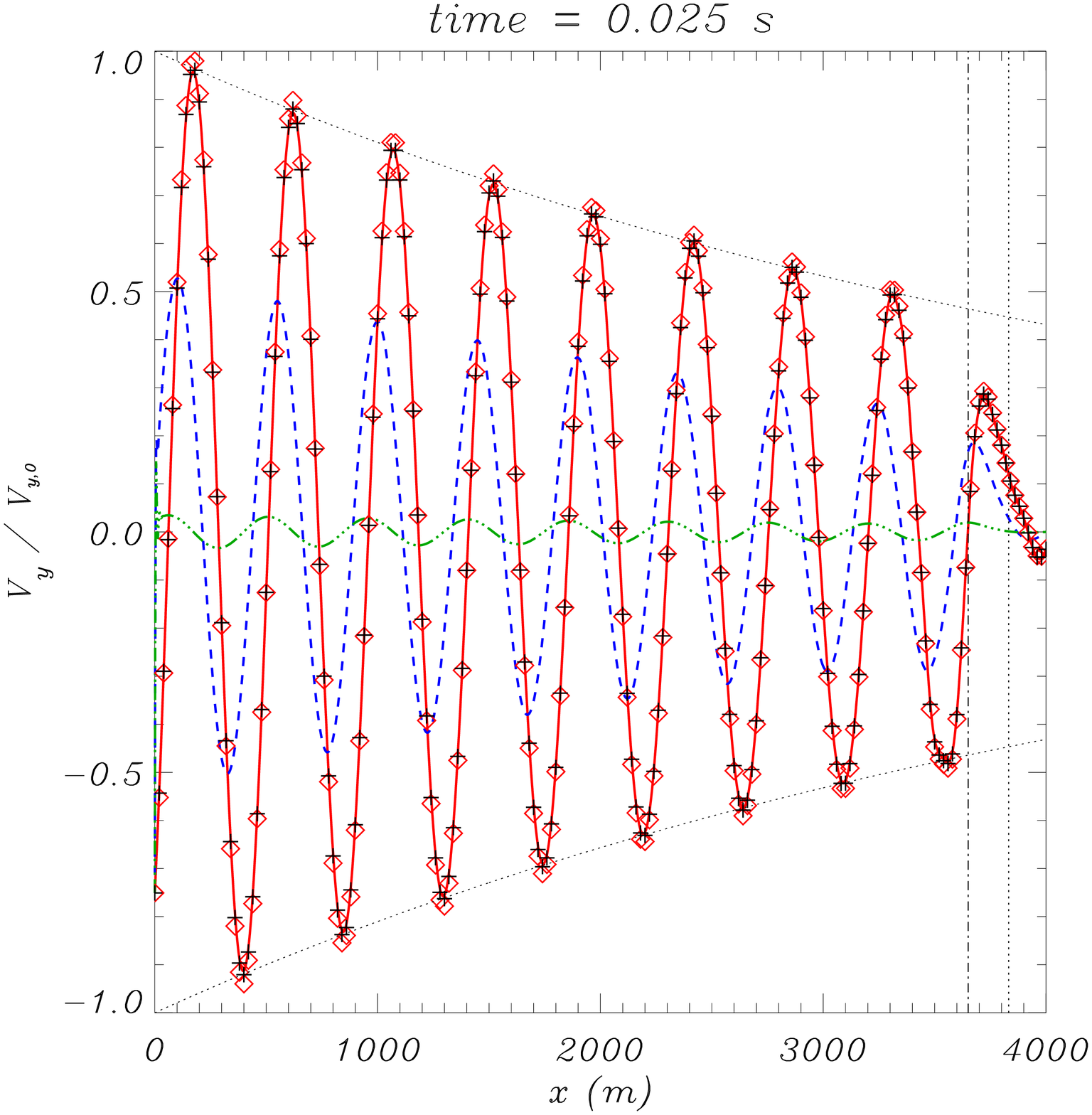} \label{subfig:shp2_c}}
	    \subfigure{\includegraphics[width=0.4\hsize]{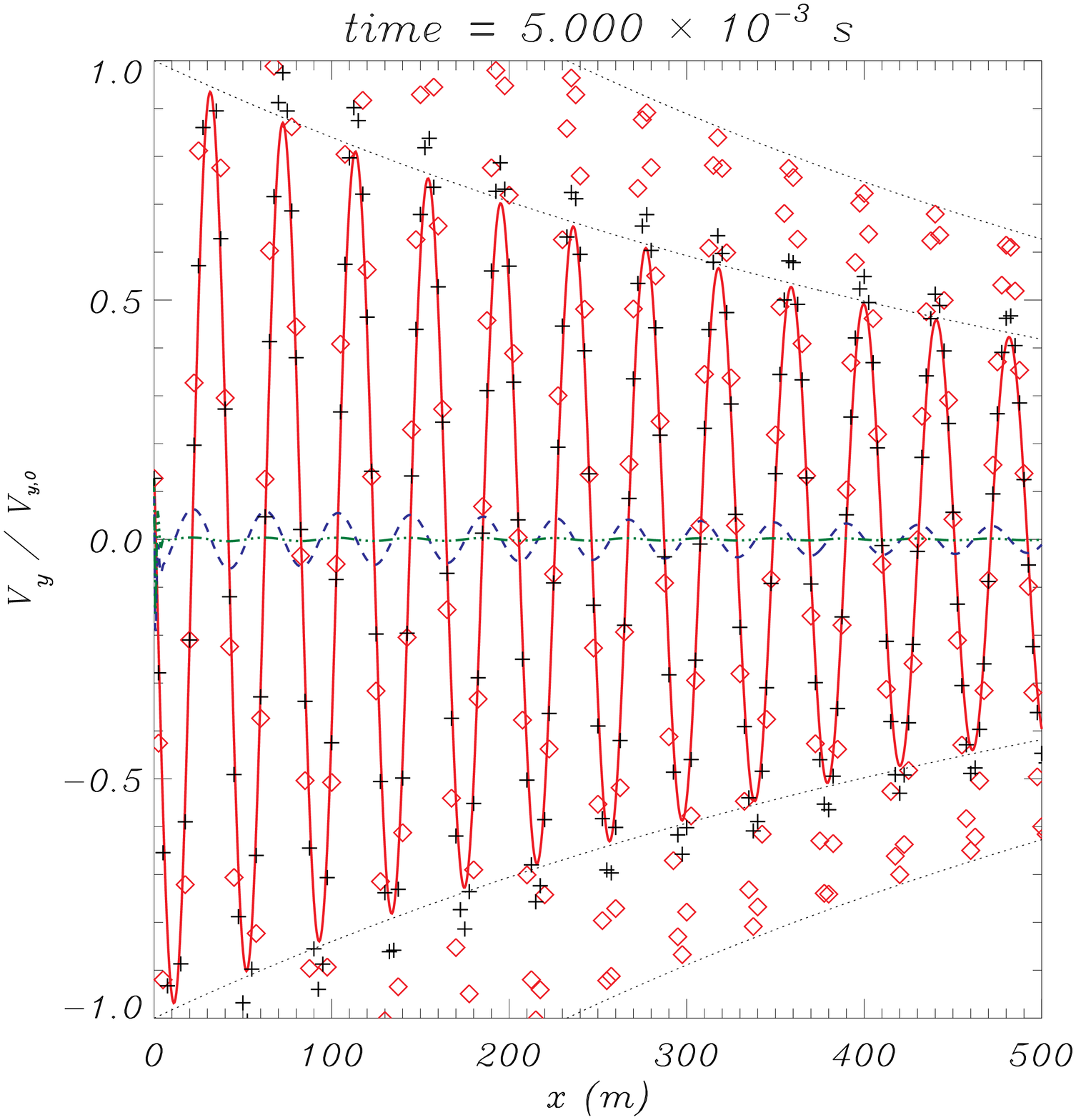} \label{subfig:shp2_d}}
		\end{center} \vspace{-0.5cm}
		\caption{Simulations of waves generated by a periodic driver with different frequencies in a region of the solar chromosphere at a height of 2016 km over the photosphere. The red solid, blue dashed and green dotted-dashed lines represent the $y$-component of the velocity of protons, neutral hydrogen and neutral helium, respectively. The red diamonds and the black crosses represent the singly and the doubly ionized helium, respectively. The vertical dotted and dotted-dashed lines represent the position of points moving at the Alfvén and the modified Alfvén speeds, respectively. The remaining black dotted lines show the damping computed from the dispersion relation.}
		\label{fig:sim_highchrom_per}
		
	\end{figure}

	In Figure \ref{fig:sim_highchrom_per}, we show the results of several simulations of waves excited by a periodic driver using the physical conditions that correspond to the region I of Table \ref{tab:plasmas}, i.e., the high chromosphere. The driver, applied to the point $x = 0$, is given by
	\begin{equation} \label{eq:driver_V}
		\bm{V_{s}}(x=0,t)=\left( \begin{array}{c}
		0 \\
		V_{0} \cos \left(\omega t\right) \\
		-V_{0} \sin \left(\omega t\right)
		\end{array} \right)
	\end{equation}
	and
	\begin{equation} \label{eq:driver_B}
		\bm{B_{1}}(x=0,t)=\left( \begin{array}{c}
		0 \\
		B_{1,0} \cos \left(\omega t\right) \\
		-B_{1,0} \sin \left(\omega t\right)
		\end{array} \right),
	\end{equation}
	which produces the excitation of the $L$ mode. The amplitude of the magnetic field perturbation is given by $B_{1,0}=-|\bm{B_{0}}|V_{0}/c_{\Rm{A}}$. To assure that the simulations correspond to the linear regime, we use the value $V_{0}=10^{-3}c_{\Rm{A}}$.
	
	Figure \ref{fig:sim_highchrom_per}(a) shows that, at a driving frequency of $\omega = 10^{-4} \Omega_{p}$, there is a strong coupling between all the species and the perturbation travels at the speed $\widetilde{c}_{\Rm{A}}$. In addition, the damping of the wave is in good agreement with that predicted by the dispersion relation (see the envelope of the wave, represented by the black dotted lines). When the frequency of the driver is increased, it can be found that the species begin to uncouple. This behavior can be noticed in Figure \ref{fig:sim_highchrom_per}(b), where the first species to uncouple from the others is the neutral helium. The reason is that the driving frequency, $\omega=10^{-3} \Omega_{p} \approx 210 \ \Rm{rad \ s^{-1}}$, is larger than the collision frequencies of the helium, for instance, $\nu_{p\Rm{He}} \approx 2.5 \ \Rm{Hz}$, $\nu_{\Rm{He}p} \approx 43 \ \Rm{Hz}$ or $\nu_{\Rm{HHe}} \approx 3.5 \ \Rm{Hz}$, but is of the order of or smaller than the collision frequencies of the other species, e.g., $\nu_{p\Rm{H}} \approx 120 \ \Rm{Hz}$ or $\nu_{\Rm{H}p} \approx 1400 \ \Rm{Hz}$. Thus, during one period of the oscillation, neutral helium particles do not collide frequently enough with the other species for the neutral helium fluid to completely follow the magnetic field oscillations.
	
	If the driving frequency is increased up to $\omega = 10^{-2}\Omega_{p}$, the hydrogen fluid starts to exhibit a similar behavior as the one explained for helium in the previous paragraph, as it can be checked in Figure \ref{fig:sim_highchrom_per}(c). Now, the interaction of neutral helium with the other species is even weaker than before and the amplitude of its oscillation is greatly reduced. Finally, Figure \ref{fig:sim_highchrom_per}(d) shows that at frequencies of the order of $\omega = 0.1 \Omega_{p} \approx 21000 \ \Rm{rad \ s^{-1}}$ even the ionized species begin to uncouple from each other, due to the fact that $\omega$ is in this case larger than $\nu_{p\Rm{He} \ \textsc{ii}}$, $\nu_{\Rm{He} \ \textsc{ii}p}$, $\nu_{p\Rm{He} \ \textsc{iii}}$ or $\nu_{\Rm{He} \ \textsc{iii}p}$. At these frequencies, there is almost no interaction with the neutral species, and at even higher frequencies the perturbation behaves as if the medium were fully ionized ignoring the presence of neutrals, situation that has already been studied in Paper I. Hence, there is almost no propagation of the perturbation in the neutral fluids at the high-frequency range.

\begin{figure}
		\centering
		\includegraphics[width=0.5\hsize]{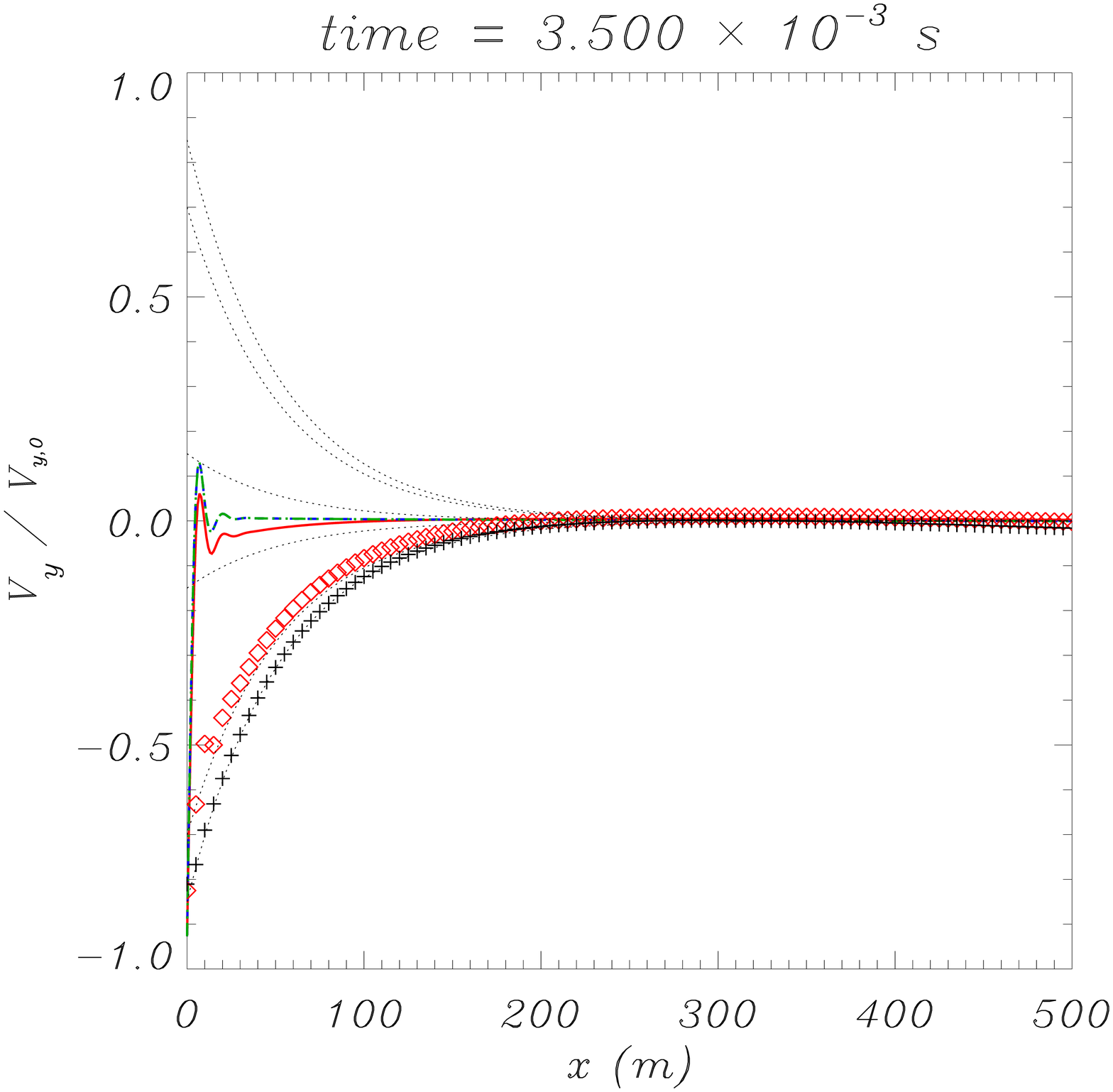}
		\caption{Simulation of waves generated by a periodic driver with frequency $\omega=10^{-3} \ \Omega_{p}$ in a region of the solar chromosphere at a height of 500 km over the photosphere. The meaning of the colors and styles of the lines is the same as in Figure \ref{fig:sim_highchrom_per}. \\
		(An animation of this figure is available)}
		\label{fig:sim_lowchrom_per}
	\end{figure}

	Now, we turn briefly our attention to the region of the lower chromosphere, where, according to Figure \ref{fig:qfactor_per}(c), the $L$ mode is overdamped at frequencies larger than $\omega_{H}$. Figure \ref{fig:sim_lowchrom_per} represents a simulation with a driver given by
	\begin{equation} \label{eq:driver_V2}
		\bm{V_{s}}(x=0,t)=\left( \begin{array}{c}
		0 \\
		V_{0} \cos \left(\omega t\right) \\
		-V_{0} \sin \left(\omega t\right)
		\end{array} \right),
	\end{equation}
	and a frequency of $\omega=10^{-3} \Omega_{p}$. A total of $N = 2001$ points have been used to cover the domain $x \in [0,2] \ \Rm{km}$, but only the section $x \in [0,0.5] \ \Rm{km}$ is shown in the plot. While the relevant physical behavior occurs in the displayed section, the larger domain is used to avoid possible unwanted numerical effects caused by the rightmost boundary. It can be seen that there is a weak coupling between the motion of ions and neutrals. Neutrals stay almost at rest, except close to $x = 0$, while the driver causes a motion in the ions that is spatially overdamped. We note that, as in the previous cases, the results of the simulation are in good agreement with the solutions obtained from the dispersion relation analyzed in Section \ref{sec:dr}. However, a small oscillation appears in the left area that does not correspond to any of the normal modes given by the dispersion relation. Such oscillation is due to a numerical effect: we have checked that its extension and amplitude depends on the spatial resolution used in the simulation.
		
\section{Conclusions}
	In this paper, we have continued the theoretical investigation of high-frequency waves in multi-fluid plasmas that was started in Paper I. The main difference with Paper I is that here we have allowed the plasma to be partially ionized, so we have added the dynamics of neutrals. This enables us to apply our multi-fluid model to partially ionized plasmas of the solar atmosphere. We have studied three specific cases which correspond to similar physical properties than those which can be found in the higher chromosphere (where $\chi < 1$), a prominence (with $\chi > 1$) and the lower chromosphere (a weakly ionized plasma with $\chi \gg 1$). The investigation and comparison of environments with such different degrees of ionization lead to a better understanding of the influence of neutral species on the propagation of small-amplitude perturbations. In addition, we have also incorporated the effect of Ohm's diffusion or magnetic resistivity in a more complete version of the generalized Ohm's law, which improves the description of the physics involved in the damping and dissipation of the waves. This improvement has allowed us to check that resistivity has a negligible influence on the properties of low frequency waves but that it becomes an important effect when the frequency is increased, provoking a remarkably rise of the damping of the $R$ modes.
	
	Previous works \citep[see, e.g.,][]{1956MNRAS.116..314P,1992Natur.360..241H,2007A&A...461..731F,2009ApJ...699.1553S,2011A&A...529A..82Z,2011A&A...534A..93Z,2012A&A...537A..84S} have already demonstrated that the inclusion of neutral components in the plasma modifies the oscillation period of the low frequency Alfvén waves and produces a damping on the perturbations. We have confirmed those findings and, moreover, we have explored a larger range of frequencies to investigate the effects of elastic collisions on the ion-cyclotron and the $\bm{R}$ modes. For instance, in the case of the higher chromosphere, we have found that the damping of the Alfvén waves is dominated by collisions with neutrals, while the damping of the higher-frequency $L$ and $R$ modes is dominated by collisions with ions and electrons. Such behavior is explained by the fact that the damping is more efficient when the collision frequency is of the order of the oscillation frequency \citep{2011A&A...529A..82Z,2013ApJ...767..171S} and that collisions with neutrals have lower frequencies than those between ions.
	
	At high enough frequencies, the properties of the $L$ modes clearly diverge from those of the $R$ modes due to the effect of Hall's current \citep{1960RSPTA.252..397L,2001paw..book.....C}. For instance, the quality factors of the $R$ modes are larger than those of the $L$ modes. In a fully ionized plasma, this separation occurs at oscillation frequencies of the order of the lower cyclotron frequency. However, if the plasma is weakly ionized and the ion-neutral collision frequency is comparable to or larger than the cyclotron frequency, the effective gyrofrequency of ions is greatly reduced \citep{2006astro.ph..8008P,2008MNRAS.385.2269P} and Hall's term must be taken into account for frequencies of the order of $\omega_{\Rm{H}}$, which is much smaller than $\Omega_{i}$. Here, we have found that, for the case of the lower chromosphere, Hall's current plays a very important role in the propagation of waves even for frequencies as low as 20 $\Rm{rad \ s^{-1}}$, approximately.
	
	In Paper I we showed that momentum transfer collisions between ions remove the cyclotron resonances and cutoffs that appear in the collisionless case when a periodic driver is considered, a result which coincides with the findings of \citet{2010PhPl...17c2102R}. In the present paper, we have found that friction due to collisions with neutral species causes a similar effect. Hence, waves can propagate at any frequency, although they are greatly damped at frequencies higher than $\omega_{\Rm{H}}$.
	
	Furthermore, in good agreement with \citet{2015A&A...573A..79S}, we have shown that in weakly ionized plasmas there are no strict cutoffs for Alfvén waves generated by an impulsive driver, but that there may be intervals of wavenumbers where the perturbations are overdamped.
	
	The simulations of waves excited by an impulsive driver have shown that, with the physical conditions considered for a solar prominence and for a region in the lower chromosphere, there is a strong coupling between all the species that compose the plasma. Therefore, a perturbation generated by an initial Gaussian profile propagates at the modified Alfvén speed, $\widetilde{c}_{\Rm{A}}$, which in the case of the lower chromosphere is approximately two orders of magnitude smaller than the classic Alfvén speed. This behavior agrees with the results obtained from the analysis of the dispersion relation. The simulations also reflect that the friction force caused by collisions dissipates a fraction of the kinetic energy of the initial perturbation and transforms it into internal energy, which implies an increment of the temperature of the plasma. However, this heating is a nonlinear effect and, due to the small amplitude of the perturbations investigated in the present paper, the temperature rise obtained is negligible compared to the background value. The subject of the heating of the solar atmosphere by MHD waves is of enormous interest \citep[see, e.g.,][]{2001ApJ...558..859D,2004A&A...422.1073K,2006AdSpR..37..447K,2010SSRv..151..243L,2010MmSAI..81..654H,2015RSPTA.37340261A,2015ASSL..415..157G,2016A&A...592A..28S} and in the forthcoming third and final paper of this series we will apply our model to the investigation of the heating caused by large-amplitude perturbations in partially ionized plasmas.
	
	Finally, the simulations of waves generated by a periodic driver have shown that as the frequency of the driver is increased, the different species begin to uncouple. Neutral species only remain strongly coupled to the ions at low frequencies. In the high-frequency limit, the plasma behaves almost as if it were fully ionized, with a minimal influence from the neutral components. In addition, we have checked numerically the prediction from the dispersion relation for the region of the lower chromosphere that the $L$ mode is strongly overdamped if $\omega > \omega_{\Rm{H}}$: the kinetic energy of the perturbation is completely dissipated in a few hundreds of meters from its point of origin.
	
	It would have been interesting to study the case of the photosphere, but in that environment the number densities are extremely large, what leads to huge collision frequencies. Due to the CFL condition, our numerical code would require the use of extraordinarily tiny time steps. Moreover, when such high frequencies are involved it is more appropriate to resort to hybrid fluid-kinetic or fully kinetic models. Another issue worth of consideration in forthcoming investigations is the effect of inhomogeneities, which may provoke a different response in each species of the multi-component plasmas. 

\acknowledgements
We acknowledge the support from MINECO and FEDER funds through grant AYA2014-54485-P. DM acknowledges support from MINECO through an “FPI” grant. RS acknowledges support from MINECO and UIB through a ``Ramón y Cajal" grant (RYC-2014-14970). JT acknowledges support from MINECO and UIB through a ``Ramón y Cajal" grant.

\bibliographystyle{aasjournal}
\bibliography{mybib2}

\begin{thebibliography}{}
\expandafter\ifx\csname natexlab\endcsname\relax\def\natexlab#1{#1}\fi
\providecommand{\url}[1]{\href{#1}{#1}}

\bibitem[{{Alfv{\'e}n}(1942)}]{1942Natur.150..405A}
{Alfv{\'e}n}, H. 1942, \nat, 150, 405

\bibitem[{{Allen} {et~al.}(1998){Allen}, {Habbal}, \&
  {Hu}}]{1998JGR...103.6551A}
{Allen}, L.~A., {Habbal}, S.~R., \& {Hu}, Y.~Q. 1998, \jgr, 103, 6551

\bibitem[{{Arber} {et~al.}(2016){Arber}, {Brady}, \&
  {Shelyag}}]{2016ApJ...817...94A}
{Arber}, T.~D., {Brady}, C.~S., \& {Shelyag}, S. 2016, \apj, 817, 94

\bibitem[{{Arregui}(2015)}]{2015RSPTA.37340261A}
{Arregui}, I. 2015, Philosophical Transactions of the Royal Society of London
  Series A, 373, 20140261

\bibitem[{{Balsara}(1996)}]{1996ApJ...465..775B}
{Balsara}, D.~S. 1996, \apj, 465, 775

\bibitem[{{Black} \& {Scott}(1982)}]{1982ApJ...263..696B}
{Black}, D.~C., \& {Scott}, E.~H. 1982, \apj, 263, 696

\bibitem[{{Bona} {et~al.}(2009){Bona}, {Bona-Casas}, \&
  {Terradas}}]{2009JCoPh.228.2266B}
{Bona}, C., {Bona-Casas}, C., \& {Terradas}, J. 2009, Journal of Computational
  Physics, 228, 2266

\bibitem[{{Braginskii}(1965)}]{1965RvPP....1..205B}
{Braginskii}, S.~I. 1965, Reviews of Plasma Physics, 1, 205

\bibitem[{Callen(2006)}]{Callen2006a}
Callen, J.~D. 2006, in Fundamentals of Plasma Physics, draft edn. (Madison,
  Wisconsin)

\bibitem[{{Cowling}(1956)}]{1956MNRAS.116..114C}
{Cowling}, T.~G. 1956, \mnras, 116, 114

\bibitem[{{Cramer}(2001)}]{2001paw..book.....C}
{Cramer}, N.~F. 2001, {The Physics of Alfv{\'e}n Waves} (Wiley-VCH)

\bibitem[{{De Pontieu} {et~al.}(2001){De Pontieu}, {Martens}, \&
  {Hudson}}]{2001ApJ...558..859D}
{De Pontieu}, B., {Martens}, P.~C.~H., \& {Hudson}, H.~S. 2001, \apj, 558, 859

\bibitem[{{Dickinson} {et~al.}(1999){Dickinson}, {Lee}, \&
  {Viehland}}]{1999JPhB...32.4919D}
{Dickinson}, A.~S., {Lee}, M.~S., \& {Viehland}, L.~A. 1999, Journal of Physics
  B Atomic Molecular Physics, 32, 4919

\bibitem[{{Draine}(1986)}]{1986MNRAS.220..133D}
{Draine}, B.~T. 1986, \mnras, 220, 133

\bibitem[{{Fiedler} \& {Mouschovias}(1992)}]{1992ApJ...391..199F}
{Fiedler}, R.~A., \& {Mouschovias}, T.~C. 1992, \apj, 391, 199

\bibitem[{{Fiedler} \& {Mouschovias}(1993)}]{1993ApJ...415..680F}
---. 1993, \apj, 415, 680

\bibitem[{{Fontenla} {et~al.}(1993){Fontenla}, {Avrett}, \&
  {Loeser}}]{1993ApJ...406..319F}
{Fontenla}, J.~M., {Avrett}, E.~H., \& {Loeser}, R. 1993, \apj, 406, 319

\bibitem[{{Forteza} {et~al.}(2007){Forteza}, {Oliver}, {Ballester}, \&
  {Khodachenko}}]{2007A&A...461..731F}
{Forteza}, P., {Oliver}, R., {Ballester}, J.~L., \& {Khodachenko}, M.~L. 2007,
  \aap, 461, 731

\bibitem[{{Gilbert}(2015)}]{2015ASSL..415..157G}
{Gilbert}, H. 2015, in Astrophysics and Space Science Library, Vol. 415, Solar
  Prominences, ed. J.-C. {Vial} \& O.~{Engvold}, 157

\bibitem[{{Goodman}(2011)}]{2011ApJ...735...45G}
{Goodman}, M.~L. 2011, \apj, 735, 45

\bibitem[{{Haerendel}(1992)}]{1992Natur.360..241H}
{Haerendel}, G. 1992, \nat, 360, 241

\bibitem[{{Heinzel}(2015)}]{2015ASSL..415..103H}
{Heinzel}, P. 2015, in Astrophysics and Space Science Library, Vol. 415, Solar
  Prominences, ed. J.-C. {Vial} \& O.~{Engvold}, 103

\bibitem[{{Heinzel} {et~al.}(2010){Heinzel}, {Anzer}, \&
  {Gun{\'a}r}}]{2010MmSAI..81..654H}
{Heinzel}, P., {Anzer}, U., \& {Gun{\'a}r}, S. 2010, \memsai, 81, 654

\bibitem[{{Heinzel} {et~al.}(2015){Heinzel}, {Gun{\'a}r}, \&
  {Anzer}}]{2015A&A...579A..16H}
{Heinzel}, P., {Gun{\'a}r}, S., \& {Anzer}, U. 2015, \aap, 579, A16

\bibitem[{{Jephcott} \& {Stocker}(1962)}]{1962JFM....13..587J}
{Jephcott}, D.~F., \& {Stocker}, P.~M. 1962, Journal of Fluid Mechanics, 13,
  587

\bibitem[{{Khodachenko} {et~al.}(2004){Khodachenko}, {Arber}, {Rucker}, \&
  {Hanslmeier}}]{2004A&A...422.1073K}
{Khodachenko}, M.~L., {Arber}, T.~D., {Rucker}, H.~O., \& {Hanslmeier}, A.
  2004, \aap, 422, 1073

\bibitem[{{Khodachenko} {et~al.}(2006){Khodachenko}, {Rucker}, {Oliver},
  {Arber}, \& {Hanslmeier}}]{2006AdSpR..37..447K}
{Khodachenko}, M.~L., {Rucker}, H.~O., {Oliver}, R., {Arber}, T.~D., \&
  {Hanslmeier}, A. 2006, Advances in Space Research, 37, 447

\bibitem[{{Khomenko} \& {Collados}(2012)}]{2012ApJ...747...87K}
{Khomenko}, E., \& {Collados}, M. 2012, \apj, 747, 87

\bibitem[{{Khomenko} {et~al.}(2014){Khomenko}, {Collados}, {D{\'{\i}}az}, \&
  {Vitas}}]{2014PhPl...21i2901K}
{Khomenko}, E., {Collados}, M., {D{\'{\i}}az}, A., \& {Vitas}, N. 2014, Physics
  of Plasmas, 21, 092901

\bibitem[{{Kulsrud} \& {Pearce}(1969)}]{1969ApJ...156..445K}
{Kulsrud}, R., \& {Pearce}, W.~P. 1969, \apj, 156, 445

\bibitem[{{Kumar} \& {Roberts}(2003)}]{2003SoPh..214..241K}
{Kumar}, N., \& {Roberts}, B. 2003, \solphys, 214, 241

\bibitem[{{Kuridze} {et~al.}(2016){Kuridze}, {Zaqarashvili}, {Henriques},
  {Mathioudakis}, {Keenan}, \& {Hanslmeier}}]{2016ApJ...830..133K}
{Kuridze}, D., {Zaqarashvili}, T.~V., {Henriques}, V., {et~al.} 2016, \apj,
  830, 133

\bibitem[{{Labrosse} {et~al.}(2010){Labrosse}, {Heinzel}, {Vial}, {Kucera},
  {Parenti}, {Gun{\'a}r}, {Schmieder}, \& {Kilper}}]{2010SSRv..151..243L}
{Labrosse}, N., {Heinzel}, P., {Vial}, J.-C., {et~al.} 2010, \ssr, 151, 243

\bibitem[{{Leake} \& {Arber}(2006)}]{2006A&A...450..805L}
{Leake}, J.~E., \& {Arber}, T.~D. 2006, \aap, 450, 805

\bibitem[{{Leake} {et~al.}(2005){Leake}, {Arber}, \&
  {Khodachenko}}]{2005A&A...442.1091L}
{Leake}, J.~E., {Arber}, T.~D., \& {Khodachenko}, M.~L. 2005, \aap, 442, 1091

\bibitem[{{Lewkow} {et~al.}(2012){Lewkow}, {Kharchenko}, \&
  {Zhang}}]{2012ApJ...756...57L}
{Lewkow}, N.~R., {Kharchenko}, V., \& {Zhang}, P. 2012, \apj, 756, 57

\bibitem[{{Lighthill}(1960)}]{1960RSPTA.252..397L}
{Lighthill}, M.~J. 1960, Philosophical Transactions of the Royal Society of
  London Series A, 252, 397

\bibitem[{{Mart{\'{\i}}nez-G{\'o}mez}
  {et~al.}(2016){Mart{\'{\i}}nez-G{\'o}mez}, {Soler}, \&
  {Terradas}}]{2016ApJ...832..101M}
{Mart{\'{\i}}nez-G{\'o}mez}, D., {Soler}, R., \& {Terradas}, J. 2016, \apj,
  832, 101

\bibitem[{{Mestel} \& {Spitzer}(1956)}]{1956MNRAS.116..503M}
{Mestel}, L., \& {Spitzer}, Jr., L. 1956, \mnras, 116, 503

\bibitem[{Mitchner \& Kruger(1973)}]{1973MitchnerKruger}
Mitchner, M., \& Kruger, Charles~H, j.~a. 1973, Partially ionized gases (New
  York : Wiley), "A Wiley-Interscience publication."

\bibitem[{{Mouschovias} {et~al.}(2011){Mouschovias}, {Ciolek}, \&
  {Morton}}]{2011MNRAS.415.1751M}
{Mouschovias}, T.~C., {Ciolek}, G.~E., \& {Morton}, S.~A. 2011, \mnras, 415,
  1751

\bibitem[{{Mueller}(1974)}]{1974PlPh...16..813M}
{Mueller}, G. 1974, Plasma Physics, 16, 813

\bibitem[{{Oliver} {et~al.}(2016){Oliver}, {Soler}, {Terradas}, \&
  {Zaqarashvili}}]{2016ApJ...818..128O}
{Oliver}, R., {Soler}, R., {Terradas}, J., \& {Zaqarashvili}, T.~V. 2016, \apj,
  818, 128

\bibitem[{{Pandey} \& {Dwivedi}(2015)}]{2015MNRAS.447.3604P}
{Pandey}, B.~P., \& {Dwivedi}, C.~B. 2015, \mnras, 447, 3604

\bibitem[{{Pandey} \& {Wardle}(2006)}]{2006astro.ph..8008P}
{Pandey}, B.~P., \& {Wardle}, M. 2006, ArXiv Astrophysics e-prints,
  astro-ph/0608008

\bibitem[{{Pandey} \& {Wardle}(2008)}]{2008MNRAS.385.2269P}
---. 2008, \mnras, 385, 2269

\bibitem[{{Parnell} \& {De Moortel}(2012)}]{2012RSPTA.370.3217P}
{Parnell}, C.~E., \& {De Moortel}, I. 2012, Philosophical Transactions of the
  Royal Society of London Series A, 370, 3217

\bibitem[{{Piddington}(1956)}]{1956MNRAS.116..314P}
{Piddington}, J.~H. 1956, \mnras, 116, 314

\bibitem[{{Pinto} {et~al.}(2008){Pinto}, {Galli}, \&
  {Bacciotti}}]{2008A&A...484....1P}
{Pinto}, C., {Galli}, D., \& {Bacciotti}, F. 2008, \aap, 484, 1

\bibitem[{{Pudritz}(1990)}]{1990ApJ...350..195P}
{Pudritz}, R.~E. 1990, \apj, 350, 195

\bibitem[{{Rahbarnia} {et~al.}(2010){Rahbarnia}, {Ullrich}, {Sauer}, {Grulke},
  \& {Klinger}}]{2010PhPl...17c2102R}
{Rahbarnia}, K., {Ullrich}, S., {Sauer}, K., {Grulke}, O., \& {Klinger}, T.
  2010, Physics of Plasmas, 17, 032102

\bibitem[{{Scalo}(1977)}]{1977ApJ...213..705S}
{Scalo}, J.~M. 1977, \apj, 213, 705

\bibitem[{{Schunk}(1977)}]{1977RvGSP..15..429S}
{Schunk}, R.~W. 1977, Reviews of Geophysics and Space Physics, 15, 429

\bibitem[{{Soler} {et~al.}(2012){Soler}, {Andries}, \&
  {Goossens}}]{2012A&A...537A..84S}
{Soler}, R., {Andries}, J., \& {Goossens}, M. 2012, \aap, 537, A84

\bibitem[{{Soler} {et~al.}(2015{\natexlab{a}}){Soler}, {Ballester}, \&
  {Zaqarashvili}}]{2015A&A...573A..79S}
{Soler}, R., {Ballester}, J.~L., \& {Zaqarashvili}, T.~V. 2015{\natexlab{a}},
  \aap, 573, A79

\bibitem[{{Soler} {et~al.}(2013{\natexlab{a}}){Soler}, {Carbonell}, \&
  {Ballester}}]{2013ApJS..209...16S}
{Soler}, R., {Carbonell}, M., \& {Ballester}, J.~L. 2013{\natexlab{a}}, \apjs,
  209, 16

\bibitem[{{Soler} {et~al.}(2015{\natexlab{b}}){Soler}, {Carbonell}, \&
  {Ballester}}]{2015ApJ...810..146S}
---. 2015{\natexlab{b}}, \apj, 810, 146

\bibitem[{{Soler} {et~al.}(2013{\natexlab{b}}){Soler}, {Carbonell},
  {Ballester}, \& {Terradas}}]{2013ApJ...767..171S}
{Soler}, R., {Carbonell}, M., {Ballester}, J.~L., \& {Terradas}, J.
  2013{\natexlab{b}}, \apj, 767, 171

\bibitem[{{Soler} {et~al.}(2009{\natexlab{a}}){Soler}, {Oliver}, \&
  {Ballester}}]{2009ApJ...699.1553S}
{Soler}, R., {Oliver}, R., \& {Ballester}, J.~L. 2009{\natexlab{a}}, \apj, 699,
  1553

\bibitem[{{Soler} {et~al.}(2009{\natexlab{b}}){Soler}, {Oliver}, \&
  {Ballester}}]{2009ApJ...707..662S}
---. 2009{\natexlab{b}}, \apj, 707, 662

\bibitem[{{Soler} {et~al.}(2016){Soler}, {Terradas}, {Oliver}, \&
  {Ballester}}]{2016A&A...592A..28S}
{Soler}, R., {Terradas}, J., {Oliver}, R., \& {Ballester}, J.~L. 2016, \aap,
  592, A28

\bibitem[{{Song} \& {Vasyli{\= u}nas}(2011)}]{2011JGRA..116.9104S}
{Song}, P., \& {Vasyli{\= u}nas}, V.~M. 2011, Journal of Geophysical Research
  (Space Physics), 116, A09104

\bibitem[{{Spitzer}(1962)}]{1962pfig.book.....S}
{Spitzer}, L. 1962, {Physics of Fully Ionized Gases} (Interscience)

\bibitem[{{Stix}(1992)}]{1992wapl.book.....S}
{Stix}, T.~H. 1992, {Waves in plasmas} (American Institute of Physics)

\bibitem[{{Terradas} {et~al.}(2015){Terradas}, {Soler}, {Oliver}, \&
  {Ballester}}]{2015ApJ...802L..28T}
{Terradas}, J., {Soler}, R., {Oliver}, R., \& {Ballester}, J.~L. 2015,
  Astrophysical Journal Letters, 802, L28

\bibitem[{{Tu} \& {Song}(2013)}]{2013ApJ...777...53T}
{Tu}, J., \& {Song}, P. 2013, \apj, 777, 53

\bibitem[{{Vranjes} \& {Kono}(2014)}]{2014PhPl...21a2110V}
{Vranjes}, J., \& {Kono}, M. 2014, Physics of Plasmas, 21, 012110

\bibitem[{{Vranjes} \& {Krstic}(2013)}]{2013AA...554A..22V}
{Vranjes}, J., \& {Krstic}, P.~S. 2013, \aap, 554, A22

\bibitem[{{Watanabe}(1961{\natexlab{a}})}]{1961CaJPh..39.1197W}
{Watanabe}, T. 1961{\natexlab{a}}, Canadian Journal of Physics, 39, 1197

\bibitem[{{Watanabe}(1961{\natexlab{b}})}]{1961CaJPh..39.1044W}
---. 1961{\natexlab{b}}, Canadian Journal of Physics, 39, 1044

\bibitem[{{Watts} \& {Hanna}(2004)}]{2004PhPl...11.1358W}
{Watts}, C., \& {Hanna}, J. 2004, Physics of Plasmas, 11, 1358

\bibitem[{{Woods}(1962)}]{1962JFM....13..570W}
{Woods}, L.~C. 1962, Journal of Fluid Mechanics, 13, 570

\bibitem[{{Zaqarashvili} {et~al.}(2012){Zaqarashvili}, {Carbonell},
  {Ballester}, \& {Khodachenko}}]{2012A&A...544A.143Z}
{Zaqarashvili}, T.~V., {Carbonell}, M., {Ballester}, J.~L., \& {Khodachenko},
  M.~L. 2012, \aap, 544, A143

\bibitem[{{Zaqarashvili} {et~al.}(2011{\natexlab{a}}){Zaqarashvili},
  {Khodachenko}, \& {Rucker}}]{2011A&A...529A..82Z}
{Zaqarashvili}, T.~V., {Khodachenko}, M.~L., \& {Rucker}, H.~O.
  2011{\natexlab{a}}, \aap, 529, A82

\bibitem[{{Zaqarashvili} {et~al.}(2011{\natexlab{b}}){Zaqarashvili},
  {Khodachenko}, \& {Rucker}}]{2011A&A...534A..93Z}
---. 2011{\natexlab{b}}, \aap, 534, A93

\bibitem[{{Zaqarashvili} {et~al.}(2013){Zaqarashvili}, {Khodachenko}, \&
  {Soler}}]{2013A&A...549A.113Z}
{Zaqarashvili}, T.~V., {Khodachenko}, M.~L., \& {Soler}, R. 2013, \aap, 549,
  A113

\bibitem[{{Zweibel}(1989)}]{1989ApJ...340..550Z}
{Zweibel}, E.~G. 1989, \apj, 340, 550

\end{thebibliography}

\appendix
%\section{Collisional cross-sections} \label{app:cross-sections}
%	The following table shows the values of the cross-sections we have used in this paper to compute the friction coefficients of collisions that involve at least one neutral species. The column ``Model" refers to the paper where the data has been found, except when it reads ``Hard sphere"; in that case the cross-section has been computed through the following formula:
%	\begin{equation}
%		\sigma_{sn}=\pi \left(r_{s}+r_{n}\right)^2,
%	\end{equation}
%	where $r_{s}$ is the radius of the species $s$.
%	According to \citet{2013AA...554A..22V}, the cross sections $\sigma_{p\Rm{H}}$, $\sigma_{p\Rm{He}}$, $\sigma_{e\Rm{H}}$, and $\sigma_{e\Rm{He}}$ computed by those authors are between one or two orders of magnitude larger than if they were calculated using the hard sphere model. Hence, it may possible that more realistic values of $\sigma_{\Rm{HHe} \ \textsc{ii}}$, $\sigma_{\Rm{HHe} \ \textsc{iii}}$, and $\sigma_{\Rm{HeHe} \ \textsc{iii}}$ are also larger than the ones employed in this paper. However, we do not expect that those larger cross sections would modify significantly the results we have explained, because the dominant ion in the studied plasmas is the proton and not the singly or doubly ionized helium.

\section{Coefficients of the matrices $\Rm{A_{\pm}}$} \label{app:coefficients}
	\begin{equation} \label{eq:a11}
		A_{11,\pm}=(\omega \mp \Omega_{p}) \pm \frac{Z_{p}n_{p}\Omega_{p}}{n_{e}}+i(\nu_{p\Rm{H}}+\nu_{p\Rm{He}}+\nu_{p\Rm{He} \ \textsc{ii}}+\nu_{p\Rm{He} \ \textsc{iii}}+\nu_{pe})+i\frac{Z_{p}^{2}n_{p}^{2}e^{2}}{\rho_{p}}\eta-i\frac{2 Z_{p}n_{p}}{n_{e}}\nu_{pe}
	\end{equation}
	
	\begin{equation} \label{eq:a12}
		A_{12,\pm}=\pm \frac{Z_{\Rm{He} \textsc{ii}}n_{\Rm{He} \ \textsc{ii}}\Omega_{p}}{n_{e}}-i\nu_{p\Rm{He} \ \textsc{ii}}+i\frac{Z_{p}n_{p}Z_{\Rm{He} \ \textsc{ii}}n_{\Rm{He} \ \textsc{ii}}e^{2}}{\rho_{p}}\eta-i\frac{Z_{\Rm{He} \textsc{ii}}n_{\Rm{He} \ \textsc{ii}}}{n_{e}}\nu_{pe}-i\frac{Z_{p}n_{p}}{n_{e}}\frac{\alpha_{\Rm{He} \ \textsc{ii}}e}{\rho_{p}}
	\end{equation}
	
	\begin{equation} \label{eq:a13}
		A_{13,\pm}=\pm \frac{Z_{\Rm{He} \textsc{iii}}n_{\Rm{He} \ \textsc{ii}}\Omega_{p}}{n_{e}}-i\nu_{p\Rm{He} \ \textsc{iii}}+i\frac{Z_{p}n_{p}Z_{\Rm{He} \ \textsc{iii}}n_{\Rm{He} \ \textsc{iii}}e^{2}}{\rho_{p}}\eta-i\frac{Z_{\Rm{He} \textsc{iii}}n_{\Rm{He} \ \textsc{iii}}}{n_{e}}\nu_{pe} -i\frac{Z_{p}n_{p}}{n_{e}}\frac{\alpha_{\Rm{He} \ \textsc{iii}}e}{\rho_{p}}
	\end{equation}
	
	\begin{equation} \label{eq:a14}
		A_{14,\pm}=-i\nu_{p\Rm{H}}-i\frac{Z_{p}n_{p}}{n_{e}}\frac{\alpha_{e\Rm{H}}}{\rho_{p}}
	\end{equation}
	
	\begin{equation} \label{eq:a15}
		A_{15,\pm}=-i\nu_{p\Rm{He}}-i\frac{Z_{p}n_{p}}{n_{e}}\frac{\alpha_{e\Rm{He}}}{\rho_{p}}
	\end{equation}
	
	\begin{equation} \label{eq:a16}
		A_{16,\pm}=\frac{k_{x}\Omega_{p}}{en_{e}\mu_{0}} \pm i\frac{ek_{x}Z_{p}n_{p}}{\rho_{p}\mu_{0}}\eta \mp i\frac{k_{x}}{en_{e}\mu_{0}}\nu_{pe}
	\end{equation}
	
	\begin{equation} \label{eq:a21}
		A_{21,\pm}=\pm \frac{Z_{p}n_{p}\Omega_{\Rm{He} \ \textsc{ii}}}{n_{e}}-i\nu_{\Rm{He} \ \textsc{ii}p}+i\frac{Z_{p}n_{p}Z_{\Rm{He} \ \textsc{ii}}n_{\Rm{He} \ \textsc{ii}}e^{2}}{\rho_{\Rm{He} \ \textsc{ii}}}\eta-i\frac{Z_{p}n_{p}}{n_{e}}\nu_{\Rm{He} \ \textsc{ii}e}-i\frac{Z_{\Rm{He} \ \textsc{ii}}n_{\Rm{He} \ \textsc{ii}}}{n_{e}}\frac{\alpha_{pe}}{\rho_{\Rm{He} \ \textsc{ii}}}
	\end{equation}
	
	\begin{eqnarray} \label{eq:a22}
		A_{22,\pm}&=&(\omega \mp \Omega_{\Rm{He} \ \textsc{ii}}) \pm \frac{Z_{\Rm{He} \ \textsc{ii}} n_{\Rm{He} \ \textsc{ii}}}{n_{e}}+i(\nu_{\Rm{He} \ \textsc{ii}p}+\nu_{\Rm{He} \ \textsc{ii}\Rm{H}}+\nu_{\Rm{He} \ \textsc{ii}\Rm{He}}+\nu_{\Rm{He} \ \textsc{ii}\Rm{He} \ \textsc{iii}}+\nu_{\Rm{He} \ \textsc{ii}e}) \nonumber \\
		&+&i\frac{Z_{\Rm{He} \ \textsc{ii}}^{2}n_{\Rm{He} \ \textsc{ii}}^{2}e^{2}}{\rho_{\Rm{He} \ \textsc{ii}}}\eta-i\frac{2Z_{\Rm{He} \ \textsc{ii}}n_{\Rm{He} \ \textsc{ii}}}{n_{e}}\nu_{\Rm{He} \ \textsc{ii}e}
	\end{eqnarray}
	 
	\begin{eqnarray} \label{eq:a23}
		A_{23,\pm}&=&\pm \frac{Z_{\Rm{He} \ \textsc{iii}}n_{\Rm{He} \ \textsc{iii}}\Omega_{\Rm{He} \ \textsc{ii}}}{n_{e}}-i\nu_{\Rm{He} \ \textsc{ii}\Rm{He} \ \textsc{iii}}+ i\frac{Z_{\Rm{He} \ \textsc{ii}}n_{\Rm{He} \ \textsc{ii}}Z_{\Rm{He} \ \textsc{iii}}n_{\Rm{He} \ \textsc{iii}}e^{2}}{\rho_{\Rm{He} \ \textsc{ii}}}\eta \nonumber \\
		&-&i \frac{Z_{\Rm{He} \ \textsc{ii}}n_{\Rm{He} \ \textsc{ii}}}{n_{e}}\frac{\alpha_{\Rm{He} \ \textsc{iii}e}}{\rho_{\Rm{He} \ \textsc{ii}}}-i\frac{Z_{\Rm{He} \ \textsc{iii}}n_{\Rm{He} \ \textsc{iii}}}{n_{e}}\nu_{\Rm{He} \ \textsc{ii}e}
	\end{eqnarray}
	
	\begin{equation} \label{eq:a24}
		A_{24,\pm}=-i\nu_{\Rm{He} \ \textsc{ii}\Rm{H}}-i\frac{Z_{\Rm{He} \ \textsc{ii}}n_{\Rm{He} \ \textsc{ii}}}{n_{e}}\frac{\alpha_{e\Rm{H}}}{\rho_{\Rm{He} \ \textsc{ii}}}
	\end{equation}
	
	\begin{equation} \label{eq:a25}
	A_{25,\pm}=-i\nu_{\Rm{He} \ \textsc{ii}\Rm{He}}-i\frac{Z_{\Rm{He} \ \textsc{ii}}n_{\Rm{He} \ \textsc{ii}}}{n_{e}}\frac{\alpha_{e\Rm{He}}}{\rho_{\Rm{He} \ \textsc{ii}}}
	\end{equation}
	
	\begin{equation} \label{eq:a26}
		A_{26,\pm}=\frac{k_{x}\Omega_{\Rm{He} \ \textsc{ii}}}{e n_{e} \mu_{0}} \pm i\frac{ek_{x}Z_{\Rm{He} \ \textsc{ii}}n_{\Rm{He} \ \textsc{ii}}}{\mu_{0}\rho_{\Rm{He} \ \textsc{ii}}}\eta \mp i\frac{k_{x}}{en_{e}\mu_{0}}\nu_{\Rm{He} \ \textsc{ii}e}
	\end{equation}
	
	\begin{equation} \label{eq:a31}
		A_{31,\pm}=\pm \frac{Z_{p}n_{p}\Omega_{\Rm{He} \ \textsc{iii}}}{n_{e}}-i\nu_{\Rm{He} \ \textsc{iii}p} +i\frac{Z_{p}n_{p}Z_{\Rm{He} \ \textsc{iii}}n_{\Rm{He} \ \textsc{iii}}e^{2}}{\rho_{\Rm{He} \ \textsc{iii}}}\eta -i\frac{Z_{p}n_{p}}{n_{e}}\nu_{\Rm{He} \ \textsc{iii}e} -i\frac{Z_{\Rm{He} \ \textsc{iii}}n_{\Rm{He} \ \textsc{iii}}}{n_{e}}\frac{\alpha_{pe}}{\rho_{\Rm{He} \ \textsc{iii}}}
	\end{equation}
	
	\begin{eqnarray} \label{eq:a32}
		A_{32,\pm}&=& \pm \frac{Z_{\Rm{He} \ \textsc{ii}}n_{\Rm{He} \ \textsc{ii}}\Omega_{\Rm{He} \ \textsc{iii}}}{n_{e}}-i\nu_{\Rm{He} \ \textsc{iii} \Rm{He} \ \textsc{ii}} +i\frac{Z_{\Rm{He} \ \textsc{ii}}n_{\Rm{He} \ \textsc{ii}}Z_{\Rm{He} \ \textsc{iii}}n_{\Rm{He} \ \textsc{iii}}e^{2}}{\rho_{\Rm{He} \ \textsc{iii}}}\eta \nonumber \\
		&-&i\frac{Z_{\Rm{He} \ \textsc{ii}}n_{\Rm{He} \ \textsc{ii}}}{n_{e}}\nu_{\Rm{He} \ \textsc{iii}e} -i\frac{Z_{\Rm{He} \ \textsc{iii}}n_{\Rm{He} \ \textsc{iii}}}{n_{e}}\frac{\alpha_{\Rm{He} \ \textsc{ii}}e}{\rho_{\Rm{He} \ \textsc{iii}}}
	\end{eqnarray}
	
	\begin{eqnarray} \label{eq:a33}
		A_{33,\pm}&=&(\omega \mp \Omega_{\Rm{He} \ \textsc{iii}}) \pm \frac{Z_{\Rm{He} \ \textsc{iii}}n_{\Rm{He} \ \Rm{He} \ \textsc{iii}}\Omega_{\Rm{He} \ \textsc{iii}}}{n_{e}}+i(\nu_{\Rm{He} \ \textsc{iii}p}+\nu_{\Rm{He} \ \textsc{iii}\Rm{H}}+ \nu_{\Rm{He} \ \textsc{iii}\Rm{He}}+\nu_{\Rm{He} \ \textsc{iii} \Rm{He} \ \textsc{ii}}+\nu_{\Rm{He} \ \textsc{iii}e}) \nonumber \\
		&+&i\frac{Z_{\Rm{He} \ \textsc{iii}}^{2}n_{\Rm{He} \ \textsc{iii}}^{2}e^{2}}{\rho_{\Rm{He} \ \textsc{iii}}}\eta -i\frac{2Z_{\Rm{He} \ \textsc{iii}}n_{\Rm{He} \ \textsc{iii}}}{n_{e}}\nu_{\Rm{He} \ \textsc{iii}e}
	\end{eqnarray}
	
	\begin{equation} \label{eq:a34}
		A_{34,\pm}=-i\nu_{\Rm{He} \ \textsc{iii}\Rm{H}}-i\frac{Z_{\Rm{He} \ \textsc{iii}}n_{\Rm{He} \ \textsc{iii}}}{n_{e}}\frac{\alpha_{e\Rm{H}}}{\rho_{\Rm{He} \ \textsc{iii}}}
	\end{equation}

	\begin{equation} \label{eq:a35}
		A_{35,\pm}=-i\nu_{\Rm{He} \ \textsc{iii}\Rm{He}}-i\frac{Z_{\Rm{He} \ \textsc{iii}}n_{\Rm{He} \ \textsc{iii}}}{n_{e}}\frac{\alpha_{e\Rm{He}}}{\rho_{\Rm{He} \ \textsc{iii}}}
	\end{equation}
	
	\begin{equation} \label{eq:a36}
		A_{36,\pm}=\frac{k_{x}\Omega_{\Rm{He} \ \textsc{iii}}}{e n_{e}\mu_{0}} \pm i\frac{ek_{x}Z_{\Rm{He} \ \textsc{iii}}n_{\Rm{He} \ \textsc{iii}}}{\mu_{0}\rho_{\Rm{He} \ \textsc{iii}}}\eta \mp i\frac{k_{x}}{en_{e}\mu_{0}}\nu_{\Rm{He} \ \textsc{iii}e}
	\end{equation}
	
	\begin{equation} \label{eq:a41}
		A_{41,\pm}=-i\nu_{\Rm{H}p}-i\frac{Z_{p}n_{p}}{n_{e}}\nu_{\Rm{H}e}
	\end{equation}
	
	\begin{equation} \label{eq:a42}
		A_{42,\pm}=-i\nu_{\Rm{H}\Rm{He} \ \textsc{ii}}-i\frac{Z_{\Rm{He} \ \textsc{ii}}n_{\Rm{He} \ \textsc{ii}}}{n_{e}}\nu_{\Rm{H}e}
	\end{equation}
	
	\begin{equation} \label{eq:a43}
		A_{43,\pm}=-i\nu_{\Rm{H}\Rm{He} \ \textsc{iii}}-i\frac{Z_{\Rm{He} \ \textsc{iii}}n_{\Rm{He} \ \textsc{iii}}}{n_{e}}\nu_{\Rm{H}e}
	\end{equation}
	
	\begin{equation} \label{eq:a44}
		A_{44,\pm}=\omega+i(\nu_{\Rm{H}p}+\nu_{\Rm{H}\Rm{He}}+\nu_{\Rm{HHe} \ \textsc{ii}}+\nu_{\Rm{HHe} \ \textsc{iii}}+\nu_{\Rm{H}e})
	\end{equation}
	
	\begin{equation} \label{eq:a45}
		A_{45,\pm}=-i\nu_{\Rm{HHe}}
	\end{equation}
	
	\begin{equation} \label{eq:a46}
		A_{46,\pm}=\mp i\frac{k_{x}}{en_{e}\mu_{0}}\nu_{\Rm{H}e}
	\end{equation}
	
	\begin{equation} \label{eq:a51}
		A_{51,\pm}=-i\nu_{\Rm{He}p}-i\frac{Z_{p}n_{p}}{n_{e}}\nu_{\Rm{He}e}
	\end{equation}

	\begin{equation} \label{eq:a52}
		A_{52,\pm}=-i\nu_{\Rm{He}\Rm{He} \ \textsc{ii}}-i\frac{Z_{\Rm{He} \ \textsc{ii}}n_{\Rm{He} \ \textsc{ii}}}{n_{e}}\nu_{\Rm{He}e}
	\end{equation}

	\begin{equation} \label{eq:a53}
		A_{53,\pm}=-i\nu_{\Rm{He}\Rm{He} \ \textsc{iii}}-i\frac{Z_{\Rm{He} \ \textsc{iii}}n_{\Rm{He} \ \textsc{iii}}}{n_{e}}\nu_{\Rm{He}e}
	\end{equation}

	\begin{equation} \label{eq:a54}
		A_{54,\pm}=-i\nu_{\Rm{HeH}}
	\end{equation}

	\begin{equation} \label{eq:a55}
		A_{55,\pm}=\omega+i(\nu_{\Rm{He}p}+\nu_{\Rm{HeH}}+\nu_{\Rm{HeHe} \ \textsc{ii}}+\nu_{\Rm{HeHe} \ \textsc{iii}}+\nu_{\Rm{He}e})
	\end{equation}

	\begin{equation} \label{eq:a56}
		A_{56,\pm}=\mp i\frac{k_{x}}{en_{e}\mu_{0}}\nu_{\Rm{H}e}
	\end{equation}
	
	\begin{equation} \label{eq:a61}
		A_{61,\pm}=\frac{k_{x}B_{x}Z_{p}n_{p}}{n_{e}} \pm iek_{x}Z_{p}n_{p}\eta \mp i\frac{k_{x}}{en_{e}}\alpha_{pe}
	\end{equation}
	
	\begin{equation} \label{eq:a62}
		A_{62,\pm}=\frac{k_{x}B_{x}Z_{\Rm{He} \ \textsc{ii}}n_{\Rm{He} \ \textsc{ii}}}{n_{e}} \pm iek_{x}Z_{\Rm{He} \ \textsc{ii}}n_{\Rm{He} \ \textsc{ii}}\eta \mp i\frac{k_{x}}{en_{e}}\alpha_{\Rm{He} \ \textsc{ii}e}
	\end{equation}
	
	\begin{equation} \label{eq:a63}
		A_{63,\pm}=\frac{k_{x}B_{x}Z_{\Rm{He} \ \textsc{iii}}n_{\Rm{He} \ \textsc{iii}}}{n_{e}} \pm iek_{x}Z_{\Rm{He} \ \textsc{iii}}n_{\Rm{He} \ \textsc{iii}}\eta \mp i\frac{k_{x}}{en_{e}}\alpha_{\Rm{He} \ \textsc{iii}e}
	\end{equation}
	
	\begin{equation} \label{eq:a64}
		A_{64,\pm}=\mp i\frac{k_{x}}{en_{e}}\alpha_{e\Rm{H}}
	\end{equation}
	
	\begin{equation} \label{eq:a65}
		A_{65,\pm}=\mp i\frac{k_{x}}{en_{e}}\alpha_{e\Rm{He}}
	\end{equation}
	
	\begin{equation} \label{eq:a66}
		A_{66,\pm}=\omega \pm \frac{k_{x}^{2}B_{x}}{en_{e}\mu_{0}}+i\frac{k_{x}^{2}}{\mu_{0}}\eta
	\end{equation}
	
\end{document}